\documentclass[a4paper,showpacs,twocolumn,amsmath,amssymb,floatfix,prb,preprintnumbers,superscriptaddress,footinbib]{revtex4}

\usepackage[dvips]{graphicx}
\usepackage{amssymb}
\usepackage{dcolumn}
\usepackage{color}
\usepackage{hyperref}

\newcommand{\comment}[1]{}

\newcommand{\M}{\text{M}}
\newcommand{\W}{\text{W}}
\newcommand{\T}{\text{T}}
\newcommand{\D}{\text{D}}
\newcommand{\TD}{\text{TD}}
\newcommand{\SD}{\text{SD}}
\newcommand{\ST}{\text{ST}}
\newcommand{\DT}{\text{DT}}
\newcommand{\s}{\text{S}}
\newcommand{\Ts}{\text{TS}}
\newcommand{\G}{\text{G}}
\newcommand{\MT}{\text{M(T)}}
\newcommand{\ii}{\text{i}}
\newcommand{\ee}{\text{e}}
\newcommand{\A}{\text{A}}
\newcommand{\MM}{\text{M}}

\begin{document}

\title{Electron tunneling into a quantum wire in the Fabry-P\'erot regime}

\author{Stefano Pugnetti}
\author{Fabrizio Dolcini}
\affiliation{Scuola Normale Superiore and NEST CNR-INFM, I-56126
Pisa, Italy}
\author{Dario Bercioux}
\author{Hermann Grabert}
\affiliation{Physikalisches Institut and Freiburg Institute for
Advanced Studies, Universit\"at Freiburg, D-79104 Freiburg,
Germany}

\date{\today}

\begin{abstract}
We study a gated quantum wire contacted to source and drain 
electrodes in the Fabry-P\'erot regime. The wire is also coupled 
to a third terminal (tip), and we allow for an asymmetry of the tip tunneling amplitudes 
of right and left moving electrons. We analyze 
configurations~where~the~tip acts as an electron injector or as a 
voltage-probe, and show that the transport properties of this three-terminal set-up   exhibit   very rich physical behavior.  For a non-interacting wire we find that a tip in 
the voltage-probe configuration affects the source-drain 
transport in different ways, namely by suppressing the 
conductance, by modulating the Fabry-P\'erot oscillations, and by 
reducing their visibility. The combined effect of electron electron interaction and   finite length of the wire, accounted for by the inhomogeneous Luttinger liquid model, leads to significantly modified predictions as compared to  models based on infinite  wires. We show that when~the~tip injects electrons asymmetrically the charge fractionalization induced by interaction cannot be inferred from the asymmetry of the currents flowing in source and drain.
Nevertheless interaction effects are visible as oscillations in the non-linear 
tip-source and tip-drain conductances. 
Important differences with respect to a two-terminal set-up emerge, suggesting new strategies for the experimental investigation of Luttinger liquid behavior.

\end{abstract}


\maketitle
\section{Introduction}
Electron scanning of a conductor with a probe terminal is a
customary technique to investigate its local properties. The local
density of states can be gained from the dependence of the
tunneling current on the applied bias. Nowadays, atomically
resolved images are obtained both with scanning tunnel microscopes
(STM) and atomic force microscopes (AFM).\cite{NW-1} So far, most
of the efforts of the scientific community have focused on
improving the resolution power of the probe terminal. For
instance, the recent realization of stable and sharp
superconducting STM tips exploits the singularity in the
quasiparticle density of states to this purpose.\cite{NW-2} A
probe terminal, however, may also be used as a ``handle",
\emph{i.e.}\ as an active component to tune the transport
properties of the conductor. Recent works in this direction have shown that the sign of the
supercurrent can be changed when a third terminal injects
electrons into a Josephson junction under appropriate
conditions,\cite{NW-3} that the conductance of a quantum dot can be
tuned by moving an AFM tip over the sample,\cite{NW-4}  or that a single-electron transistor can be used to cool down  a nanomechanical resonator, or to drive it into a squeezed state.\cite{Schwab}

The promising applications of scanning probes in the study of
transport properties of nanodevices require a theoretical analysis
of electron transport in a three-terminal set-up, a subject which
has been explored only partly so far. In particular, most of the
available investigations are restricted to the case of
non-interacting conductors,\cite{NW-5,voltage-probe-refs} whereas relatively little
attention has been devoted to those nanodevices in which
electronic correlations play a dominant role. This is the case for
one-dimensional (1D) conductors, such as semiconductor heterostructure
quantum wires\cite{NW-6} and single-walled carbon
nanotubes.\cite{NW-7,liang:2001} There, electron-electron
interaction dramatically affects the dynamics of charge injection.
The response of the system to the scanning probe is quite
different from that of ordinary three dimensional metals, since in
1D electronic correlations lead to a breakdown of the
Fermi liquid picture. Semiconductor quantum wires and carbon
nanotubes rather exhibit Luttinger liquid (LL)
behavior.\cite{NW-8,yacoby,nanotube-LL,yamamoto:2007} While for
this type of systems two-terminal electron transport has been
widely analyzed in the last 15 years,\cite{NW-6,NW-7,NW-8,yacoby,nanotube-LL,liang:2001,yamamoto:2007} the electric
current and noise in a three-terminal set-up, including source and
drain electrodes and a tip, have remained mostly unexplored.

There are, however, a few notable works in this direction. The
case where a bias is applied between a tip and a semi-infinite LL
was investigated by Eggert,\cite{NW-9a} and by Ussishkin and
Glazman.\cite{NW-9b} Martin and co-workers\cite{martin:2003,martin:2005} have
recently analyzed the electric noise of the current injected from
a tip into a nanotube  adiabatically  contacted at each end  to grounded 
metallic leads.

In this paper we extend these investigations to a quite general three-terminal set-up. We shall thus explore the non-equilibrium current in all three terminals in presence of a transport voltage between the source
and drain electrodes, an applied tip voltage, and also a tunable gate voltage. This enables us to address various physical phenomena that are of relevance for recent experiments. Among other effects, we discuss the influence of the tip
on the transport along the interacting wire,  even when no net current is injected from the tip   into the wire. 
In particular, we focus on the Fabry-P\'erot transport regime of the wire,   which could be recently   observed   in carbon nanotubes\cite{liang:2001,yamamoto:2007,hakonen:2007,kontos:2007}, and analyze how Fabry-P\'erot oscillations are modified by both the presence of the tip  and the electron-electron interaction. To this purpose, the finite length of the wire,  the contact
resistances at the interfaces between the wire and the side
electrodes, as well as an arbitrary position of the tip
along the 1D wire are taken into account in our model. Furthermore, inspired by recent experiments on semiconductor quantum wires\cite{yacoby,yacoby:2}, we allow for an asymmetry in  electron tunneling from the tip, and investigate how the presence of side electrodes  affects the fractionalization of charges injected by the tip into an interacting  wire. Finally, regarding the experimental observation of interaction effects, we discuss  the advantages of a three-terminal set-up over a two-terminal one.

The paper is organized as follows. In Sec.~\ref{Sec2} we describe
the model that we adopt for the set-up. In Sec.~\ref{Sec3} we
provide results about the electric current in the case of a
non-interacting wire, while Sec.~\ref{Sec4} is devoted to the
effects of electron-electron interaction. Finally, we shall discuss the results in
Sec.~\ref{Sec5} and present our conclusions. Some more technical
details are given in the appendices.

\section{The model}\label{Sec2}
\begin{figure}
\centering
  \includegraphics[scale=0.4]{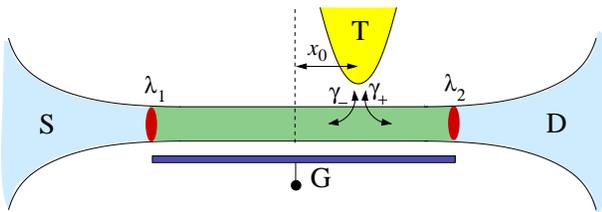}
\caption{\label{Fig-setup} (Color online) Sketch of the set-up. A
quantum wire is connected to two metallic electrodes, denoted as
source (S) and drain (D) at voltages $V_\s$ and $V_\D$,
respectively. A third sharp electrode, denoted as tip (T), at
voltage $V_\T$, injects electrons into the wire at position $x_0$.
A gate (G) is also present and held at a gate bias voltage $V_\G$.
The contact resistances are accounted for by two delta-like
scatterers with strengths $\lambda_1$ and $\lambda_2$, and the
electron tunneling amplitudes between the tip and the wire are
denoted by $\gamma_\pm$.}
\end{figure}
%
%

We consider a single channel spinless quantum wire connected, as
sketched in Fig.~\ref{Fig-setup}, to two metallic electrodes,
source (S) and drain (D), as well as to a third sharp electrode,
henceforth denoted as tip (T). The wire has a finite length $L$
and for the $x$ coordinate along it we choose the origin in the
middle of the wire so that the interfaces to the S and D
electrodes are located at $x_1=-L/2$ and $x_2=+L/2$, respectively.
Electron backscattering at the side contacts due to non-adiabatic
coupling is modeled by two delta-like scatterers. The tip is
described as a semi-infinite non-interacting Fermi liquid, and $y \le
0$ denotes the coordinate axis along the tip orthogonal to the
wire, the origin corresponding to the injection point on the tip.
The latter is located at position $x_0$ with respect to the middle
of the wire, and electron injection is modeled by a tunnel
Hamiltonian. We also envisage the presence of a metallic gate (G),
biased at a voltage $V_\G$. Screening by this gate yields an
effectively short-ranged electron-electron interaction potential
within the wire, for which the LL model applies.\cite{safi-schulz,maslov-stone,nota-short-range} The
total Hamiltonian of the system reads
\begin{equation}
\mathcal{H} = \mathcal{H}_\W \, + \, \mathcal{H}_\T \, + \,
\mathcal{H}_\text{tun} \label{HTOT}
\end{equation}
where the first term describes the wire and its coupling to the S
and D electrodes as well as to the gate. The second term accounts
for the tip, and the last one describes wire-tip
tunneling.

As far as the wire is concerned, we shall address here the
low-energy regime, where the wire electron band can be linearized
around the Fermi level. Then the wire electron operator $\Psi(x)$
can be decomposed into right- and left-moving components
$\Psi_{+}(x)$ and $\Psi_{-}(x)$
\begin{equation}
\Psi(x) = \ee^{+\ii k_\W x} \Psi_{+}(x) \, + \, \ee^{-\ii k_\W x}
\Psi_{-}(x) \label{Psi-split}
\end{equation}
where $k_\W$ denotes the equilibrium Fermi momentum of the wire.
By definition, this is the Fermi momentum in case that the
electrochemical potentials of all electrodes, source, drain, tip
and gate, are identical. This corresponds to vanishing applied voltages. Explicitly the Hamiltonian of the
wire reads
\begin{equation}
\mathcal{H}_\W = \mathcal{H}_\text{kin,W} \, + \,
\mathcal{H}_\lambda \, + \, \mathcal{H}_{\mu_\W} \, + \,
\mathcal{H}_{U} \label{HW}\,.
\end{equation}
In Eq.~(\ref{HW}) the first term
\begin{eqnarray}
\mathcal{H}_\text{kin,W} &=& -\ii \hbar v_\W \! \int_{-\infty}^{\infty} \!\!\! dx \left[ : \Psi^{\dagger}_{+}(x)\partial_x \Psi^{}_{+}(x) : - \right. \nonumber \\
& & \hspace{2cm} \left. - :\Psi^{\dagger}_{-}(x) \partial_x
\Psi^{}_{-}(x) : \right] \label{Hkin-W}
\end{eqnarray}
describes the band energy linearized around the wire Fermi points
$\pm k_\W$ and characterized by a Fermi velocity $v_\W$. The
symbol $: \, \, \, :$ stands for normal ordering with respect to
the equilibrium ground state. The second term models scatterers at
the interfaces~\cite{NW-15,recher-PRB} with the S and D electrodes
\begin{equation}
\mathcal{H}_{\lambda} = \hbar v_\W \sum_{i=1,2} \lambda_i
\rho(x_i)
\end{equation}
where the dimensionless parameters $\lambda_i \ge 0$ denote the
impurity strengths at the contacts $x_i$, and the term $\rho(x)$=
:$\Psi^{\dagger}(x)\Psi^{}(x)$: is the electron density
fluctuation with respect to the equilibrium value. The third term
in Eq.~(\ref{HW}),
\begin{equation}
\mathcal{H}_{\mu_\W} = \int_{-\infty}^{+\infty} \mu_\W(x) \rho(x)
\, dx \label{HmuW}
\end{equation}
with
\begin{equation}
\mu_\W(x)=
\begin{cases}
e V_\s & \text{for } x <-L/2 \\
e V_\G & \text{for } -L/2 < x < L/2 \\
e V_\D & \text{for } x > L/2
\end{cases}\, ,
\end{equation}
accounts for the bias $V_\s$ and $V_\D$ of the source and drain
electrodes, as well as for the gate voltage $V_\G$. The applied transport voltage is then $V=V_\s-V_\D$. Finally, the
last term
\begin{equation}
\mathcal{H}_{U} = \frac{U}{2} \int_{-L/2}^{L/2} \! \!
dx\sum_{r,r'=\pm} :\rho_{r}(x) \rho_{r'}(x): \label{HU}
\end{equation}
describes the screened Coulomb interaction in the
wire,\cite{safi-schulz,maslov-stone} where $\rho_r(x)$= :$\Psi^{\dagger}_r(x)
\Psi^{}_r(x)$: is the density fluctuation of $r$-moving electrons.
As it is customary in LL theory, in the sequel, we characterize
the interaction strength by the dimensionless coupling constant
%
%
\begin{equation}\label{g:deg}
g = \left(1+\frac{U}{\pi\hbar v_\W}\right)^{-\frac{1}{2}}\,.
\end{equation}
%
%

The Hamiltonian of the tip, the second term in Eq.~(\ref{HTOT}),
reads
\begin{equation}
\mathcal{H}_{\T}= \mathcal{H}_\text{kin,T}+\mathcal{H}_{\mu_\T}\,.
\label{HT}
\end{equation}
Here
\begin{equation}
\mathcal{H}_\text{kin,T}= -\ii \hbar v_\T \int_{-\infty}^{\infty}
\!\!\! dy \, \,: c^{\dagger}(y) \partial_y c^{}(y) :
\label{Hkin-T}
\end{equation}
describes the (linearized) band energy with respect to the
equilibrium Fermi points $\pm k_\T$ of the tip, and $v_\T$ denotes
the Fermi velocity. Notice that the integral runs also over the
positive $y$-axis, since right and left moving electron operators
along the physical tip axis $y<0$ have been unfolded into one
chiral (right-moving) operator $c^{}(y)$ defined on the whole
$y$-axis. The second term in Eq.~(\ref{HT}) describes the bias
$V_\T$ applied to the tip which affects the incoming electrons
according to
\begin{equation}
\mathcal{H}_{\mu_\T}= e V_\T \int_{-\infty}^{0} \!\!\! dy \, :
c^{\dagger}(y) c^{}(y): \quad . \label{Hmu-T}
\end{equation}
Finally, the third term in Eq.~(\ref{HTOT}) accounts for the
wire-tip electron tunneling and reads
\begin{equation}
\mathcal{H}_\text{tun}= \hbar \sqrt{v_\W v_\T} \sum_{r=\pm}
\gamma_r \left(\ee^{-\ii r k_\W x_0} \Psi^\dagger_r(x_0) c^{}(0)
+\rm{h.c.} \right) \label{Hgamma}
\end{equation}
where $\gamma_r$ is the  dimensionless tunneling amplitude for $r$-moving
electrons, and $x=x_0$ [$y=0$] is the coordinate of the injection
point along the wire [tip]. Here, we have allowed for a right/left
asymmetry of electron tunneling between the tip and the wire, 
which can arise from the presence of a  magnetic 
field\cite{yacoby,yacoby:2,yacoby-lehur}. Note that for $\gamma_+ 
\neq \gamma_{-}$ the Hamiltonian is not invariant under 
time-reversal symmetry.

In the following sections the electron current will be evaluated
in the three terminals of the described set-up. Explicitly we
shall compute
\begin{eqnarray}
\lefteqn{I(x,t) = e v_\W } & & \label{I-W} \\
& & \left\langle : \Psi^\dagger_+(x,t)
\Psi^{}_{+}(x,t):-:\Psi^\dagger_{-}(x,t) \Psi^{}_{-}(x,t) :
\right\rangle \nonumber
\end{eqnarray}
where $x$, with $|x|>L/2$, is a measurement point located in the S
or D leads. As far as the tip is concerned, due to the unfolding
procedure described above, the electron current flowing in the tip
at a point $y\le 0$ acquires the form
\begin{equation}
I(y,t) = e v_\T \left\langle :c^\dagger(y,t) c^{}(y,t):
-:c^\dagger(-y,t) c^{}(-y,t) : \right\rangle\,. \label{I-T}
\end{equation}
In Eqs.~(\ref{I-W}) and (\ref{I-T}) the averages are computed with
respect to the stationary state in presence of the applied dc
voltages $V_\s$, $V_\D$, $V_\T$ and $V_\G$. \\

Under these conditions, the current in each electrode is actually 
independent of the measurement point. We thus denote by $I_\s$ 
and $I_\D$ the currents flowing in the source and drain 
electrodes. The current $I_\s$ is positive when flowing into the 
wire, while $I_\D$ is positive when flowing out of the wire.  The 
current $I_\T$ flowing in the tip is  positive when flowing in 
the direction of the tip-wire tunnel contact. Current 
conservation then implies $I_\D=I_\s+I_\T$, so that all currents 
can be expressed in terms of two independent quantities. One can 
write
\begin{eqnarray}
I_{\s} &=&I_{\M}-I_{\T}/2 \label{ISDT-1}\\
I_{\D} &=&I_{\M}+I_{\T}/2 \label{ISDT-2}
\end{eqnarray}
where $I_{\M}$ describes the current flowing in the wire under 
the condition that no net current flows through the tip (voltage 
probe configuration). Importantly, $I_{\M}$ should {\it not} be 
identified with the two-terminal current flowing in the absence 
of the tip. Indeed, while $\gamma_{\pm}=0$ implies that $I_\T=0$, 
the opposite does not hold, so that $I_{\M}$ needs to be 
evaluated by accounting for the whole three-terminal set-up.

\section{The non-interacting case}\label{Sec3}

In this section we first discuss results for the case that the
electron interaction (\ref{HU}) is neglected. Then the Hamiltonian
(\ref{HTOT}) of the whole system is quadratic in the fields
$\Psi_{\pm}(x)$ and $c(y)$, and transport properties can be
determined within the Landauer-B\"uttiker formalism. In the
three-terminal set-up that we are considering, the scattering
matrix $\mathsf{S}(E)$ is a $3 \times 3$ matrix which depends on
the energy $E$ measured with respect to the equilibrium wire-lead
Fermi level. The currents
$I_{\M}$ and $I_{\T}$ defined through Eq.~(\ref{ISDT-1}) and
(\ref{ISDT-2}) read
\begin{eqnarray}
I_{\M} &=&\frac{e}{h} \left[ \frac{1}{2} \int_{-\infty}^{\infty}
\! \! \!
\left( |\mathsf{S}_{12}|^{2} + |\mathsf{S}_{21}|^{2}\right) \left( f_{\s}(E) - f_{\D}(E) \right) \, dE \, \nonumber \right. \\
& & + \frac{1}{2} \int_{-\infty}^{\infty} \! \! \! \,
|\mathsf{S}_{13} |^{2} \left( f_{\s}(E) \, - \, f_{\T}(E) \right)
\,
dE \label{IM} \\
& & \left. + \frac{1}{2} \int_{-\infty}^{\infty} \! \! \! \,
|\mathsf{S}_{23} |^{2} \left( f_{\T}(E) \, - \, f_{\D}(E) \right)
\, dE \right] \nonumber
\end{eqnarray}
and
\begin{eqnarray}
I_{\T} &=& \frac{e}{h} \int_{-\infty}^{\infty} \! \! dE \left[
|\mathsf{S}_{31}|^{2} \left(f_\T(E)-f_\s(E) \right) \right. \nonumber \\
& & \left. \hspace{1cm} +|\mathsf{S}_{32}|^{2} \left(f_\T(E)-
f_\D(E) \right) \right] \label{IT} .
\end{eqnarray}
In the $\mathsf{S}$-matrix elements appearing in Eqs.~(\ref{IM}) 
and (\ref{IT}) the source, drain and tip electrodes are 
identified as 1, 2, and 3 respectively, whereas their Fermi 
functions are denoted as $f_\s$, $f_\D$ and $f_\T$. Note that  
the $\mathsf{S}$-matrix is in general not symmetric, because 
time-reversal symmetry is broken for $\gamma_{+} \neq \gamma_{-}$.
The $\mathsf{S}$-matrix can straightforwardly be evaluated with
standard techniques by combining the transfer matrices
$\mathsf{M}_{x_i}$ ($i=1,2$) of the two side contacts
\begin{widetext}
%
%
\begin{align}
\mathsf{M}_{x_1,x_2} =  \label{Mi}
\begin{pmatrix}
\ee^{- \frac{\ii}{2}u_\G} (1-\ii \lambda_{1,2}) & \pm\ii \ee^{\pm\frac{\ii}{2}( 2\varepsilon-u_\G +2\kappa_\W)} \lambda_{1,2} & 0 \\
&&\\
\ii \ee^{\mp\frac{\ii}{2} (2\varepsilon-u_\G + 2\kappa_\W)} \lambda_{1,2} & \ee^{\frac{\ii}{2}u_\G} (1+\ii \lambda_{1,2}) & 0 \\
&&\\ 0 & 0 & 1
\end{pmatrix}
\end{align}
%
%
with the one, $\mathsf{M}_{x_0}$, at the tip injection point
\begin{equation}
\mathsf{M}_{x_0} = \frac{1}{1+ (\gamma_{+}^2 - \gamma_{-}^2)/4 }
\begin{pmatrix}
1- (\gamma_{+}^2 + \gamma_{-}^2)/4 &
- \ee^{-2\ii (\varepsilon + \kappa_\W - u_\G) \xi_0} \gamma_+
\gamma_{-} /2 &
\ \ - \ii \ee^{-\ii(\varepsilon + \kappa_\W - u_\G) \xi_0}
\gamma_+
\\&&\\
\ee^{2\ii (\varepsilon + \kappa_\W - u_\G) \xi_0} \gamma_{+}
\gamma_{-}/2 &
1+ (\gamma_{+}^2 + \gamma_{-}^2)/4 &
\ii \ee^{\ii (\varepsilon + \kappa_\W - u_\G) \xi_0} \gamma_- \\
&&\\
- \ii \ee^{\ii(\varepsilon + \kappa_\W - u_\G) \xi_0} \gamma_{+} &
- \ii \ee^{-\ii (\varepsilon + \kappa_\W - u_\G) \xi_0} \gamma_{-}
&
1-(\gamma_{+}^2 - \gamma_{-}^2)/4 \\
\end{pmatrix}\,.
\label{M0}
\end{equation}
\end{widetext}
Here, we have introduced the ballistic frequency
\begin{equation}\label{omegal}
\omega_L= \frac{v_\text{F}}{L}
\end{equation}
associated with the length of the wire, and the following
dimensionless quantities
\begin{subequations}
\begin{align}
\xi_0 & = \frac{x_0}{L}\,, \\
\kappa_\W & = k_\W L\,, \\
u_\G & = \frac{e V_\G}{\hbar \omega_L}\,, \\
\varepsilon & = \frac{E}{\hbar \omega_L}\,.
\end{align}
\end{subequations}
The scattering matrix is obtained as a combination of the elements
of the transmission matrix $\mathsf{M} = \mathsf{M}_{x_2}
\mathsf{M}_{x_0} \mathsf{M}_{x_1}$ in the form
\begin{eqnarray}
\mathsf{S}&=&\mathsf{M}_{22}^{-1} \label{S-matrix}\\
&& \nonumber \\
&\times& \begin{pmatrix}
- \mathsf{M}_{21} & 1 & - \mathsf{M}_{23} \\&&\\
\mathsf{M}_{11}\mathsf{M}_{22}-\mathsf{M}_{12}\mathsf{M}_{21} & \
\ \mathsf{M}_{12}\ \ &
\mathsf{M}_{13}\mathsf{M}_{22}-\mathsf{M}_{12}\mathsf{M}_{23}
\\&&\\
\mathsf{M}_{31}\mathsf{M}_{22}-\mathsf{M}_{21}\mathsf{M}_{32} &
\mathsf{M}_{32} &
\mathsf{M}_{33}\mathsf{M}_{22}-\mathsf{M}_{23}\mathsf{M}_{32}
\end{pmatrix}\nonumber
\end{eqnarray}
where $\mathsf{M}_{ij}$ are the matrix elements of $\mathsf{M}$.


\subsection{Fabry-P\'erot oscillations  in a two-terminal set-up}

Before discussing the influence of the STM tip, we shortly
describe the transport properties in the absence of the tip,
\emph{i.e.} for $\gamma_{\pm}=0$. In this case we have a
two-terminal set-up with $I_\T=0$ and $I_\s=I_\D=I_{\MM}$. The
solid line in Fig.~\ref{Fig-4} shows the two-terminal conductance
$dI_{\MM}/dV$ at zero temperature plotted in units of $e^2/h$ as a function of the
(dimensionless) source-drain bias
\begin{equation}
u=\frac{e(V_\s-V_\D)}{\hbar \omega_L} \label{u-def}
\end{equation}
for identical contact impurity strengths $\lambda_1=\lambda_2$. For $\lambda_i \ll 1$ the conductance shows the
typical Fabry-P\'erot oscillations with maximum values close to
one.
For carbon nanotubes the
Fabry-P\'erot regime of highly transparent contacts   could be reached experimentally only recently
due to progress achieved in device
contacting.\cite{liang:2001,yamamoto:2007,hakonen:2007,kontos:2007}
In the sequel, we will focus on this regime.

The electron current $I_\s=I_\D=I_{\M}$ can be written as
\begin{equation}
I_{\M}= I_{0} + I_{\rm imp} \label{IM_0+imp}
\end{equation}
where $I_{0}=(e^2/h) V$ represents the current of a perfectly
contacted wire, and $I_{\rm imp}$ characterizes the (negative)
correction due to the contact resistances. The exact expression
for $I_{\rm imp}$, which can be gained from the
$\mathsf{S}$-matrix, is not easily tractable for arbitrary
impurity strengths and temperature. In the Fabry-P\'erot regime at
zero temperature, however, a simpler expression is obtained by
expanding in terms of the impurity strengths. To third order in
the $\lambda_i$'s one obtains
\begin{equation}
I_{\rm imp}= \frac{e \omega_L}{2 \pi} \, ( j_{\rm inc} + j_{\rm
coh}) \label{Iimp}
\end{equation}
where $j_{\rm inc} $ and $j_{\rm coh}$ are dimensionless
quantities describing the incoherent and coherent contributions,
respectively, to the reduction of the current by the contact
impurities. The term
\begin{equation}
j_{\rm inc}= -\sum_{i=1,2} \lambda_i^2 \, u \,, \label{jinc}
\end{equation}
is linear in the applied bias voltage, and the coefficient of
proportionality is the ``classical" series resistance of two
impurities. In contrast, the term $j_{\rm coh}$ stems from quantum
interference between scattering processes. This interference leads
to the Fabry-P\'erot oscillations of $j_{\rm coh}$. Explicitly,
\begin{equation}
j_{\rm coh}= j_{\rm coh}^{(2)} \, + j_{\rm coh}^{(3)} \,
\label{Icoh}
\end{equation}
where
\begin{equation}
j_{\rm coh}^{(2)} = - 2 \lambda_1 \lambda_2 \cos\left[2(u_\W
+\kappa_\W -u_\G) \right] \sin(u) \label{j-coh-L2}
\end{equation}
and
\begin{equation}
j_{\rm coh}^{(3)} = - 2 \left( \lambda_1 \lambda_2^2 + \lambda_1^2
\lambda_2 \right) \sin\left[ 2(u_\W + \kappa_\W - u_\G)
\right]\sin(u)\, , \label{j-coh-L3}
\end{equation}
where we have introduced
\begin{equation}
u_\W=\frac{e(V_\s+V_\D)}{2\hbar \omega_L}\, \label{uW-def}.
\end{equation}
From Eqs. (\ref{j-coh-L2})-(\ref{j-coh-L3}) one can see that Fabry-P\'erot oscillations arise both as a function of the source-drain bias $u$ and as a function of the gate voltage $u_\G$. Note that for a non-interacting system the period in the former case is twice as large as the period in the latter case.

We also emphasize that $j_{\rm coh}^{(3)}$ originates from impurity
forward-scattering processes (more precisely from second order in
backward scattering and first order in forward scattering).
Forward scattering processes are typically neglected in single
impurity problems, where they can be gauged away. However, when
two or more impurities are present they affect the coherent part
of transport. Although this contribution is in general smaller
than $j_{\rm coh}^{(2)}$, it becomes the dominant term for the
Fabry-P\'erot oscillations when $j^{(2)}_{\rm coh}$ vanishes,
which is the case for
\begin{equation}
\frac{4}{\pi}\left( k_\W L + \frac{e (V_\s+V_\D-2 V_\G)}{2 \hbar
\omega_L} \right) \, \simeq \, \mbox{odd integer}\,.
\label{condition}
\end{equation}
Thus, the third order term is crucial for certain values of the
biasing voltage.

We conclude the discussion of the two-terminal case by emphasizing
that for a non-interacting wire in the Fabry-P\'erot regime the
current depends not only on the difference $V_\s-V_\D$, but in
general on $V_\s$ and $V_\D$ separately. This is simply due to the
fact that Fabry-P\'erot interference effects lead to an
energy-dependent transmission coefficient and, hence, to
non-linearity in the applied bias. Notice that Eqs.~(\ref{jinc}),
(\ref{j-coh-L2}) and (\ref{j-coh-L3}) fulfill the gauge-invariance
condition emphasized by B\"uttiker,\cite{buttiker:1993} since they
are invariant under an overall shift of the potentials $V_p
\rightarrow V_p + \mbox{const}$ ($p=\s,\D,\G$).

\subsection{Effect of the tip on Fabry-P\'erot oscillations}
In this section we shall address, within the non-interacting
electron approximation, the effect of the STM tip on the
Fabry-P\'erot oscillations. When $\gamma_{\pm}
\neq 0$, the currents $I_\MM$  and $I_\T$ are non-vanishing for
arbitrary values of the applied voltages $V_\s$, $V_\D$ and
$V_\T$.
We analyze the effects of the tip as a function of  the total
tunneling strength $\gamma$, defined through
\begin{equation}
\label{gamma-def}
\gamma^2= \frac{\gamma^2_{+}+\gamma^2_{-}}{2} \quad,
\end{equation}
the tunneling asymmetry coefficient
\begin{equation}
\label{chi-def}
\chi=\frac{\gamma^2_{+}-\gamma^2_{-}}{\gamma^2_{+}+\gamma^2_{-}}\, \hspace{1cm} |\chi| \le 1 \, \,, 
\end{equation}
and the position $x_0$ of the tip.\\
\begin{figure}
              \centering
            \includegraphics[scale=0.3]{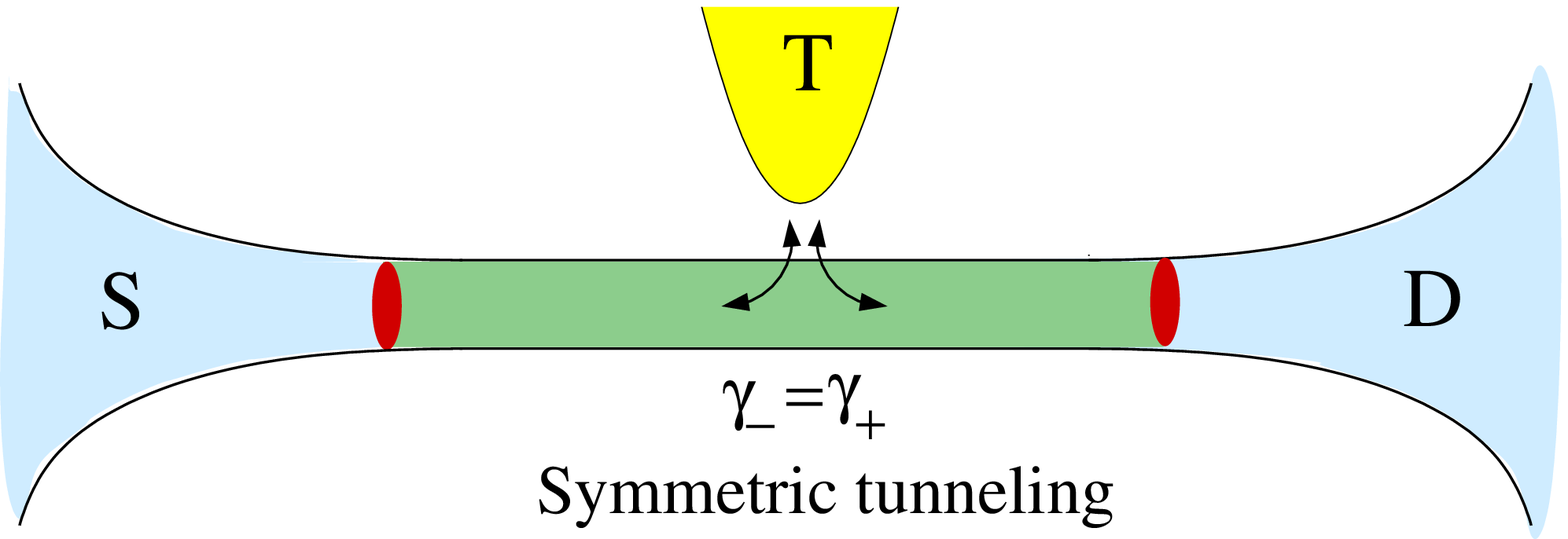}
          \includegraphics[width=\columnwidth,clip]{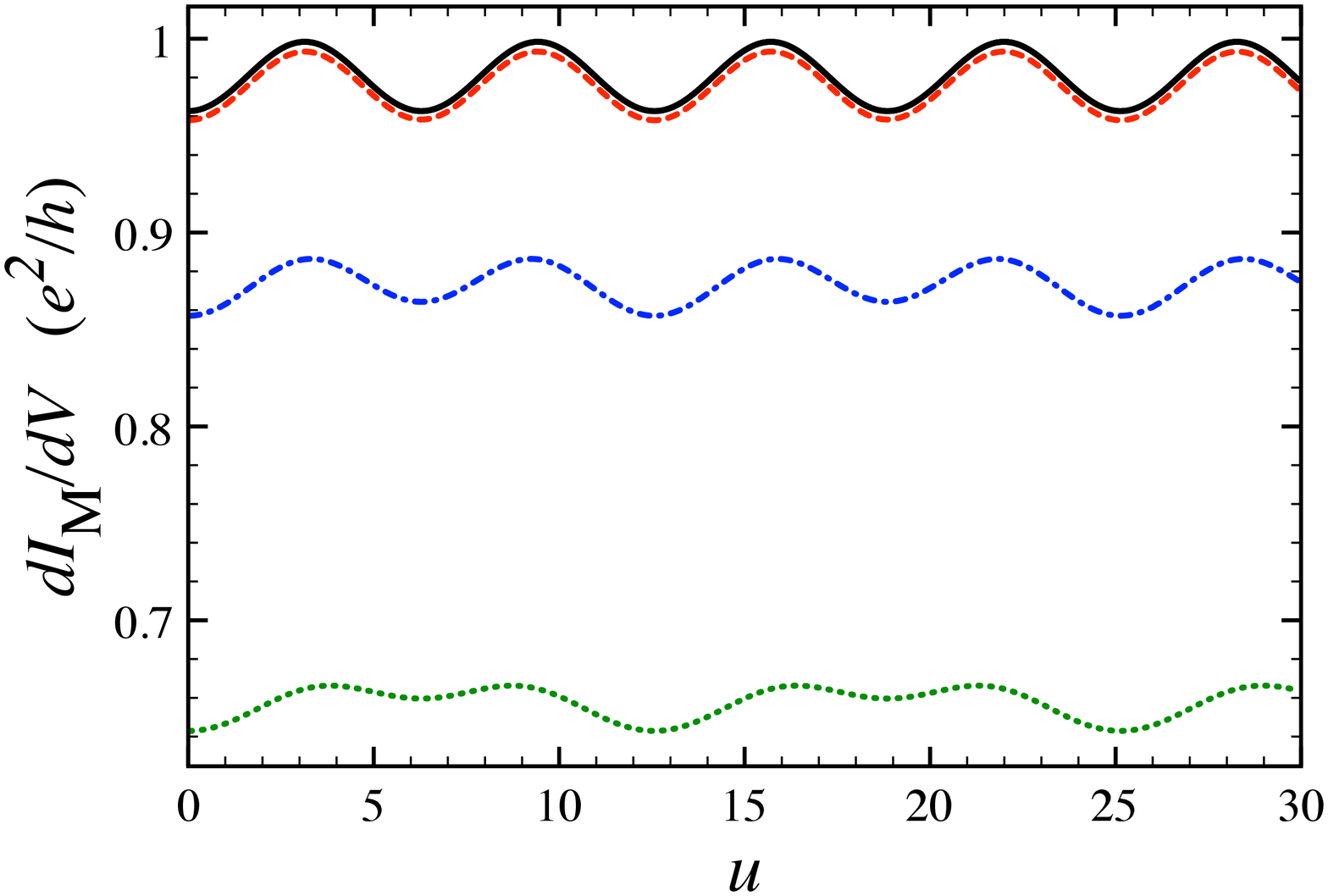}
\caption{\label{Fig-4} (Color online)  Zero temperature
differential conductance as a function of the source-drain bias
for a non-interacting wire characterized by contact impurity strengths 
$\lambda_1=\lambda_2=0.1$ and a Fermi wavevector   
$\kappa_\W=0.3$. The  tip is located in the middle of the wire 
and the tip voltage $V_\T$ is adjusted to fulfill the condition 
$I_\T=0$. Tunneling is symmetric ($\chi=0$) and the tunneling 
strength has the values: $\gamma=0$ (solid line), $\gamma=0.1$ 
(dashed-line), $\gamma=0.5$ (dashed-dotted line) and $\gamma=1$ 
(dotted line).  The gate is grounded ($V_\G=0$), and the bias is applied symmetrically ($V_{\s/\D}=\pm V/2$).}
\end{figure}
%
%

We start by considering the situation where the tip behaves as an 
electron injector: a bias is applied between the tip and the 
source and drain electrodes, which, for simplicity, are assumed 
to be at the same electrochemical 
potential. A quite standard calculation applies to the case of 
fully symmetric tunneling ($\chi=0$), allowing, {\em e.g.},  to 
relate the local density of states in the wire to the non-linear 
conductance as a function of the tip-wire bias. Here, we shall 
instead focus on the case of fully asymmetric tunneling 
($\chi=\pm 1$), which has become of particular interest due to 
recent experiments where only right-moving and/or only 
left-moving electrons could be selectively tunneled into a  
semiconductor quantum wire due to the presence of a magnetic field normal 
to the plane of the wire and the tip.\cite{yacoby:2} We find that 
novel physical aspects emerge from a tunneling asymmetry. In the 
first instance, a direct inspection of  the scattering matrix 
(\ref{S-matrix}) shows that its elements $\mathsf{S}_{ij}$ are 
independent of $x_0$, implying that, differently from the case of 
symmetric tunneling $\chi=0$, the lead currents $I_\D$ and $I_\s$ 
{\it do not} depend on the position of the tip. Furthermore, 
asymmetric tunneling can be used to extract the transmission 
coefficient  of each contact. Indeed evaluating   the asymmetry
\begin{equation}\label{cur-asym-def}
\mathcal{A}(\chi) \doteq \left.\frac{|I_\D|-|I_\s|}{|I_\D|+|I_\s|}\right|_{\chi}
\end{equation}
between $I_\D$ and $I_\s$ in the two cases of totally asymmetric 
injection only to the right ($\chi=1$) and only to the left 
($\chi=-1$), one  obtains
%
%
\begin{equation}\label{r:plus}
\mathcal{A}_+   = \frac{1+\lambda_1^2-\lambda_2^2}{1+\lambda_1^2+\lambda_2^2}\,
\end{equation}
%
and
\begin{equation}\label{r:minus}
\mathcal{A}_-  =  \frac{1+\lambda_2^2-\lambda_1^2}{1+\lambda_1^2+\lambda_2^2}\, .
\end{equation}
%
where $\mathcal{A}_{\pm}=\pm \mathcal{A}(\pm1)$.  From these coefficients it is straightforward to extract the strengths of the contact impurities
\begin{eqnarray}
\lambda_1^2&=&\frac{1-\mathcal{A}_-}{\mathcal{A}_+ + \mathcal{A}_-} \nonumber \\
& & \label{imp:exp} \\
\lambda_2^2&=&\frac{1-\mathcal{A}_+}{\mathcal{A}_+ + \mathcal{A}_-}  \nonumber \,.
\end{eqnarray}
as well as the transmission coefficients
\begin{equation}
\mathcal{T}_{1,2} \doteq \frac{1}{1+\lambda_{1,2}^2} = \frac{\mathcal{A}_{+}+\mathcal{A}_{-}}{1+\mathcal{A}_{\pm}} \label{trans-coeff-from-asym}
\end{equation}
related to each of the two contacts.
\\Notice that, while $I_{\s}$ and $I_{\D}$ depend on the temperature~$T$,
Eqs.~(\ref{r:plus}) and (\ref{r:minus}) are independent of $T$
within the approximation of a linearized band. 
 Interestingly, these equations also enable one to identify the relation between  the current asymmetry coefficients $\mathcal{A}_{\pm}$ and the two-terminal conductance $G_\text{2t} = \partial I_\M /\partial V |_{\gamma=0}$. In Ref. \onlinecite{yacoby-lehur}, the equality $\mathcal{A}_{\pm} = G_\text{2t}/(e^2/h)$ is
claimed to hold for a set-up with symmetric contacts to the leads,
even in the presence of interactions. However,
Eqs.~(\ref{r:plus}) and (\ref{r:minus}) show that for a
quantum wire in the Fabry-P\'erot regime, even in the absence of
interactions and with perfectly symmetric contacts $\lambda_1=\lambda_2$, one has
\begin{equation}
\mathcal{A}_+ =\mathcal{A}_{-} \neq G_\text{2t}/(e^2/h) \label{inequality}
\end{equation}
since $\mathcal{A}_\pm =1/(1+2\lambda_1^2)$ is a constant,
whereas $G_\text{2t}$ depends on temperature, source-drain bias and gate voltage.
The equality sign in Eq.~(\ref{inequality}) holds only
under the specific circumstances of
perfectly transmitting  contacts  ($\lambda_{1,2}=0$), or of
a  perfectly symmetric set-up  ($\lambda_1=\lambda_2 \neq 0$)  at sufficiently high temperatures $k_\text{B} T \gg \hbar \omega_L$, where Fabry-P\'erot oscillations of $G_\text{2t}$ wash out.   \\

The second situation that we want to investigate is
when the tip voltage $V_\T$ is set to an appropriate value
$\bar{V}_\T$ so that no net current flows through the tip. This corresponds to a situation where the tip behaves as  a voltage probe\cite{NOTA-voltageprobe}. Notice that, even under the condition $I_\T=0$,
electrons can tunnel from the tip to the wire and  vice
versa, and therefore the tip {\em does} affect the electron transport
between source and drain.\\
%

\begin{figure}
\centering
\includegraphics[scale=0.3]{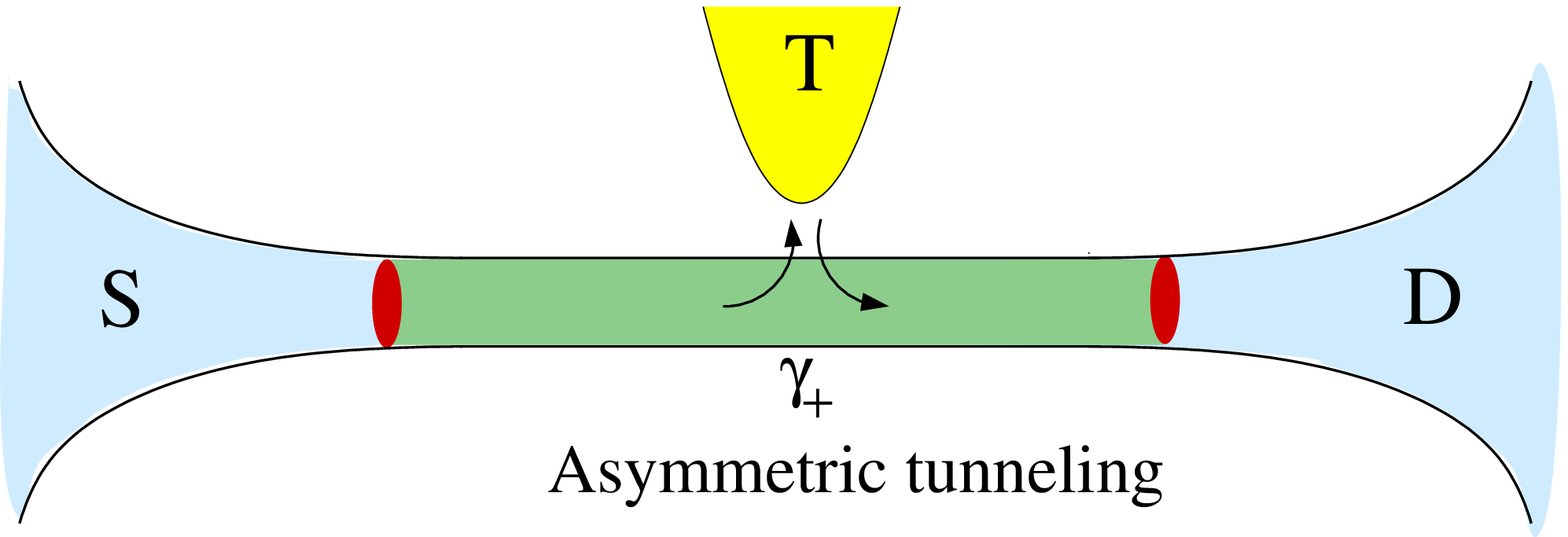}
\includegraphics[width=\columnwidth,clip]{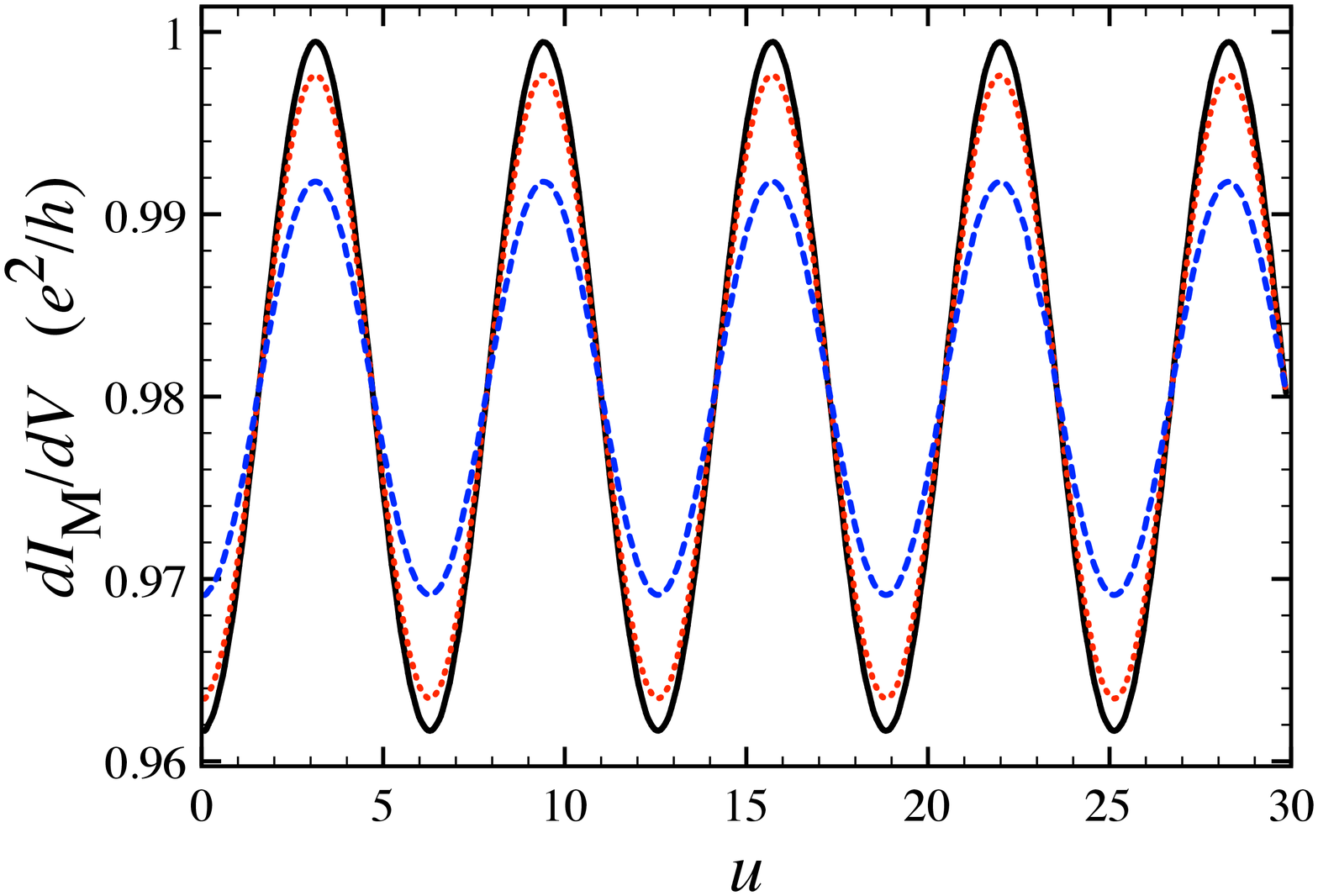}
\caption{\label{fig:four} (Color online) Zero temperature
differential conductance as a function of the source-drain bias
for a  non-interacting wire with  contact impurity strengths $\lambda_1=\lambda_2=0.1$ in
presence of a tip with an applied voltage $V_\T$ adjusted to
fulfill the condition $I_\T=0$. Tunneling is totally asymmetric
($\chi=1$) and the tunneling strength has the values
$\gamma=0.3$ (dotted line) and $\gamma = 0.7$ (dashed line).
The solid line represents the case with $\gamma=0$. The result is independent of
the tip position $x_0$.  The gate is grounded ($V_\G=0$), and the bias is applied symmetrically ($V_{\s/\D}=\pm V/2$).}
\end{figure}

We start by describing the  case of symmetric tunneling 
($\chi=0$) with the tip located in the middle of the wire 
($x_0=0$). The differential conductance $dI_\MM/dV$,  evaluated 
under the condition $I_\T=0$, is depicted in Fig.~\ref{Fig-4} as 
a function of the source-drain bias (\ref{u-def}), for different 
values of $\gamma$, ranging from weak to strong tunneling. The 
tip has three main effects on the Fabry-P\'erot oscillations: i) 
an overall suppression of the conductance,   ii) a modulation of 
the  maxima and minima, and iii) a reduction of the   visibility 
of the oscillations. 

The origin of the first effect can  be illustrated already in 
the  case of a clean wire ($\lambda_i=0$),  where  it is  easy to 
show that the condition $I_\T=0$ is fulfilled for a tip voltage 
$\bar{V}_\T=(V_\s+V_\D)/2$, and that
\begin{equation}
I_{\M} = \frac{e^2}{h}  \frac{V_\s - V_\D}{1+ \gamma^2/2} \, \, < \, \frac{e^2}{h} (V_\s - V_\D), \label{dep:zero}
\end{equation}
Notice that a reduction of the conductance already shows up to 
order $\gamma^2$ in the tunneling strength. The reason for this 
suppression of the current is that a fraction of the electron 
flow originating from the source is diverted into the tip due to 
the tip-wire coupling. While the condition $I_\T=0$ ensures that 
the same electron current is re-injected into the wire, for 
symmetric tunneling the tip injects with equal probabilities   
right and  left moving electrons. Hence half of the injected 
current  flows back to the source electrode, causing the 
reduction of the two-terminal conductance. As we shall see below, 
the situation is different in the case of asymmetric tunneling. 

The second feature that can be observed   in Fig.~\ref{Fig-4} is 
an alternating depth of the Fabry-P\'erot minima. This modulation 
originates from the interference between different paths that are 
possible for an electron ejected from the tip. For instance, the 
path of an electron ejected as  right mover towards the drain 
can  interfere with the path starting as left mover towards the 
source followed by an elastic backscattering at the source 
contact. The difference in length between these paths corresponds 
to a new frequency in the oscillations, which causes the 
modulation of the peaks. In the case of Fig.~\ref{Fig-4}, where 
the tip is located in the middle, this additional frequency 
equals twice the Fabry-P\'erot frequency, so that the tip affects 
every second minimum in the same way. As we shall see below, in 
general, the modulation pattern depends both on the asymmetry 
coefficient and on the  position of the tip. The modulation 
effect arises to order $\gamma^2 \lambda$ when we treat the 
impurity strength and tunneling amplitudes as perturbation parameters.  

%

The third effect of the tip consists in a reduction of the 
visibility of the the Fabry-P\'erot oscillations: in the presence 
of the tip the relative separation between  maxima and minima 
decreases. This reduction stems from the decoherence introduced 
by the tip, since   the probability of constructive interference 
between paths with two backscattering processes at the contacts 
decreases when electrons can be incoherently absorbed and 
re-ejected by the tip. Notice that the reduction of visibility is 
of order $\gamma^2 \lambda^2$, and it is therefore negligible 
with respect to the modulation effect in the Fabry-P\' erot 
regime.

\begin{figure}
  \centering
\includegraphics[scale=0.3]{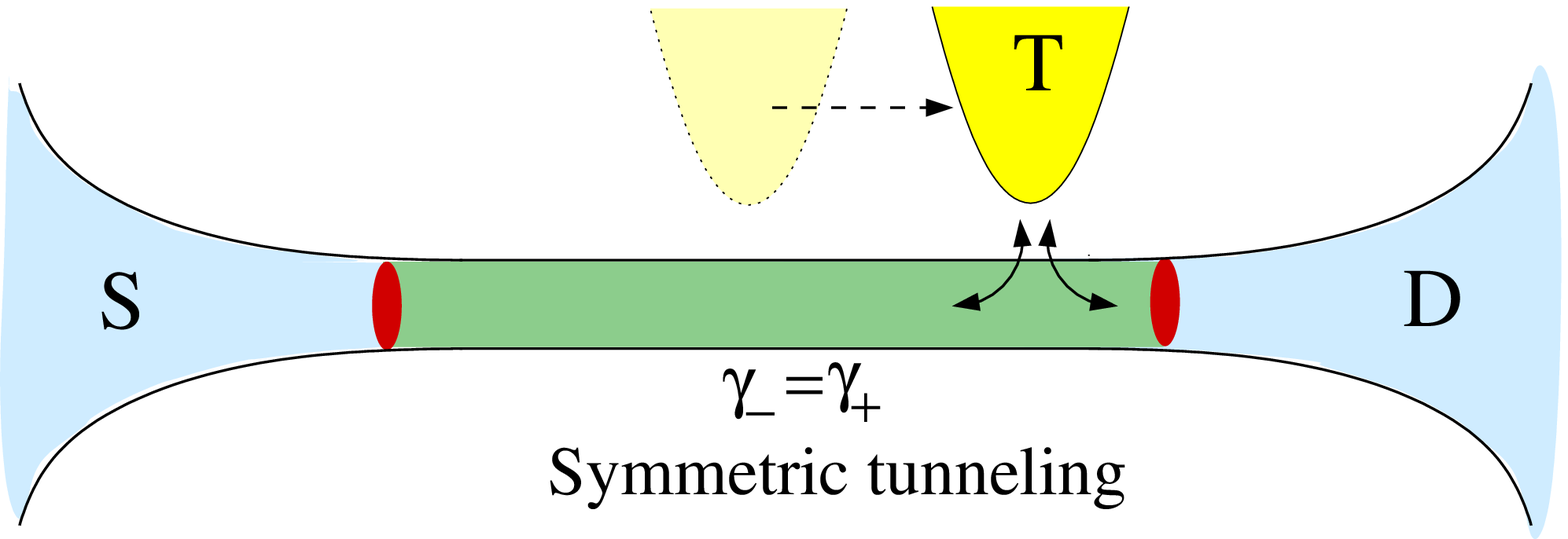}
   \includegraphics[width=\columnwidth]{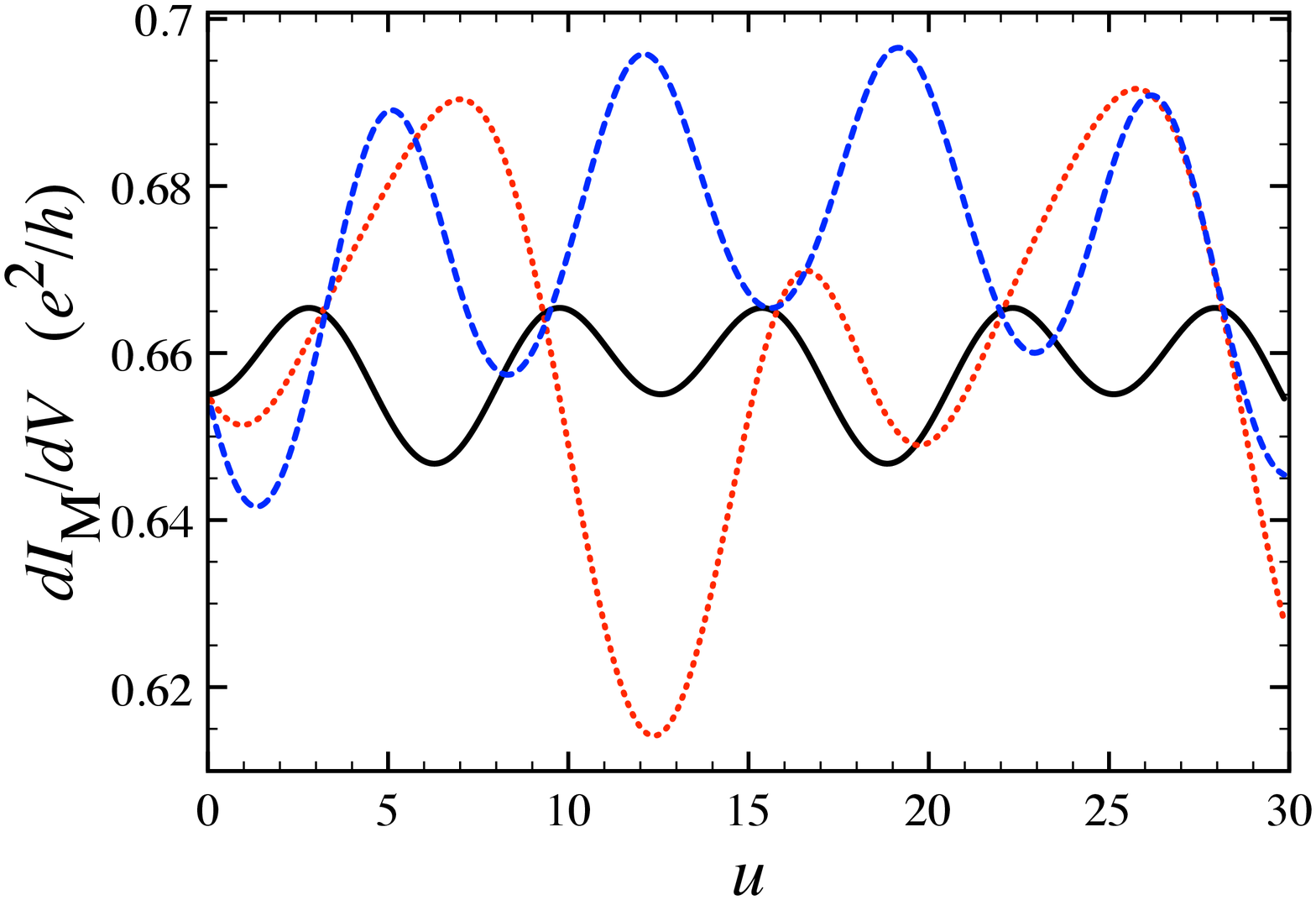}
   \caption{\label{voltage:probe:position} (Color online)
    Zero temperature  differential conductance as a function of the source-drain voltage
   for a non-interacting  wire
   for several values of the tip position $x_0=0$ (solid line),
   $x_0=0.17$ (dotted line) and $x_0 = 0.41$ (dashed line).
   Tunneling has amplitude $\gamma=1$ and is symmetric ($\chi=0$).
   The contact impurities have equal strengths $\lambda_1=\lambda_2=0.1$.  The gate is grounded ($V_\G=0$), and the bias is applied symmetrically ($V_{\s/\D}=\pm V/2$).}
\end{figure}

Let us now discuss the role of asymmetric tunneling in the voltage probe configuration. 
When $\chi \neq 0$, the effect of  conductance suppression is 
less pronounced then for symmetric tunneling. This can be seen 
already in the case of a clean wire ($\lambda_i=0$), where
\begin{equation}
I_{\M} = \frac{e^2}{h} (V_\s - V_\D) \frac{2+\gamma^2
\chi^2}{2+\gamma^2}\,, \label{dep:one}
\end{equation}
and the value $\bar{V}_\T$ of the tip voltage  ensuring $I_\T=0$ is given by
\begin{equation}
\bar{V}_\T = \frac{1}{2}\left[ V_\s (1+\chi) + V_\D
(1-\chi) \right]\,. \label{vtipstar}
\end{equation}
As one can see from the last factor in Eq.~(\ref{dep:one}), the 
suppression  of the current $I_\MM$ is completely absent for 
fully asymmetric tunneling $\chi=\pm1$. Importantly, this 
features persists also in the presence of realistic contacts  
($\lambda_i \neq 0$), as shown in Figure~\ref{fig:four}, where 
the differential conductance $dI_\MM/dV$  is plotted as a 
function of the source-drain voltage for several values of the 
tunneling strength~$\gamma$. Increasing the tunneling strength 
simply decreases the amplitude of the Fabry-P\'erot oscillations 
but does not change the average value of the conductance. Two 
more noteworthy features can be observed: in the fully asymmetric 
case also the modulation of the peaks is  absent, and the 
non-linear conductance is independent of the tip position. The 
reason lies in the specific tunneling conditions. For example, a 
right moving electron ejected by the tip cannot be re-adsorbed 
after scattering as a left moving one, and this rules out 
interference effects between electrons traveling through the tip 
and electrons that have undergone an odd number of backscattering 
events at the contacts. Such processes would give rise to effects 
related to the tip position, while interference phenomena with 
electrons that have undergone an even number of backscattering 
events, which continue to be present also for $\chi=\pm 1$, are 
independent of the tip position. Moreover, in the completely 
asymmetric case, electrons passing through the tip continue to 
move in the same direction, and this is the reason why, also for 
strong tunneling, the average value of the
differential conductance is independent of $\gamma$.\\

Finally, we analyze the dependence of the differential 
conductance on the tip position. For simplicity we limit this 
discussion to the  case of symmetric tunneling  illustrated in 
Fig.~\ref{voltage:probe:position}. Apart from the  conductance 
suppression discussed above, one sees that the modulation effect 
exhibits a strong dependence on the tip position. In particular, 
when the tip is close to a contact impurity, we observe 
Fabry-P\'erot-like oscillations over-imposed by an oscillation 
with large period due to coherent motion of carriers between the 
tip and the contact impurity remote from the tip.

\section{The interacting case}\label{Sec4}

In this section we discuss the three terminal set-up
in presence of electron-electron interaction. For arbitrary values
of the interaction strength, contact resistances, and tunneling
amplitudes an analytical treatment is not possible, therefore we
focus here on the Fabry-P\'erot regime. In this regime,
characterized by highly transparent contacts to the electrodes,
the role of interactions has so far only been analyzed for a two
terminal set-up.\cite{NW-15,recher-PRB} Since the impurity strengths
$\lambda_i$ are small, they can be treated perturbatively. The
electron-electron interaction~(\ref{HU}) will be accounted for
exactly using bosonization. The evaluation of the currents in the
three terminals will be based on the out-of-equilibrium Keldysh
formalism.\cite{KEL} We shall first discuss the effects of
electron-electron interaction for the two-terminal set-up in the
Fabry-P\'erot regime, \emph{i.e.}\ in the absence of the tip, and
then turn to the combined effect of tip and electronic
correlations.

\begin{figure}
  \centering
\includegraphics[scale=0.3]{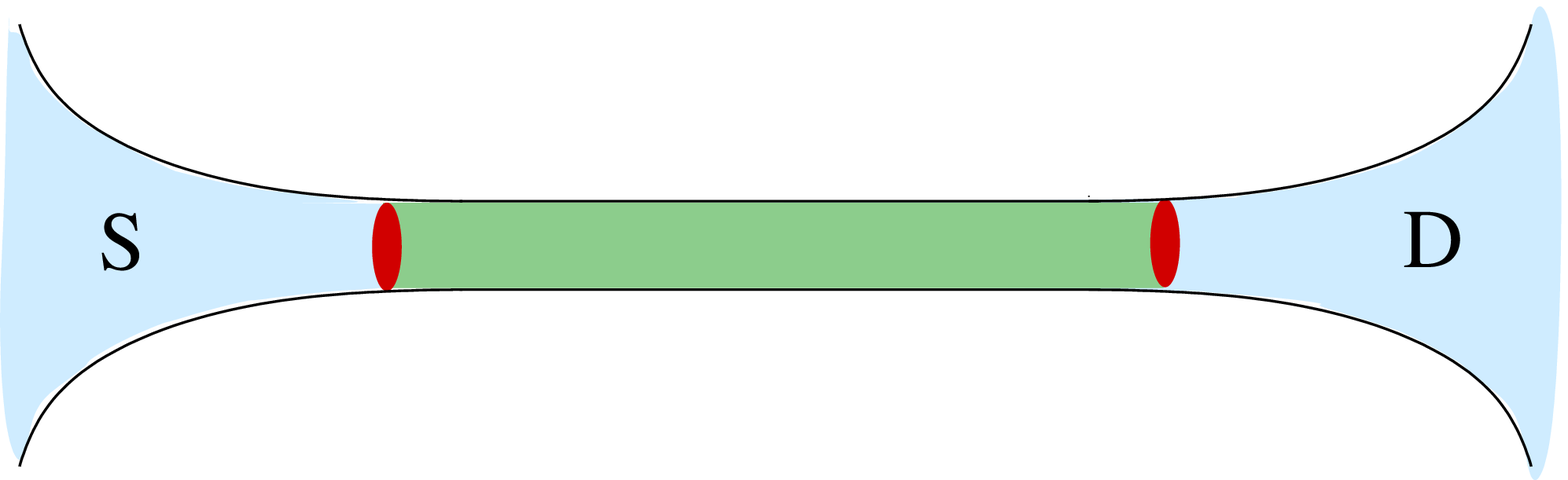}
   \includegraphics[width=\columnwidth]{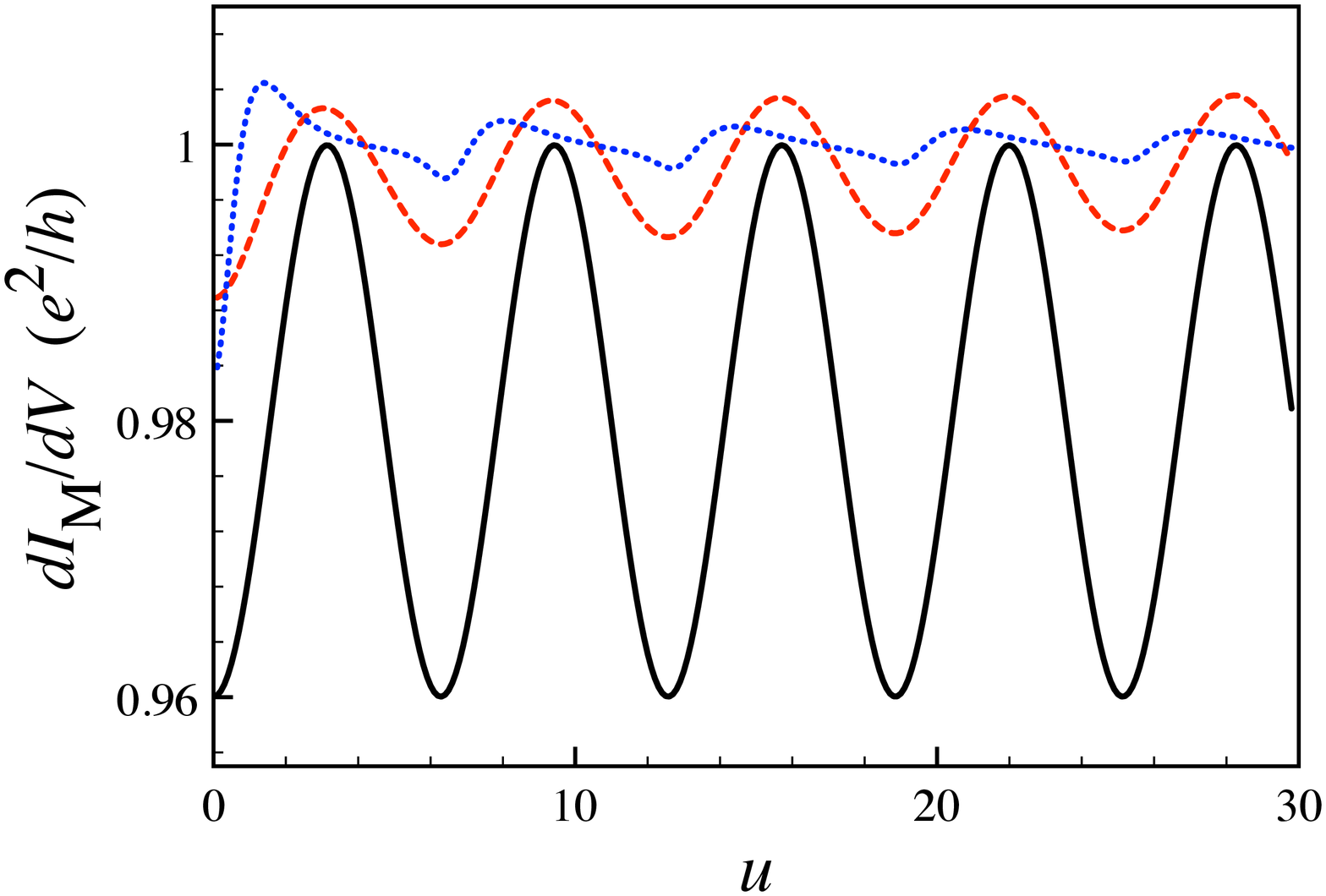}
   \caption{\label{fig:five} (Color online)
   Differential conductance of the two-terminal set-up (in the absence of a tip) as a function of the source-drain voltage
   for several values of the interaction parameter $g=1$ (solid line),
   $g=0.75$ (dashed line) and $g = 0.25$ (dotted line).
   The contact impurities have equal strengths $\lambda^*_{\rm B,1}=\lambda^*_{\rm B,2}=0.1$. The gate voltage is $V_\G=0$,  and   the bias is applied symmetrically ($V_{\s/\D}=\pm V/2$).}
\end{figure}

\begin{figure*}
 \begin{minipage}[c]{0.45\textwidth}
     \includegraphics[width=\textwidth]{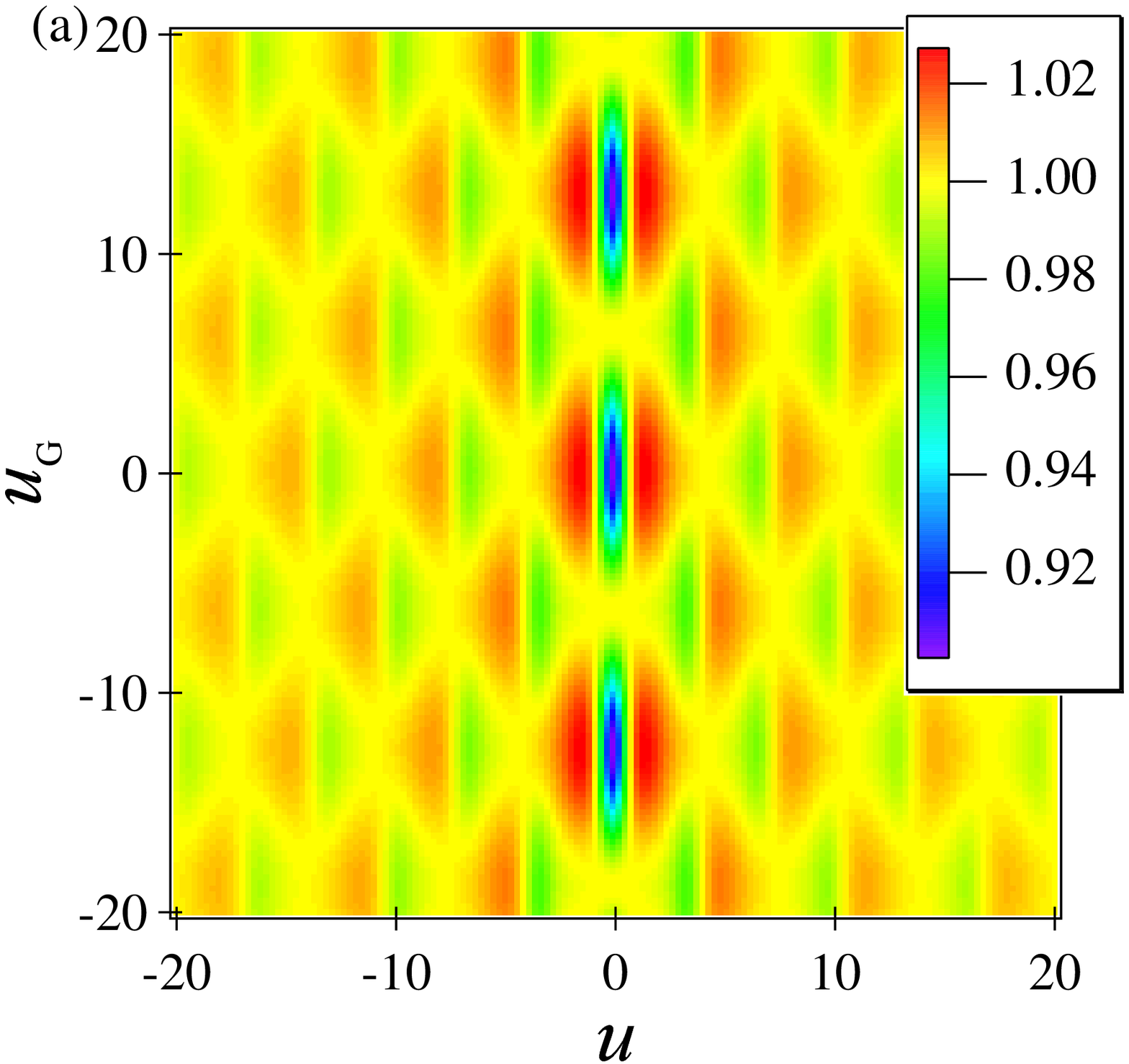}
 \end{minipage}
 \begin{minipage}[c]{0.45\textwidth}
     \includegraphics[width=\textwidth]{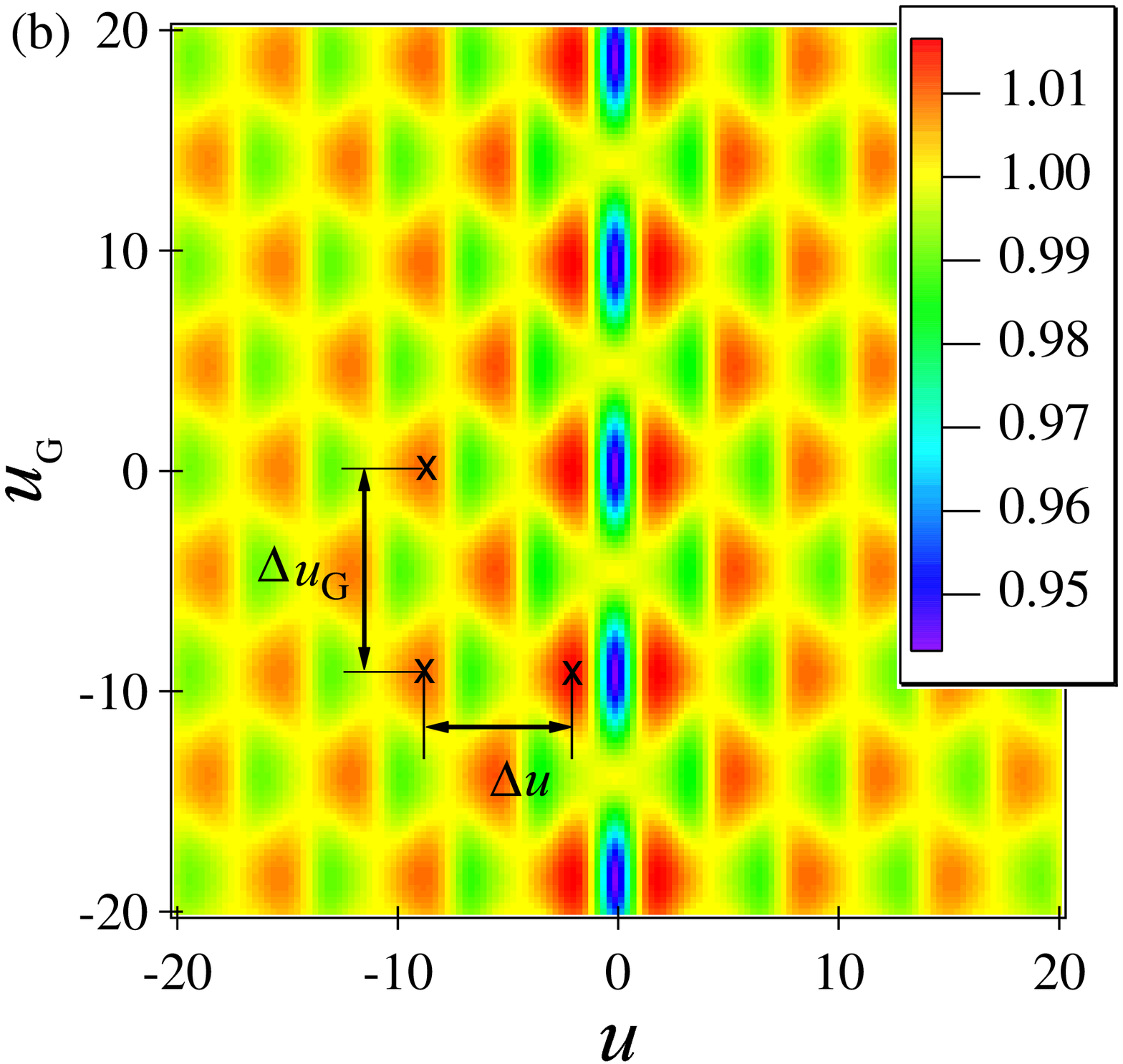}
 \end{minipage}
 \begin{minipage}[c]{0.45\textwidth}
     \includegraphics[width=\textwidth]{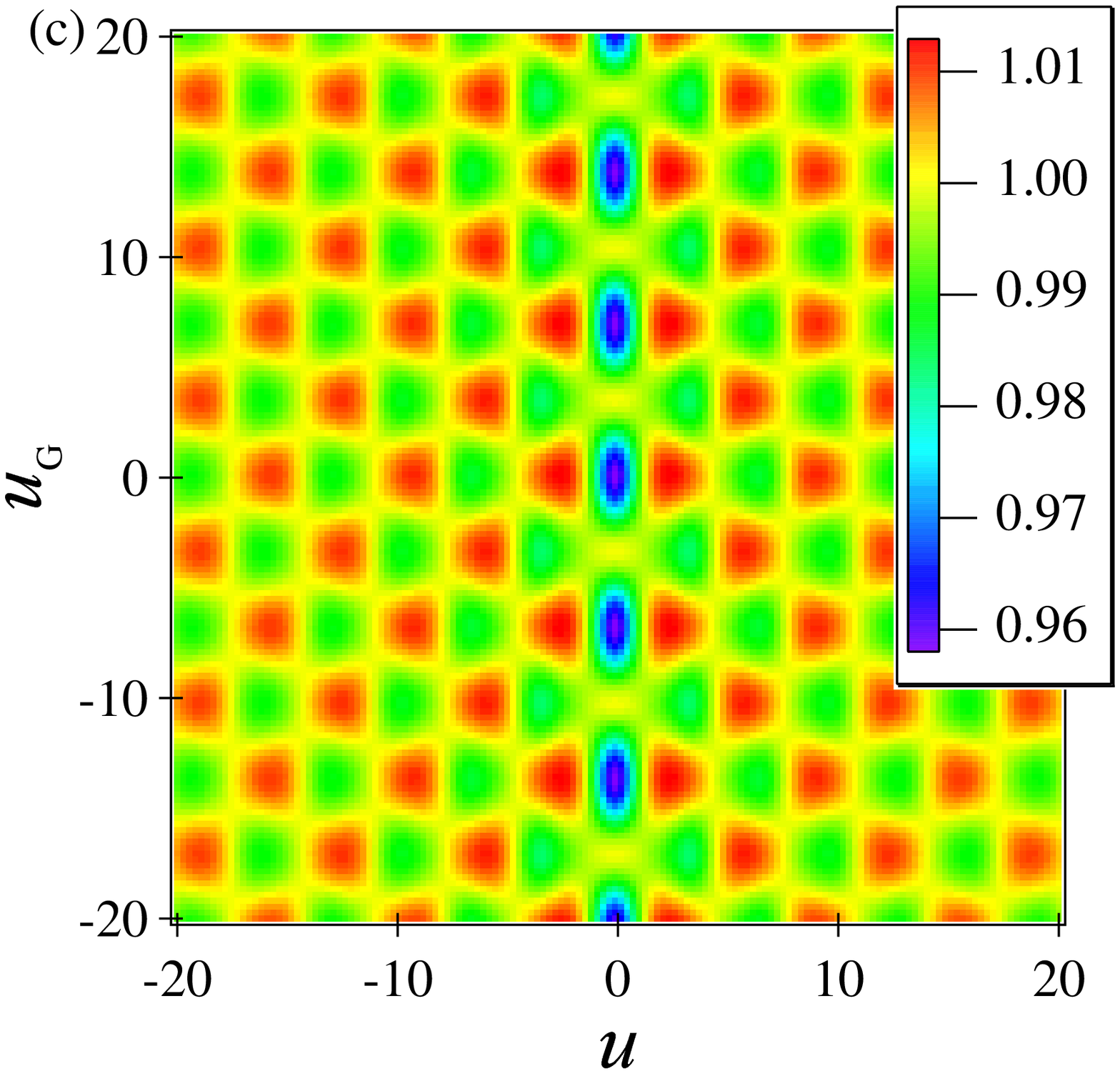}
 \end{minipage}
 \begin{minipage}[c]{0.45\textwidth}
  \includegraphics[width=\textwidth]{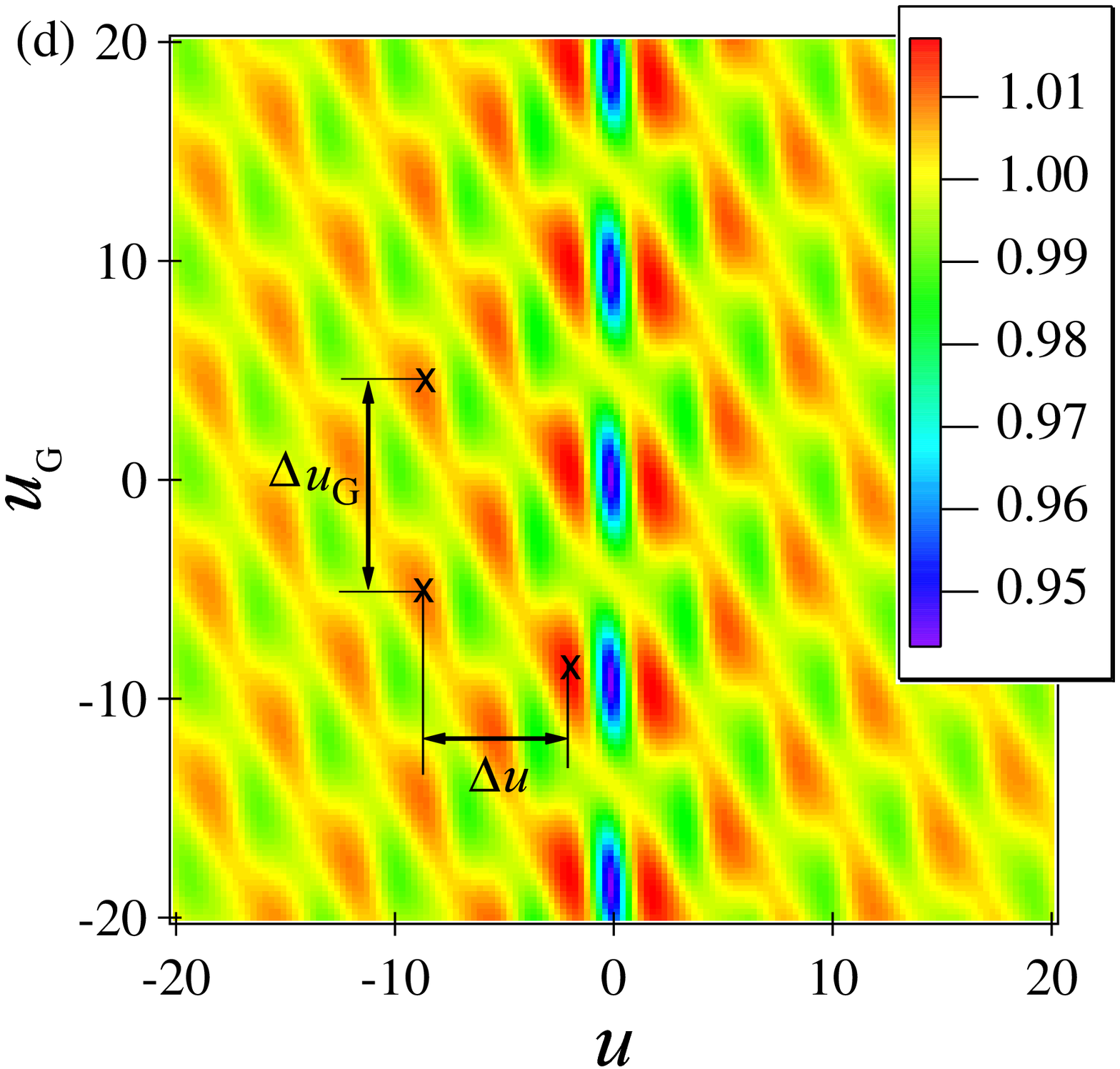}

 \end{minipage}
 \begin{minipage}[c]{0.45\textwidth}
  \includegraphics[scale=0.3]{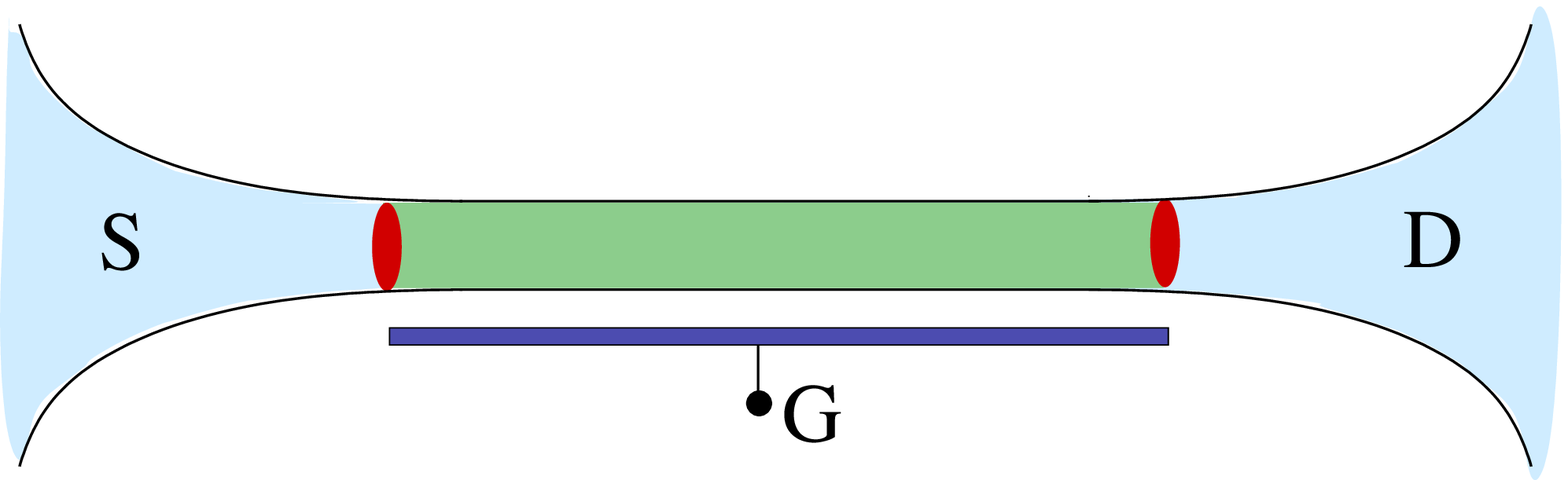}

 \end{minipage}
\begin{minipage}[c]{0.4\textwidth}
 \begin{ruledtabular}
 \begin{tabular}{r | r r}
 $g$ & $\Delta u$ & $\Delta u_\G$ \\
 \hline \\
 0.25 & 6.44 & 12.67 \\
 0.34 & 6.44 & 9.40  \\
 0.46 & 6.44 & 6.98
 \end{tabular}
 \end{ruledtabular}
\end{minipage}
\caption{\label{Fig-two-periods} (Color online) The differential 
conductance $dI_\MM/dV$ [in units of $(e\omega_L^*/2\pi)$]  of the two-terminal set-up (in the absence of a tip) is 
shown as a function of the dimensionless source-drain bias $u$ 
and the dimensionless gate voltage $u_\G$.  The strengths of the 
contact impurities are equal, and are characterized by 
$\lambda_{\text{F},i}=0.1$ and $\lambda^{*}_{\text{B},i}=0.25$. In 
panels (a), (b), and~(c) the source-drain voltage is applied 
symmetrically, $u_{\rm S/D}=\pm u/2$, and the interaction 
parameter is $g=0.25,0.34$, and $0.46$, respectively. The 
dashed-dotted lines in panels~(b) and (d) are a guide for the eye to 
identify the periodic pattern of the Fabry-P\'erot oscillations determined by 
the periods $\Delta u$ and $\Delta u_\text{G}$. Their ratio 
yields the value of $g$, as shown in the table for the three 
cases. In panel~(d) the source-drain voltage is applied 
asymmetrically ($u_{\rm S}=u$ and $u_\D=0$) to a wire with 
interaction strength $g=0.34$. When compared with panel~(b) for the same interaction strength, the asymmetric bias twists the pattern.}
\end{figure*}
%
%


\subsection{Interaction effects on Fabry-P\'erot oscillations  in a two-terminal set-up}
Let us first analyze the effects of electron-electron interaction
for a contacted wire without tip. As in the non-interacting case,
for $\gamma_{\pm}=0$ the problem is reduced to a two-terminal
set-up, where $I_\T=0$ and $I_\s=I_\D=I_{\M}$. Furthermore,
$I_{\M}$ can again be written as a sum of the current $I_0$ in a
wire with adiabatic contacts and $I_{\rm imp}$, see
Eq.~(\ref{IM_0+imp}). Importantly, while $I_{0}$ is unaffected by
the interaction in the wire,\cite{safi-schulz,maslov-stone} the current $I_{\rm
imp}$, accounting for the contact resistances, is strongly
modified by the interaction. One can still decompose
$I_{\rm imp}$ into
\begin{equation}
I_{\rm imp}= \frac{e \omega_L^*}{2 \pi} \, ( j_{\rm inc} + j_{\rm
coh}) \label{Iimp-int}\, ,
\end{equation}
where $j_{\rm inc}$ is the sum of two terms related to a single
impurity each, and $j_{\rm coh}$ describes interference between
scattering processes at the two impurities. Here, an important difference
emerges with respect to the non-interacting case. The
Fermi velocity {\em in the wire} is enhanced by the interaction
parameter $g$, leading to a higher ballistic frequency
%
%
\begin{equation}\label{new:omega}
\omega^*_L=\frac{\omega_L}{g}\,.
\end{equation}
%
%
Moreover, the interaction is also affecting the  strength of the 
contact impurities: the forward-scattering processes are left 
unchanged whereas the backscattering ones are 
renormalized\cite{kane:1992}
%
%
\begin{equation}\label{lambda:rim}
\lambda \rightarrow
\begin{cases}
\lambda^{*}_{\text{B},i} = \lambda \,\alpha_\W^{g-1} \\
\lambda_{\text{F},i} = \lambda
\end{cases}\,.
\end{equation}
%
%
where $\alpha_\W=a_\W/gL$ is a small dimensionless cutoff
parameter. The cutoff length $a_\W$, which is related to the
lattice spacing or the electronic bandwidth of order $\hbar
v_\W/a_\W$, is introduced in App.~\ref{AppB}.

In the Fabry-P\'erot regime we can again restrict ourselves to
terms up to third order in the contact impurity strengths
$\lambda_i$. Then, the incoherent and the coherent contributions
can be written as
\begin{equation}
j_\text{inc} = \sum_{i=1,2} j_{\text{inc},i}\, \label{Iinc-int}
\end{equation}
and
\begin{equation}
j_\text{coh} = j_\text{coh}^{(2)} + j_\text{coh}^{(3)}\, ,
\label{Icoh-int}
\end{equation}
with
\begin{equation}
j_{\text{inc,i}}= (\lambda_{\text{B},i}^{*})^{2} D_{ii}(u)  \label{j-inc-int}
\end{equation}
%
%
and
%
%
\begin{align}\label{j-coh-2}
j_\text{coh}^{(2)} = & 2 \lambda^{*}_{\text{B},1} \lambda^{*}_{\text{B},2} D_{12}(u)  \cos\left\{2\left[\kappa_\W + g (u_\W-u_\G)\right]\right\} \\
j_\text{coh}^{(3)}  = &  2 \lambda^{*}_{\text{B},1} \lambda^{*}_{\text{B},2} (\lambda_{\text{F},1}+\lambda_{\text{F},2})  D_{12}(u)  \nonumber\\
& \times g^2\sin\left\{2\left[\kappa_\W + g
(u_\W-u_\G)\right]\right\} \label{j-coh-3},
\end{align}
%
%
where we have introduced
%
%
\begin{align}\label{eq:45}
D_{ij}(u) =  \frac{2}{\pi  \alpha_\W^{2 g}} & \int_{0}^{\infty}   d \tau \sin(u \tau) \\
& \times \sin\left[4 \pi \mathcal{I}^{\Phi \Phi}(\xi_i;\xi_j;
\tau)\right] \, \ee^{4 \pi \mathcal{R}^{\Phi
\Phi}_\text{reg}(\xi_i;\xi_j; \tau)}\,.\nonumber
\end{align}
%
%
The dimensionless voltages $u,u_\W$ and $u_\G$ are now scaled by 
the factor $e/\hbar\omega_L^*$ compared to the physical voltages 
$V_\s-V_\D, (V_\s+V_\D)/2$ and $V_\G$, respectively. In the 
expression for $D_{ij}$, the dimensionless integration time is 
defined as $\tau= \omega_L^*t$, and the functions 
$\mathcal{R}^{\Phi\Phi}_\text{reg}(\xi;\xi';\tau)$ and 
$\mathcal{I}^{\Phi\Phi}(\xi;\xi';\tau)$ are the real and imaginary 
parts, respectively,  of the auto-correlation function of the 
bosonic phase field $\Phi$ introduced in App.~\ref{AppB}. The 
quantity $D_{ij}$ defined in Eq.~(\ref{eq:45}) is  cut-off 
independent, since the cut-off dependence of the prefactor is 
compensated by the one of the correlation functions. Explicit 
results for the phase field auto-correlation function have been 
given in a previous paper.\cite{NW-14} Further, the $\xi_i=x_i/L\ 
(i=1,2)$ are dimensionless contact impurity positions. 
Equations~(\ref{j-inc-int}-\ref{j-coh-3}) are obtained from a 
perturbative development of the current in the impurity strengths 
$\lambda_i$ employing the methods described in Apps.~\ref{AppA} 
and \ref{AppB}. The current $j_{\rm coh}^{(3)}$ in 
Eq.~(\ref{j-coh-3}) includes forward scattering processes that 
give rise to the factor 
$\lambda_{\text{F},1}+\lambda_{\text{F},2}$ and a twofold 
backscattering contribution leading to the factor 
$\lambda^{*}_{\text{B},1} \lambda^{*}_{\text{B},2}$.

Another important effect of the interaction is that the
incoherent term $j_\text{inc}$ does not depend linearly on the bias as in
the non-interacting case. Instead, it exhibits oscillations of
period $\Delta u = \pi$, due to the interplay between
backward scattering at one contact impurity and Andreev-type
reflection at the other contact.~\cite{NW-14}
On the other hand, the coherent term $j_\text{coh}$, responsible for Fabry-P\'erot oscillations, shows a power-law suppression  with increasing voltage.\cite{NW-15,recher-PRB} 
Thus, in the presence of interaction two types of oscillations are
present, namely the Fabry-P\'erot ones (already existing for a
non-interacting wire and modified by the interaction), and the
Andreev-type ones (purely due to the interaction). These
two types of oscillations are characterized by the same period in the
source-drain bias, and they are of the same order in the impurity 
strength, if we assume that the two contact transparencies are 
comparable ($\lambda_1 \simeq \lambda_2$). It is therefore difficult to distinguish the  two  phenomena  from an inspection of the two-terminal differential conductance, which is shown in
Fig.~\ref{fig:five} as a function of the source-drain bias for various values of the interaction parameter $g$. Besides  the power-law suppression of the amplitude at high applied bias, we see that for strong interaction ($g<1/2$)  the sinusoidal behavior of the oscillations is deformed into a saw-tooth-like shape. Furthermore, although the total current (\ref{IM_0+imp}) in the presence of contact resistances is always smaller than the current $I_0$ of an ideally contacted wire  ($I_{\rm imp} \le 0$), the differential conductance  may exceed $e^2/h$. This is a well known effect of non-linear transport in Luttinger liquids\cite{FLS:95}, reflecting the fact that the conductance cannot be expressed in terms of single-particle transmission coefficients.  In Sec.~\ref{Sec5}  we shall comment on how  the two types of oscillations may be experimentally distinguished in a three-terminal set-up.\\

Further interesting insights emerge from the analysis 
of the conductance $d I_\MM/dV$ as a function of both the 
source-drain bias $V=V_\s-V_\D=(\hbar\omega_L^*/e)u$ and the gate 
bias $V_\G=(\hbar\omega_L^*/e)u_\G$. Corresponding conductance plots are shown in Fig.~\ref{Fig-two-periods}. Panels (a), (b) and (c) refer to  three different values of the interaction 
strength $g$, in the case  of a symmetrically applied  
source-drain bias, $u_{\rm S/D}=\pm u/2$.  The oscillations of the conductance as a function of $V$ and 
$V_\G$ are characterized by two periods $\Delta V$ and $\Delta 
V_\G$. The period $\Delta V$ coincides with the period of the function $D_{12}(u)$ [Eq.~(\ref{eq:45})] appearing in the coherent  terms (\ref{j-coh-2}) and (\ref{j-coh-3}), since the functions $D_{11}(u)$ and $D_{22}(u)$ related to the incoherent contribution (\ref{j-inc-int}) exhibit 
the period $\Delta V/2$. We thus recover the result of Ref.~\onlinecite{recher-PRB}. On the other hand, the
period $\Delta V_{\rm G}$ in the gate voltage is determined 
by the sinusoidal factors of Eqs.~(\ref{j-coh-2}) and (\ref{j-coh-3}). 
The values of $\Delta V$ and $\Delta V_\G$ depend on the 
interaction strength $g$ and are inversely proportional 
to $g$ and $g^2$, respectively. Interestingly, the ratio of these 
periods yields the Luttinger liquid interaction strength, $\Delta 
V/\Delta V_\G=\Delta u \, / \Delta u_\G = 2 g$, as can be checked 
from the table associated with Fig.~\ref{Fig-two-periods}.  
\\Panel (d) describes the case of an  asymmetrically applied bias ($u_{\rm S}=u$ and 
$u_\D=0$), for the same interaction strength as panel~(b). In this case  $u_\W=u/2$ [see Eq.(\ref{uW-def})], so that  an additional dependence on~$V$ arises from the sinusoidal factors of Eqs.~(\ref{j-coh-2}) and (\ref{j-coh-3}), and the period in $V$ at fixed $V_{\rm G}$ changes. For this reason the two-dimensional pattern of the nonlinear conductance is twisted with respect  to panel~(b). However, the quantities $\Delta V$ and $\Delta V_\G$ related to a symmetrically applied bias can still be obtained, {\em e.g.}, by projecting the conductance maxima on the $V$-axis and measuring the distance between these projections as indicated by the arrows in panel~(d). The value of $g$ can therefore be extracted also in this case as $\Delta u \, / \Delta u_\G = 2 g$. We remark that a qualitatively similar twist of the conductance pattern has recently been observed in carbon nanotubes.\cite{hakonen:2007}\\

Conductance plots as a function of the transport and gate voltages have previously been discussed in the context of carbon nanotubes in
Refs.~\onlinecite{NW-15} and \onlinecite{recher-PRB}. We  point out that  the way we introduce the bias and gate voltages  in our model [see Eq.~(\ref{HmuW})] differs from the one adopted in the above papers. Our approach accounts for several basic physical facts. In a non-chiral quantum wire only the electrochemical potentials of the leads can be controlled experimentally, whereas the electrochemical potentials of right and left movers inside the wire are a result of the biasing of the wire and its screening properties. As a consequence, the source and drain biases, $V_\s$ and $V_\D$, are applied here only in the related leads. This is in accord with a basic hypothesis 
underlying the definition of an electrode, namely that inelastic processes in the lead 
equilibrate absorbed electrons, yielding a voltage drop at the 
contacts {\em even} in the absence of contact impurities.
On the other hand, the charge density of metallic electrodes is typically insensitive to a gate, due to  their electroneutrality. For this reason, in our model the gate voltage $V_\G$ is applied only to the interacting wire and not to the leads. 
\\The precise form of the coupling to the biasing voltages adopted in the model  has implications on the 
behavior of the current as a function of  bias and gate voltages. We find that the 
dependence on $V_\G$ and $(V_\s+V_\D)/2$ involves a factor~$g^2$, as shown, for instance, in  Eqs.~(\ref{j-coh-2}) and (\ref{j-coh-3}).  
[In the dimensionless formulation  
one factor of $g$ is contained in the definition of the 
dimensionless quantities  $u_\G$ and $u_\W$.] The difference $(V_\s+V_\D)/2-V_\G$ is proportional to the {\it bare} electron charge injected into the wire, whereas the $g^2$ factor originates from the partial 
screening occurring in a Luttinger 
liquid,\cite{Egger-Grabert-electroneutrality} and  physically describes the fraction of the bare charge that  remains unscreened. In particular, in the limit  $g \rightarrow 0$  of an electroneutral wire we obtain that the
current depends only on the difference $V_\s-V_\D$ and is independent of the gate, as it should be.\\ 

On a more formal level, these physical properties are encoded in  the zero modes $\Phi_{0,\pm}(x)$ [see Eq.~(\ref{zero-mode})]. Indeed, the transformation $\Phi_{\pm} \rightarrow \Phi_{\pm}+\Phi_{0,\pm}$ of the chiral boson fields gauges away the bias term (\ref{HmuW}). We note that, differently from the homogenous Luttinger liquid case, in the presence of  leads the zero modes   cannot  be just linear functions of the position uniformly  along the entire system. The inhomogeneity of the system leads to a  non-trivial space dependence of the zero modes  $\Phi_{0,\pm}(x)$, which can be obtained  from the boson Green function of the inhomogeneous LL model, as shown in  Eq.~(\ref{zero-mode-eq}).


\subsection{Interaction effects on electron tunneling from the tip: The case of adiabatic contacts}

We shall now consider the full three-terminal set-up, and discuss 
the effects of the wire  electron-electron interaction  on 
tunneling from the tip, both for the case of electron injection 
and in the voltage probe configuration. We start by presenting 
  results for a wire with adiabatic contacts ($\lambda_i=0$). 
For a non-interacting wire, the calculation described in 
Sec.~\ref{Sec3} yields  
\begin{equation}
\label{IT-nonint-adiab}
I_{\T}=\frac{e^2}{h} \frac{8 \gamma^2}{(2+\gamma^2)^2} \left(V_\T-\frac{1+\chi}{2}V_{\rm S}- \frac{1-\chi}{2}V_\D\right)
\end{equation}
and 
\begin{eqnarray}
I_{\M}&=&\frac{e^2}{h}\frac{1}{(2+\gamma^2)^2} \left\{ V_{\rm S} \, [4+2 \gamma^2 (1-\chi+\gamma^2 \chi^2)] \right. \label{IM-nonint-adiab} \\
& & \left. \, \, \, -V_\D [4+2 \gamma^2 (1+\chi+\gamma^2 \chi^2)] +4 \chi \gamma^2 V_\T \right\} \nonumber ,
\end{eqnarray}
where $\gamma$ is the total tunneling strength defined in 
Eq.~(\ref{gamma-def}) and $\chi$ is the tunneling asymmetry 
parameter introduced in Eq.~(\ref{chi-def}). Thus, in the absence 
of interaction, the currents depend linearly on the three applied 
voltages and are independent of the position $x_0$ of the tip.

When electron-electron interaction is taken into account, an 
exact solution of the tunneling problem is not possible for 
arbitrary values of the tunneling amplitudes~$\gamma_\pm$. We 
shall assume that~$\gamma_\pm \ll 1$, consistent with the tunnel 
Hamiltonian approach, and provide results to leading order in 
perturbation theory. The  currents in the source and drain 
leads are again written as in Eqs.~(\ref{ISDT-1}) and (\ref{ISDT-2}), where 
$I_{\M}$ and $I_{\T}$ are evaluated now to order~$\gamma^2$ 
yielding
\begin{equation}
I_{\M}= I_{0} + \, I_{\M, \gamma^2} \, \label{IM-int-G2}
\end{equation}
and
\begin{equation}
I_{\T}= I_{\T, \gamma^2}  \label{IT-int-G2}
\end{equation}
%
%
where
%
%
\begin{equation}
I_{\MT,\gamma^2}=\frac{e \omega_L^*}{2\pi}
{(\gamma^*)}^2 \, j_{\MT,\gamma^2} \label{IMT-G2}\,.
\end{equation}
%
%
Here
\begin{equation}
\gamma^*  = \gamma \, {\alpha_\W}^{\frac{g+g^{-1}-2}{4}}
\label{ren-gamma}
\end{equation}
is the tunneling amplitude renormalized by the electron-electron interaction.
The dimensionless currents $j_{\MT,\gamma^2}$ read
\begin{widetext}
\begin{equation}
j_{\MT,\gamma^2} =\frac{2}{\pi\alpha_\T \alpha_\W^{\frac{g+g^{-1}}{2}} } \int_{0}^{\infty} \!\! \! \!\!\! d \tau \,
Q_{\MT}(\tau) \, \sin\left\{4 \pi \left[\mathcal{I}^{\Phi_{+}
\Phi_{+}}(\xi_0;\xi_0; \tau)+\mathcal{I}^{\varphi \varphi}(0;0;
\tau)\right]\right\}  \ee^{4 \pi \left[\mathcal{R}^{\Phi_{+}
\Phi_{+}}_\text{reg}(\xi_0;\xi_0; \tau)+\mathcal{R}^{\varphi
\varphi}_\text{reg}(0;0; \tau)\right]} \, \label{jMT-G2}
\end{equation}
where
\begin{eqnarray}
Q_{\M}(\tau) &=& \sin ( u \tau/2 ) \cos\left[(u_\W-u_\T) \tau \right] + \chi \cos ( u \tau/2 ) \sin\left[(u_\W-u_\T) \tau \right] \nonumber \\
& & \label{QMT} \\
Q_{\T}(\tau) &=& 2 \cos ( u \tau/2 )
\sin\left[(u_\W-u_\T) \tau \right] +2 \chi \sin (u
\tau/2 ) \cos \left[(u_\W-u_\T) \tau \right] \quad. \nonumber
\end{eqnarray}
\end{widetext}
Here $\alpha_\T$ is  a small dimensionless cutoff parameter for 
the tip defined in App.~\ref{AppB}. The functions 
$\mathcal{R}_\text{reg}^{\Phi_+\Phi_+}(\xi;\xi';\tau)$ and 
$\mathcal{I}^{\Phi_+\Phi_+}(\xi;\xi';\tau)$ are the real and 
imaginary parts of the auto-correlation function of the chiral 
wire field $\Phi_+$ defined in Eqs.~(\ref{rr-ww-GS-Re}) and 
(\ref{rr-ww-GS-Im}), respectively, while 
$\mathcal{R}_\text{reg}^{\varphi \varphi}(\xi;\xi';\tau)$ and 
$\mathcal{I}^{\varphi\varphi}(\xi;\xi';\tau)$ are the real and 
imaginary parts of the correlator of the tip field $\varphi$ 
given in Eqs.~(\ref{R_T}) and (\ref{I_T}). The integral 
(\ref{jMT-G2}) is a cut-off independent quantity.

We consider two parameter domains of the three-terminal
set-up corresponding to the cases where the tip operates as an
electron injector and as voltage probe, respectively. In the
electron injection case, source and drain are at the same
electrochemical potential while a bias is applied to the tip. For
this configuration the current noise was evaluated in
Refs.~\onlinecite{martin:2003} and \onlinecite{martin:2005}. Here we shall explicitly evaluate the
non-linear tunneling conductances
\begin{equation}\label{diff:cond:one}
G_\text{ST} \doteq  \left. \frac{\partial I_\s}{\partial V_\T} \right|_{V_\s=V_\D=0}
\end{equation}
%
%
and
%
%
\begin{equation}\label{diff:cond:two}
G_\text{DT} \doteq  \left. \frac{\partial I_\D}{\partial V_\T} \right|_{V_\s=V_\D=0}\quad.
\end{equation}
%
%
\\

Conventional Luttinger liquid theory, where the presence of the 
source and drain electrodes is neglected, predicts that an 
electron charge injected by tunneling {\em e.g.}\ as a 
right-mover into an interacting wire breaks up into separate 
charge pulses moving in opposite directions, namely a fraction 
$(1+g)/2$ moving to the right and a fraction  $(1 -g)/2$ going to 
the left.\cite{Safi:1997,pham,martin:2005,yacoby-lehur,lehur,fisher-glazman} This effect 
originates from the coupling between the densities of right and 
left moving electrons, accounted for by the homogeneous LL 
Hamiltonian. As a consequence, one expects that when the tip 
injects electrons asymmetrically, {\em e.g.} only  toward the 
drain electrode on the right ($\chi =1$), the electron-electron 
interaction would cause a part of the current to flow also to the 
source electrode on the left. 

However, when the source and drain electrodes are explicitly taken into account, 
our results show that the above expectation is in fact wrong. 
Remarkably, using Eq.~(\ref{jMT-G2}), one can indeed prove that 
for $V_\s=V_\D$ the equality
\begin{equation}
\mathcal{A}(\chi) = \left. \frac{|I_\D| - |I_\s|}{|I_\D|+|I_\s|} \right|_{\chi} \equiv \chi \label{equal-to-a} 
\end{equation}
holds, indicating that for a clean wire the current
asymmetry is {\em independent} of the wire interaction strength
$g$. In particular, for  fully asymmetric tunneling ($\chi=1$),
the whole current is injected into the drain electrode, just as in
the non-interacting case. This unidirectional charge flow even in
the presence of interaction arises from the phenomenon of
Andreev-type reflections.\cite{safi-schulz,maslov-stone} Even though charge
fractionalization occurs in the bulk of the wire, the plasmonic
excitations reaching an interface with the leads experience the
mismatch of the interaction strengths in the wire and in the
electrode and are thus partly reflected as an oppositely charged
excitation. The sum of all reflected pulses at both interfaces
restores the property  that the whole current flows into the
drain, like in the non-interacting wire. This behavior is in
fact very similar to an effect occurring in a
two-terminal set-up, where the conductance of a wire adiabatically
connected to electrodes is $G_\text{2t}=e^2/h$, independent of the
interaction strength. Thus, for perfectly transmitting contacts,
it is impossible to extract the interaction constant
neither from the conductance of a two-terminal set-up nor from the
current asymmetry in three-terminal measurements.

\begin{figure*}
  \begin{minipage}{0.45\textwidth}
          \centering
          \includegraphics[scale=0.3]{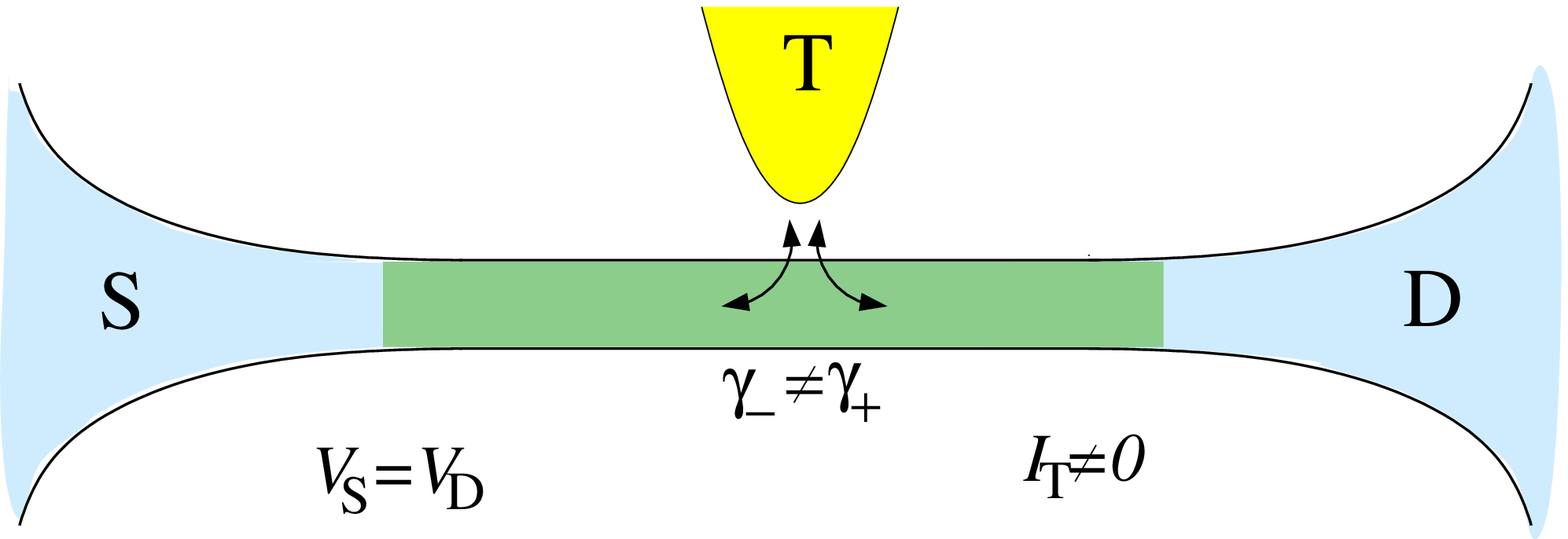}
          \includegraphics[width=\columnwidth,clip]{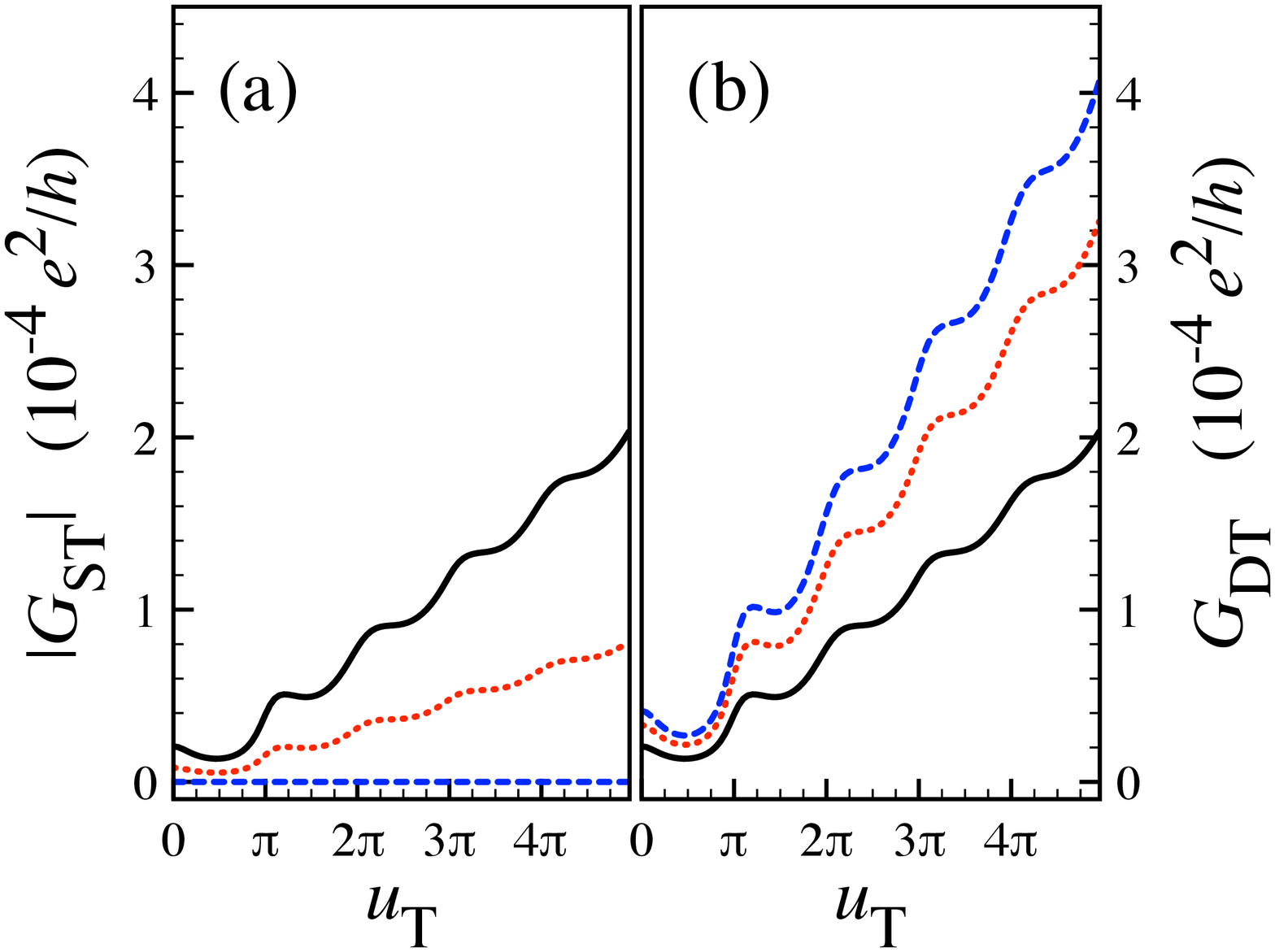}
  \end{minipage}
  \begin{minipage}{0.45\textwidth}
          \centering
          \includegraphics[scale=0.3]{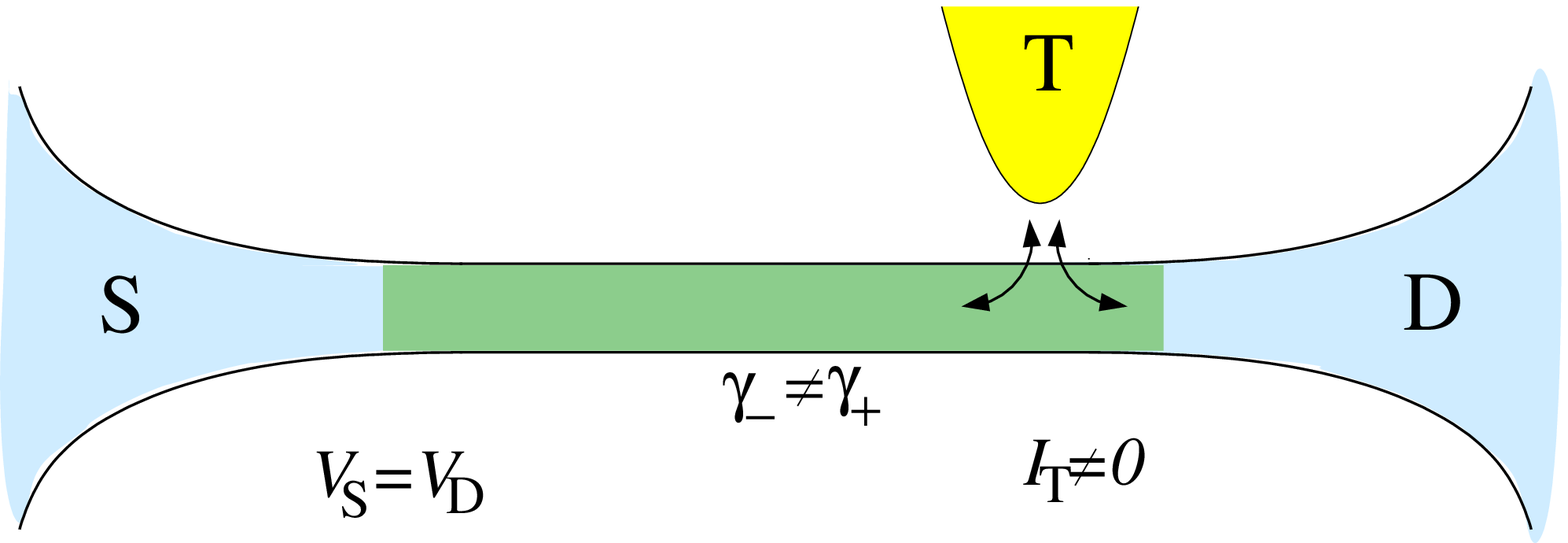}
          \includegraphics[width=\columnwidth,clip]{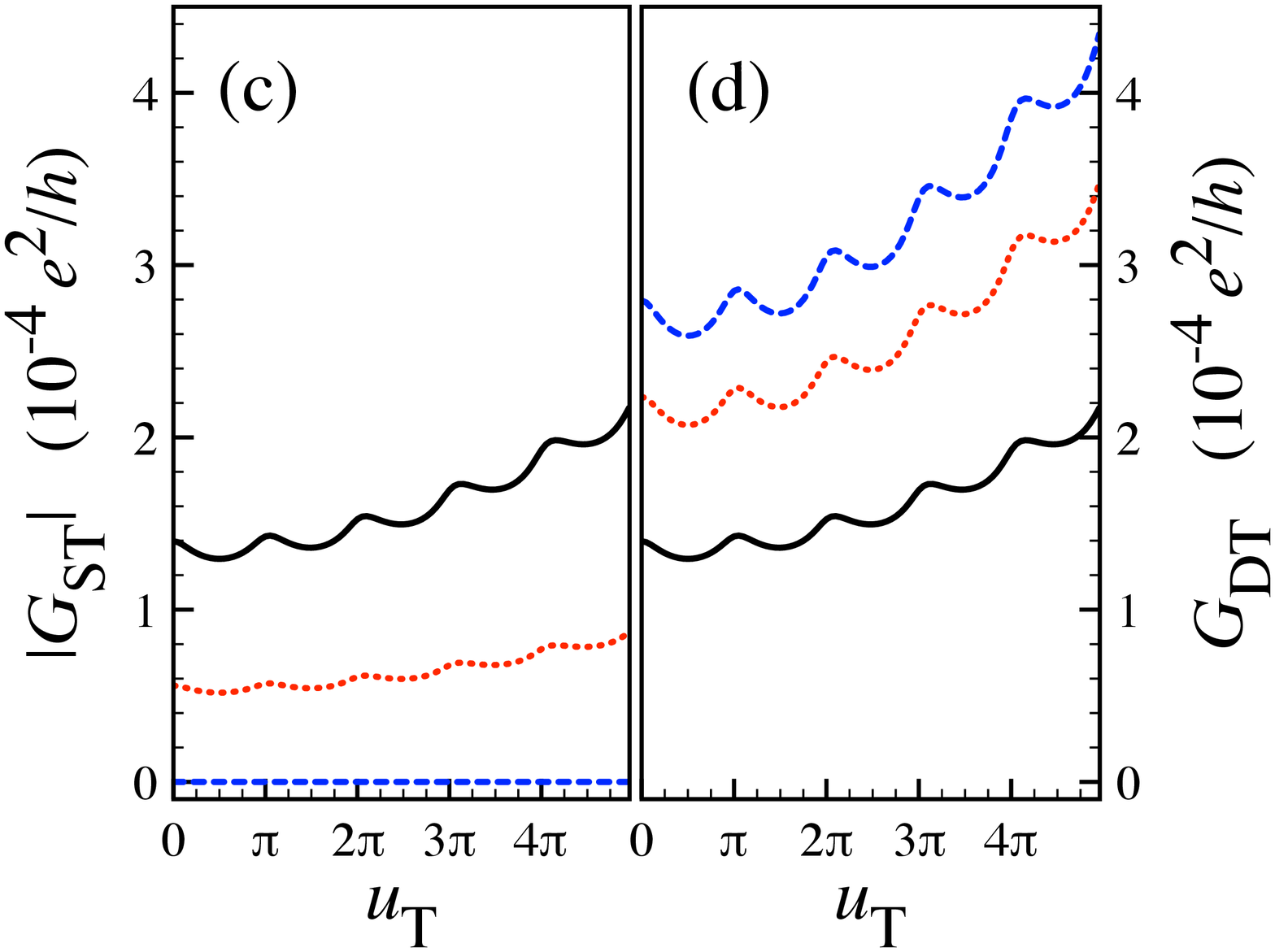}
  \end{minipage}

\caption{\label{Fig-int-G2-1} (Color online) Electron injection 
into an interacting wire. Panel (a) [(b)] shows the tunneling 
differential conductance $G_\Ts$ [$G_\TD$] between the tip and 
source [drain] electrode as a function of the dimensionless 
tip-source [tip-drain] bias for a wire with interaction strength 
$g=0.25$. The various curves refer to different values of the 
tunneling asymmetry ($\chi=0$ solid line, $\chi=0.6$ dotted line, 
and $\chi=1$ dashed line). The tunneling strength is 
$\gamma^*=0.01$, and the tip is located in the middle of the 
wire. Panels (c) and (d) are the same as panels (a) and (b) but 
the tip is located near a contact at $x_0=0.45 L$.}
\end{figure*}

Nevertheless, in a three-terminal set-up signatures of interaction
{\em do appear} in the behavior of the   differential conductances
$G_\text{ST}$ and $G_\text{DT}$ as a function of the tip-source 
and tip-drain bias. Figures~\ref{Fig-int-G2-1}(a) and 
\ref{Fig-int-G2-1}(b) show $G_\text{ST}$ and $G_\text{DT}$ for the 
case of a tip located in the middle of a wire with interaction 
strength $g=0.25$. The various curves correspond to different 
values of the asymmetry parameter $\chi$, which  unbalances the 
amount of injected right \emph{vs.} left moving electrons. The 
fully symmetric case ($\chi=0$) was discussed in 
Ref.~\onlinecite{nazarov:1997}. While for a non-interacting wire 
$G_\text{ST}$ and $G_\text{DT}$ are constant [as can easily be  
seen from Eqs.~(\ref{IT-nonint-adiab}) and 
(\ref{IM-nonint-adiab})], in the presence of interaction an 
oscillatory behavior arises. These oscillations are entirely due 
to the electron-electron interaction in the wire, which causes 
Andreev-type reflections even at adiabatic contacts.  With 
increasing $\chi$ the conductance $G_\text{ST}$ decreases until 
it vanishes for $\chi=1$, whereas the conductance $G_\text{DT}$  
increases up to the maximum value for the completely asymmetric 
case. The relation $G_\text{ST}=G_\text{DT} (1-\chi)/(1+\chi)$ 
between these two conductances is  independent of $g$.
\\
Figures~\ref{Fig-int-G2-1}(c) and \ref{Fig-int-G2-1}(d) describe 
the case of an off-centered tip located at $x_0=0.45 L$. 
Apparently, the period of the oscillations is the same as in 
panels (a) and (b) where the tip is in the middle. This is due to 
the fact that this period is related to the traversal  time of  
plasmonic excitations originating from the tip and interfering at 
the same point after an even number of Andreev-type reflections 
at the contacts. This traversal time  depends neither on~$x_0$ 
nor on the asymmetry coefficient.

\begin{figure}[h] 
  \centering
   \includegraphics[scale=0.3]{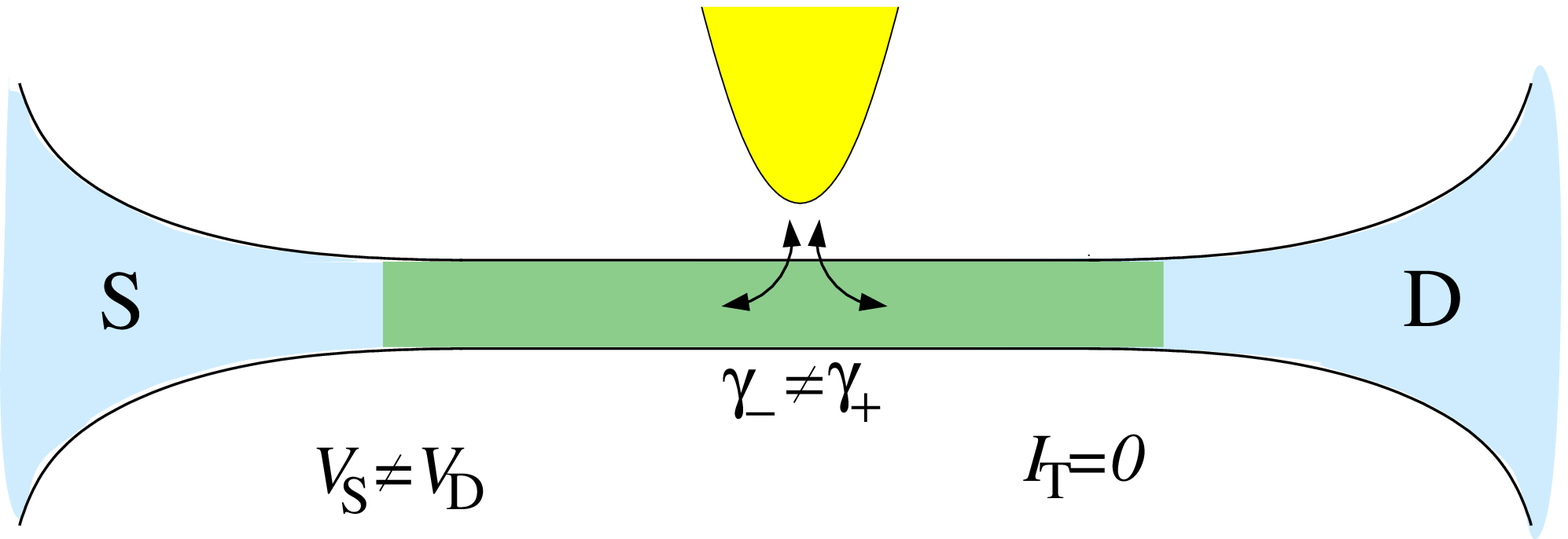}
          \includegraphics[width=\columnwidth,clip]{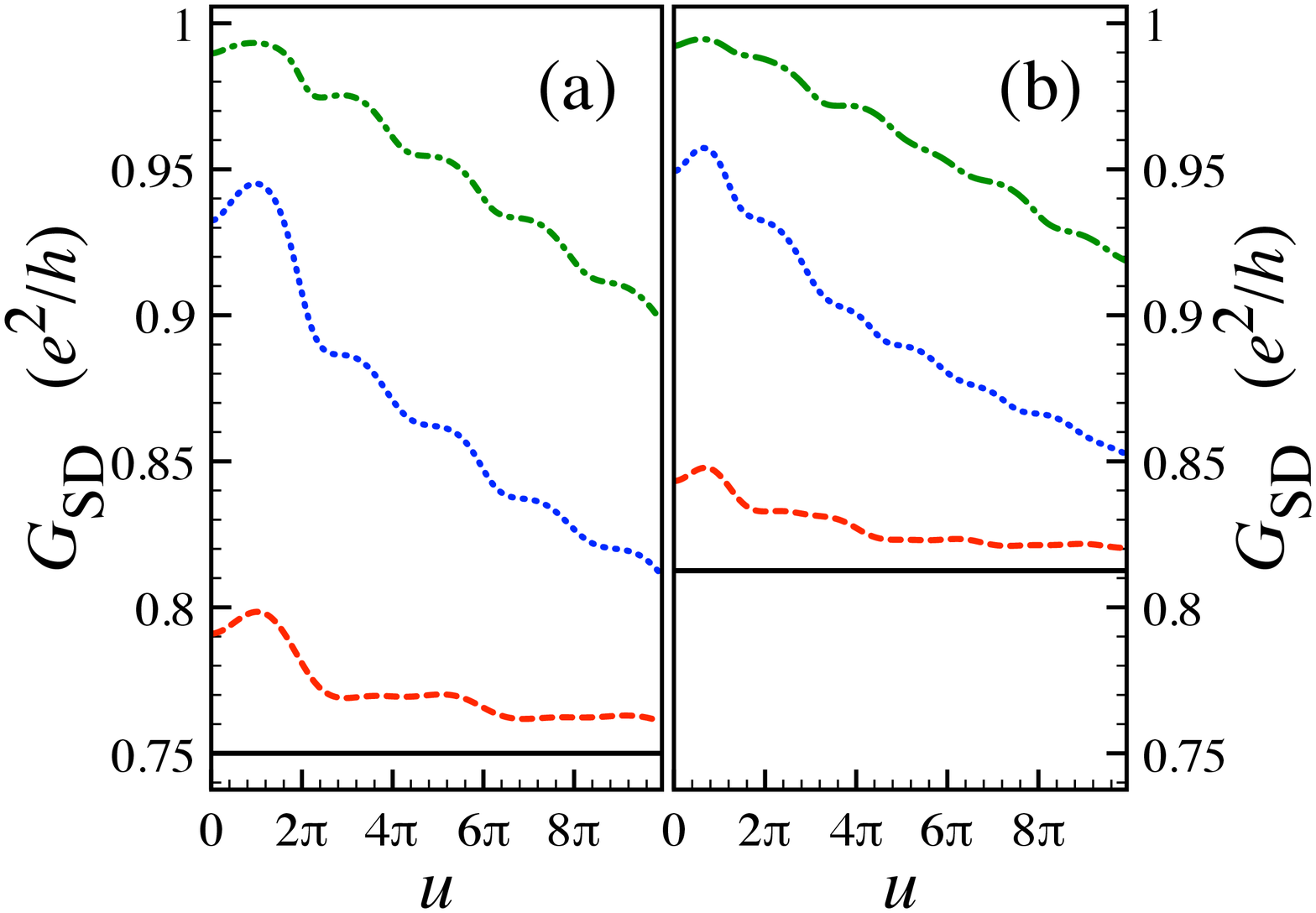}

  \caption{\label{fig:seven} (Color online) 
The effect of the tip in the voltage probe configuration on the 
zero temperature conductance is shown as a function of the 
source-drain bias. Panel (a): the case of symmetric tunneling 
($\chi=0$). Panel (b): the case of an asymmetry in tunneling 
($\chi=1/2$). Different curves refer to different values of the
interacting strength: non-interacting ($g=1$, solid curve), 
weakly interacting ($g=0.7$, dashed curve), moderately 
interacting ($g=0.4$, dotted curve), and strongly interacting 
($g=0.25$, dashed-dotted curve). The bare tunneling strength is 
$\gamma=0.5$ and the dimensionless cutoff parameter is 
$\alpha_\text{W}=10^{-3}$.}
\end{figure}

Let us now discuss the configuration where the tip acts as a
voltage probe, \emph{i.e.}\ when $V_\s \neq V_\D$ and $V_\T$ is
set to a value such that $I_\T=0$ is fulfilled. \comment{In this
configuration the conductances defined in
Eqs.~(\ref{diff:cond:one}) and (\ref{diff:cond:two}) are by
definition vanishing, and so is} In this configuration the 
quantity on the left hand side of Eq.~(\ref{cur-asym-def}) is 
vanishing, due to Eqs.~(\ref{ISDT-1}) and (\ref{ISDT-2}). By  
applying a source-drain bias, one can analyze how the  
source-drain conductance
\begin{equation}\label{diff:cond:three}
G_\SD = \left.\frac{\partial I_{\M}}{\partial(V_\s-V_\D)}\right|_{I_\text{T}=0} 
\end{equation}
is affected by the interaction strength. It is worth emphasizing 
that in a two-terminal set-up, \emph{i.e.}\ in the absence of the 
tip ($\gamma_{\pm}=0$), one obtains for a clean wire 
$G_\SD=G_\text{2t}=e^2/h$ , independent of the interaction strength. As 
already mentioned previously,  this is due to the fact that,  
although the electron charge injected by the source splits up in 
fractions through the interaction-induced Andreev-type 
reflections at the contacts,  in a clean wire the series of these 
fractions  always sums up to $e$, disguising the interaction 
effects in the dc average current.\cite{safi-schulz} Our results 
show that a quite different behavior emerges for a three-terminal 
set-up, even in the configuration where the tip does not inject 
any net current into the wire. Figure~\ref{fig:seven}(a) shows 
$G_\SD$ as a function of the source-drain bias, for different 
values of the interaction strength, ranging from a 
non-interacting  to a strongly interacting wire. The left panel 
refers to the case of symmetric tunneling $\chi=0$, whereas the 
right one analyzes the role of a tunneling asymmetry. As one can 
see, the effects of interaction in the wire become observable through the 
voltage probe, since oscillation of $G_\SD$ originating from   
Andreev-type reflections emerge. Notice 
that at constant bare tunneling amplitude $\gamma$ the zero bias 
conductance is higher in the presence of interaction than for a 
non-interacting wire, since the renormalization (\ref{ren-gamma}) of the tunneling 
amplitude suppresses $\gamma$. With increasing 
tunneling asymmetry [see Fig.~\ref{fig:seven}(b)],  the differences 
between interacting and non-interacting wires become less 
pronounced, and indeed the oscillations are washed out for fully 
asymmetric tunneling $\chi=\pm1$.

%
%

The dimensionless tip voltage $\bar{u}_\T=(e/\hbar\omega_L^*)\bar{V}_\T$
ensuring $I_\text{T}$=0 shows an interesting dependence on the
source-drain bias. In the limiting cases of symmetric and
completely asymmetric tunneling this dependence coincides for
interacting and non-interacting wires  (namely $\bar{u}_\T= u_\W$ for
$\chi=0$ and $\bar{u}_\T= u_\W \pm u/2$ for $\chi=\pm 1$). For
intermediate values of the asymmetry parameter $\chi$ the tip
voltage $\bar{u}_\T$ shows an oscillatory behavior with period $\Delta
\bar{u}_\T=2\pi$ as a function of the source-drain bias. We also
see that the period of $G_\text{SD}$ in
Fig.~\ref{fig:seven} is twice as large as the period of $G_\Ts$
and $G_\TD$ in Fig.~\ref{Fig-int-G2-1} where the tip is in the
electron injection configuration. This is due to the fact that in
Fig.~\ref{fig:seven} the source-drain bias is applied
symmetrically ($u_\W=0$) while in Fig.~\ref{Fig-int-G2-1}
source and drain are both grounded and the bias is only applied to the tip. \\
Finally, we emphasize again the difference between the electron 
injection and the voltage probe configurations of the tip: While 
in the former case an asymmetry in tunneling does not spoil the 
observation of effects of electron-electron interaction (see 
Fig.~\ref{Fig-int-G2-1}), in the latter case interaction-induced 
oscillations can be best observed for symmetric tunneling and 
they are in fact vanishing for fully asymmetric tunneling.\\


\subsection{Interaction effects on electron tunneling from the tip: The case of a wire with non-ideal contacts}
\label{Sec3C} In this section we analyze the three-terminal 
transport properties in the presence of electron-electron 
interaction, contact impurity scattering and electron tunneling 
from the tip. In particular, we discuss how a finite contact 
resistance modifies the Andreev-type oscillations of the 
tunneling conductances, previously  discussed for the case of 
adiabatic contacts ($\lambda_i=0$). We present results obtained 
by perturbation theory for weak contact impurities  $\lambda_i$, 
and tunneling amplitudes $\gamma_{\pm}$. Technical details can be 
found in the Appendices. The currents may be written as
\begin{equation}
I_{\M}= I_{0} + \, I_{\rm imp} \, + \, I_{\M, \gamma^2} \, + \,
I_{\M, \gamma^2 \lambda} \label{IM-int-G2L}
\end{equation}
and
\begin{equation}
I_{\T}= I_{\T, \gamma^2} \, + I_{\T, \gamma^2 \lambda} \,
\label{IT-int-G2L}
\end{equation}
Here $I_0=(e^2/h)(V_{\rm S}-V_\D)$ is the current of an ideally 
contacted wire in the absence of the tip, whereas $I_{\rm imp}$ 
is the leading order term accounting for non-ideal contacts 
[see   Eq.~(\ref{Iimp-int})]. In general, this latter term 
involves both Fabry-P\'erot and Andreev-type oscillations. Both, 
$I_0$ and $I_{\rm imp}$, vanish when the electrochemical 
potentials for source and drain electrodes are equal ($V_{\rm 
S}=V_\D$); alternatively, they can be easily determined by 
measuring the current-voltage characteristics in the absence of 
the tip. Henceforth, we shall focus on contributions to the 
currents arising from the presence of the tip. The leading order 
terms $I_{\M, \gamma^2}$ and $I_{\T, \gamma^2}$, given by 
Eq.~(\ref{IMT-G2}), describe tunneling into an ideally contacted 
wire and contain only Andreev-type oscillations. The 
next-to-leading order terms ($\gamma^2 \lambda$,  $\gamma^2 
\lambda^2$, ...) also exhibit oscillations originating from 
interference between backscattering at the contacts and tunneling 
to/from the tip. Such oscillations,  though modified by the 
interaction, are already present in a non-interacting wire, 
unlike the Andreev-type oscillations of the leading order terms 
($\gamma^2$), that are instead entirely due to the interaction. 
We thus analyze how interaction affects the terms $I_{\T ({\rm 
M}), \gamma^2 \lambda}$, which represent  the most relevant 
correction to the Andreev-type oscillations discussed above. 
These terms describe to leading order the interplay between 
electron injection at the tip and backscattering at the S and D 
contacts.


\begin{figure*}
\begin{minipage}[t]{0.25\textwidth}
  \centering
  \includegraphics[scale=0.2]{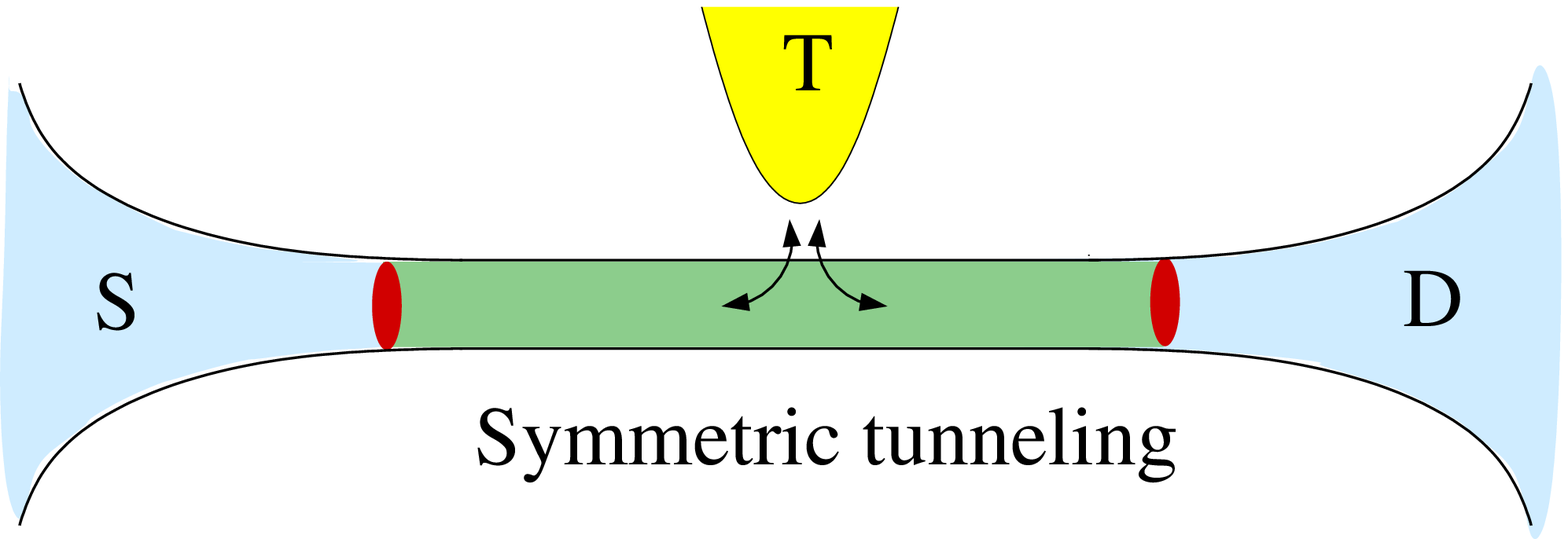}
  \end{minipage}
\begin{minipage}[c]{0.65\textwidth}
\centering
 	\includegraphics[width=\textwidth]{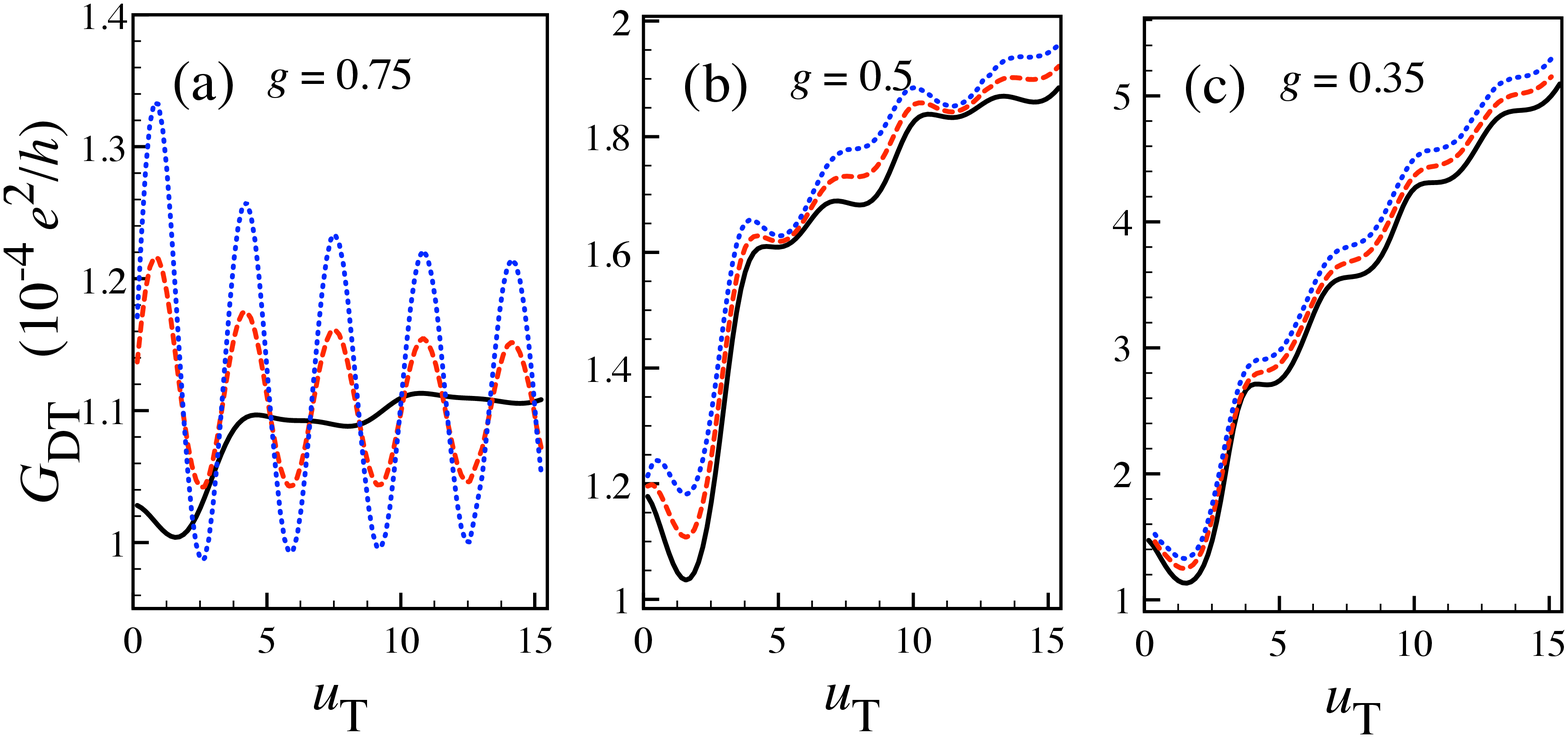}
\end{minipage}
\begin{minipage}[t]{0.25\textwidth}
  \centering
  \includegraphics[scale=0.2]{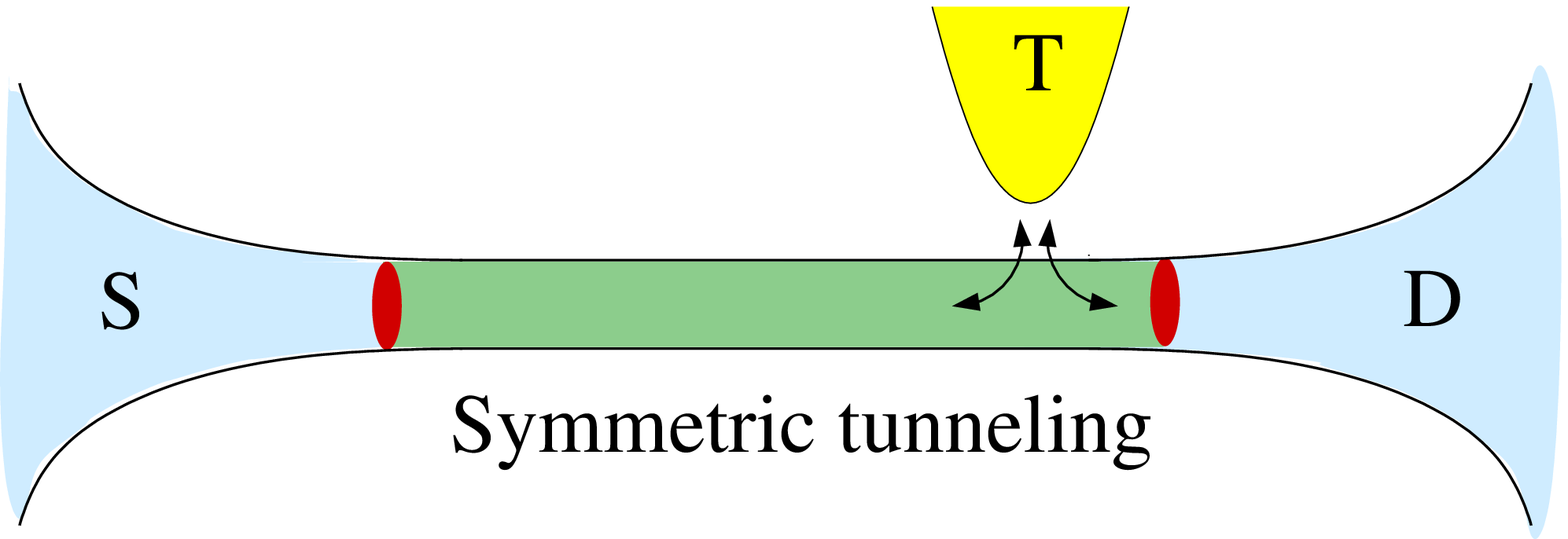}
  \end{minipage}
\begin{minipage}[c]{0.65\textwidth}
\centering
  \includegraphics[width=\textwidth]{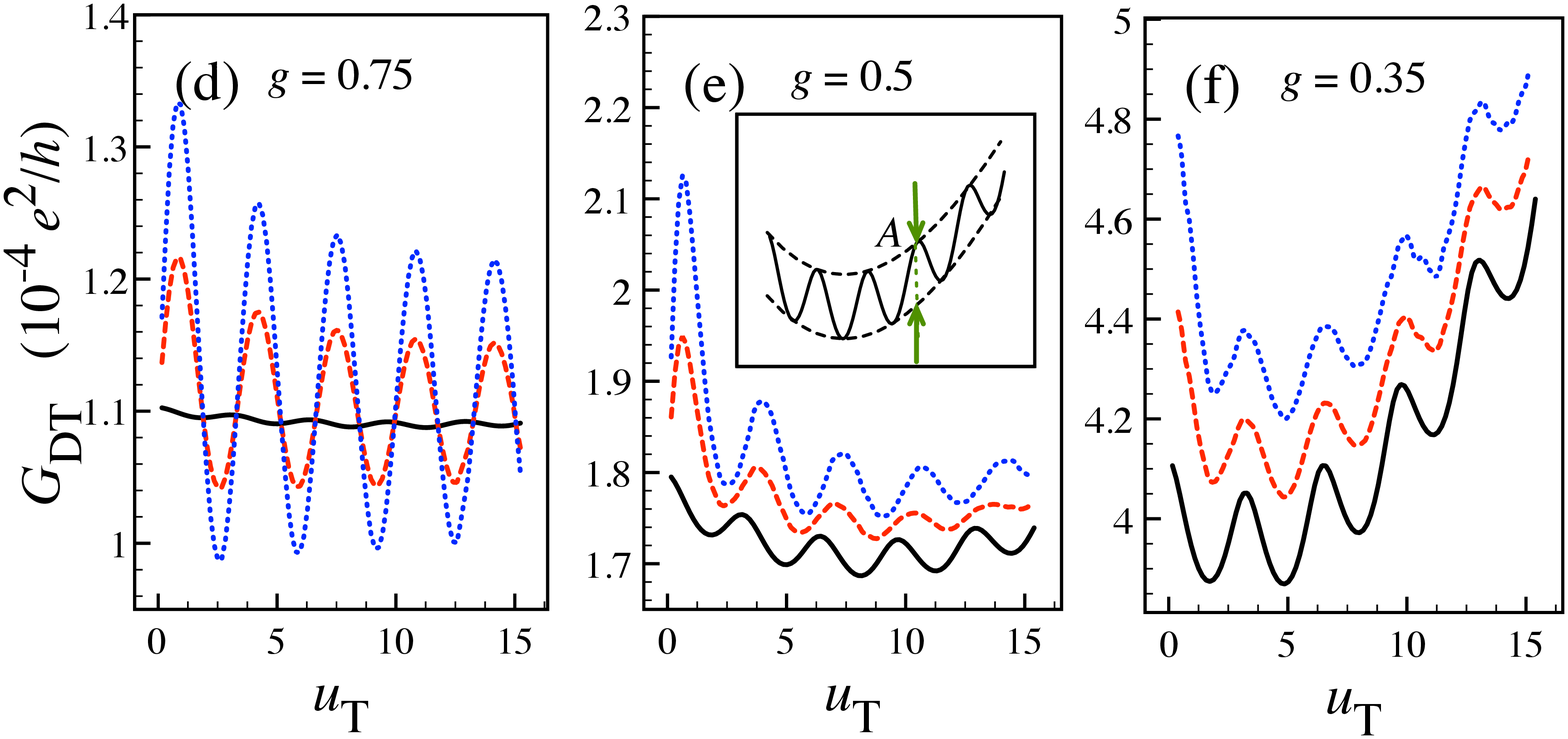}
\end{minipage}   \caption{\label{fig:nine} (Color online) Zero temperature
non-linear tip-drain conductance $G_{\rm DT}$ as a function of the tip bias 
for tunneling amplitude $\gamma^*=10^{-2}$ and symmetric tunneling $\chi=0$. The upper  panels (a), (b) and (c), are related to a tip in the middle of the wire ($x_0=0$), whereas the lower ones (d), (e) and (f) to a tip located at  $x_0=0.45L$. Panels pairs (a) and (d), (b) and (e), and (c) and (f) describe the case of a wire with weak ($g=0.75$), moderate ($g=0.5$), and strong ($g=0.35$) interaction strength, respectively. In each panel the different curves refer to different contact impurity strengths. The solid curves describe the case of ideal contacts where the oscillations are purely Andreev-type. The dashed [dotted] lines refer to finite contact impurity strength $\lambda_{\text{B},1}^*=\lambda_{\text{B},2}^*=0.1$ [$=0.2$]. For a weakly interacting wire the conductance oscillations are mostly due to the conventional interference between backscattering at the contacts and tip tunneling, and Andreev-type oscillations become visible only for extremely low contact resistance. In contrast, for stronger interaction strength a finite contact resistance is sufficient for the oscillations to be attributed to Andreev-type processes. The inset of panel (e) shows the definition of the average amplitude referred to in the text.}
\end{figure*}

Explicitly one finds
\begin{equation}\label{eq:65}
I_{\MT,\gamma^2 \lambda}= \frac{e \omega_L^*}{2\pi} 
{(\gamma^*)}^{2} \, \lambda^* \sqrt{1-\chi^2} \, \, j_{\MT,\gamma^2 
\lambda}
\end{equation}
where $\lambda^*=\lambda_{\text{B},1}^*+\lambda_{\text{B},2}^*$, 
and 
\begin{widetext}
\begin{align}
j_{\MT,\gamma^2 \lambda} =  -\frac{1}{2 \pi
\alpha_\W^{\frac{g+g^{-1}+6}{4}} \alpha_\T^{}} \sum_{i=1,2} \!
\frac{\lambda_{\text{B},i}^*}{\lambda^*} \! \iint_{-\infty}^{+\infty}
\!\!\! & d\tau_1 d\tau_2  \bigg( \cos\left[ (u_\s-u_\T) \tau_1 +(u_\T-u_\D) \tau_2 -2 (\kappa_\W+gu_\W-gu_\G)(\xi_0-\xi_i)\right]  \nonumber \\
\times & \text{e}^{4\pi [\mathcal{R}_{\W}(\xi_0;\xi_i;\tau_1;\tau_2)+\mathcal{R}_{\rm T}(\tau_1;\tau_2)]}
\sum_{\eta_1,\eta_2=\pm} P^{\eta_1 \eta_2}_{\MT} \, F^{\eta_1 \eta_2 +}_{\W, \gamma^2 \lambda}(\tau_1;\tau_2) \, F^{\eta_1 \eta_2}_{\T, \gamma^2 \lambda}(\tau_1;\tau_2) \nonumber \\
\times & \sin\left\{ 4 \pi \left[\mathcal{I}^{\eta_1
\eta_2+}_{\rm W}(\xi_0;\xi_i;\tau_1;\tau_2)+\mathcal{I}^{\eta_1
\eta_2}_{\rm T}(\tau_1;\tau_2)\right]\right\} \bigg) \label{jMT-G2L}
\end{align}
\end{widetext}
where
\begin{align}
P^{\eta_1 \eta_2}_{\M} & = \eta_2+\eta_1-2 \eta_1 \eta_2 \label{PW}\\
P^{\eta_1 \eta_2}_{\T} &= 2(\eta_2-\eta_1)\label{PT} \, .
\end{align}
The functions $F^{\eta_1 \eta_2 \eta_3}_{\W, \gamma^2 \lambda}(\tau_1;\tau_2)$
and $F^{\eta_1 \eta_2}_{\T, \gamma^2 \lambda}(\tau_1;\tau_2)$ are defined in
App.~\ref{AppB} in Eqs.~(\ref{FW-G2L}) and (\ref{FT-G2}), respectively,
and the functions $\mathcal{R}_{\W}(\xi_0;\xi_i;\tau_1;\tau_2)$,
$\mathcal{I}_{\W}(\xi_0;\xi_i;\tau_1;\tau_2)$, $\mathcal{R}_{\T}(\tau_1;\tau_2)$
and $\mathcal{I}_{\T}(\tau_1;\tau_2)$, accounting for the real and the imaginary
parts of several correlation functions in the wire and in the tip, are
defined in App.~\ref{AppC} in Eqs.~(\ref{R_W}), (\ref{I_W}), (\ref{R_T}) and (\ref{I_T}), respectively.\\

For simplicity, we limit the analysis of Eqs.~(\ref{eq:65}) and
(\ref{jMT-G2L}) to the electron injection configuration where 
source and drain are grounded. We start with the case of 
symmetric tunneling  ($\chi=0$).


%
%

As already observed in Sec.~\ref{Sec3} for a noninteracting wire, 
the term (\ref{eq:65}) leads to additional oscillations in the 
differential conductance of the three-terminal set-up. These 
conventional oscillations are characterized by two periods 
related to the distances between the tip and the contact 
impurities, so that the pattern  depends on the tip position. 
Electron-electron interaction modifies this pattern reducing the 
amplitude of the conventional oscillations  and giving rise to 
additional Andreev-type oscillations. The case of a tip in the 
middle of the wire is shown in the upper panels (a), (b) and (c) of Fig.~\ref{fig:nine}, where the 
differential conductance $G_{\DT}$ is plotted as a function of 
the tip bias $u_\text{T}$ for three different values of interaction 
strength, ranging from weak ($g=0.75$), over moderate ($g=0.5$) 
to strong interaction ($g=0.35$), as displayed in the three 
panels. In each panel the solid curve refers to the case of ideal 
contacts where the oscillations are purely of Andreev-type. The 
dotted and dashed curves describe the effect of finite contact 
resistances arising from the contribution of the term 
(\ref{eq:65}). As one can see from panel (a), for weak 
electron-electron interaction  the conventional oscillations 
dominate and mask the Andreev-type oscillations. In this case, 
only extremely good contacting might allow  to identify 
Andreev-type processes. However, for moderate interaction strength  
[panel (b)], the two types of oscillations have comparable amplitudes, and for strong interaction [panel (c)] the 
conventional oscillations are strongly suppressed while the term (\ref{eq:65}) only causes a small shift of the conductance value. The 
oscillations of $G_{\DT}$ are essentially Andreev-type.   
\\A similar effects occurs when the tip is closer to one of the 
contacts displayed in the lower panels~(d), (e) and (f) of Fig.~\ref{fig:nine}. The main difference is that in this case the pattern of 
the Andreev-type oscillations is more sinusoidal, even for weak interactions.\\

Our result indicates that, for a wire with a given interaction 
strength, there is crossover value $\lambda^*_{\rm C}$ of the (renormalized) contact resistance, 
below which the oscillations of the non-linear conductance can 
essentially be attributed to Andreev-type processes. 
We have quantified $\lambda^*_{\rm C}$ for the case of  a tip close to the contacts, where the regularity of oscillations allows for a straightforward determination of their amplitude, defined as the average distance between maxima and minima, as schematically displayed in the inset 
of Fig.~\ref{fig:nine}(e). The crossover impurity strength $\lambda^*_{\rm C}$ is then simply determined by the value of $\lambda^*$ for which the amplitude $A_{\gamma^2 \lambda}$  of the conventional oscillation term 
$I_{{\rm T},\gamma^2 \lambda}$ [see Eq.~(\ref{eq:65})] equals the amplitude  $A_{\gamma^2}$ of the Andreev-type oscillation term 
$I_{{\rm T},\gamma^2}$ [see Eq.~(\ref{IMT-G2})]. The result is given in Table 
\ref{tab:one} for different values of interaction strength. For contact impurity strength $\lambda^* \le \lambda^*_{\rm C}$ the oscillations of the non-linear 
conductance are essentially of Andreev-type. 

\begin{table}
\begin{ruledtabular}
\begin{tabular}{c|ccccc}
$g$ & 0.4 & 0.5 & 0.6 & 0.7 & 0.8 \\ \hline \\
$\lambda^*_\text{C}$ & 0.2 & 0.02 & $4 \cdot 10^{-3}$ & $4 \cdot 10^{-4}$ &   $10^{-4}$ 
\end{tabular}
\end{ruledtabular}
\caption{Crossover value of the (renormalized) contact impurity strength 
$\lambda^*_{\rm C}$, below which oscillations can be attributed to 
Andreev-type processes, for various values of the interaction 
strength~$g$. The tip is located at $x_0=0.45 L$, like in the lower panels of Fig.\ref{fig:nine}.} \label{tab:one}
\end{table}

Let us finally briefly consider the case of asymmetric tunneling 
$\chi \neq 0$. An important result is that, in view of 
Eq.~(\ref{eq:65}), the contribution to the current of order 
$\gamma^2 \lambda$ vanishes in the case of totally asymmetric 
tunneling ($\chi=\pm1$). This property is thus robust to 
electron-electron interaction within the Luttinger liquid 
picture. In fact, one can show that in this case only 
perturbative contributions of order 
$\gamma_i^{2n}(\lambda_1\lambda_2)^{n+m}$ ($n=~1,2,3,\ldots; 
m=~0,1,2,\ldots$) are nonvanishing.

\section{Discussion and Conclusions}\label{Sec5}

In the present work, we have investigated transport
properties of a quantum wire contacted to source and drain
reservoirs in the presence of a third electrode (tip) injecting
electrons into the wire. We have tailored our model to account for
various aspects of a typical experimental situation by
including finite contact resistances and the presence of a gate in
addition to electron-electron interaction, and by analyzing the
effect of the position of the tip as well as the role of a
tunneling asymmetry. Specifically, we have considered both the
situation where the tip behaves as an electron injector and the
voltage probe configuration.    We have found that the three-terminal
set-up exhibits extremely rich behaviors, determined not
only by each of the above aspects, but also by their interplay. In
order to facilitate the discussion, we propose to the reader  
different perspectives from which our results can be considered.\\ 

\noindent{\it The effects of electron-electron interaction on Fabry-P\'erot oscillations.} 
The origin of Fabry-P\'erot oscillations boils down to quantum 
interference between electron backscattering at two (or more) 
impurities. As a consequence, this phenomenon is 
present also in a non-interacting quantum wire (see Sec.~\ref{Sec3}), where the oscillations appear both as a function of the source-drain bias and as a function of the gate voltage. The interference pattern is 
modified by electron-electron interaction, which introduces a power-law suppression of the amplitude and, especially for $g < 1/2$, deforms the sinusoidal shape towards a saw-tooth-like shape (see Fig.~\ref{fig:five}).
Interaction also leads to a (partial)  screening of the charge in the wire\cite{Egger-Grabert-electroneutrality}, causing a change of 
the oscillation period as a function of the  gate bias 
with respect to the period as a function of the source-drain bias. This effect suggests an operative procedure to extract the Luttinger liquid parameter $g$ from measurements of the non-linear conductance   in the Fabry-P\'erot regime (see 
Fig.~\ref{Fig-two-periods}). The effects of an asymmetrically applied source-drain bias have also been discussed.   We emphasize that, differently from previous approaches adopted in the literature, our way to introduce the biasing voltages correctly recovers both  gauge invariance\cite{buttiker:1993} and the property that, in the limit of strong interaction $g \rightarrow 0$, the current-voltage characteristics only depends on the difference between source and drain bias $V_\s-V_\D$. \\

\noindent{\it Conventional vs.\ Andreev-type oscillations.}
Besides modifying Fabry-P\'erot oscillations, electron-electron 
interaction also yields another major effect, which is absent in 
a non-interacting wire: At the wire-electrode interfaces, plasmon 
excitations are partially reflected  due to the mismatch of 
interaction strengths in the interacting wire and the 
non-interacting electrodes. This effect, entirely due to 
interaction, occurs also for ideally contacted adiabatic 
interfaces and  gives rise to a different type of oscillations, 
which are termed  Andreev-type oscillations\cite{NW-14} since the 
incoming charge and the  fractional charge reflected  at the 
contact have opposite signs, just as at an interface between a 
normal metal and a superconductor. In real experiments with 
interacting quantum wires in the Fabry-P\'erot regime, the 
current-voltage characteristics will in general exhibit both 
conventional Fabry-P\'erot oscillations, i.e.\ oscillations that 
are already present in a non-interacting wire and that are simply 
{\em modified} by interaction, and Andreev-type oscillations, 
purely {\em originating} from interaction. The interesting  
question arises whether one can  distinguish between these two 
oscillatory phenomena in an operative way and, in particular,  
whether it is possible to determine regimes and conditions, under 
which the latter can be observed. 

Since the amplitude of Fabry-P\'erot oscillations is roughly 
proportional to the reflection coefficients of the contacts 
whereas Andreev-type processes occur even with ideal interfaces, 
one might at first think that with improving transparency of the 
contacts the non-linear conductance of a two-terminal set-up 
would exhibit a predominance of Andreev-type oscillations over 
the conventional Fabry-P\'erot ones. This is, however, not the 
case, since for an ideally contacted wire the sum of all 
Andreev-type reflection processes at the two interfaces exactly 
recovers the injected pulse, when the sign of all reflected 
charge pulses is taken into account. The transmission of an 
interacting wire adiabatically connected to non-interacting leads 
turns out to equal 1, as was pointed out in 
Refs.~\onlinecite{safi-schulz} and \onlinecite{maslov-stone}. Although 
Andreev-type oscillations of the conductance do appear in the 
presence of even a single impurity\cite{NW-14}, their amplitude 
is  proportional to the impurity reflection coefficient. This  
implies that two-terminal set-ups are not suitable to distinguish 
between Andreev-type and Fabry-P\'erot oscillations, since 
both oscillations have the same dependence on the impurity 
strengths $\lambda^*_{{\rm{B}},i}$. Furthermore they also exhibit 
the same period as a function of the source-drain bias. 

In contrast, our analysis suggests that a three-terminal set-up 
may allow one to distinguish  Andreev-type oscillations from 
conventional oscillations. As far as Andreev-type oscillations 
are concerned, three-terminal set-ups indeed offer one important 
advantage with respect to two-terminal ones: in the presence of  
a third electrode, Andreev-type oscillations appear even for 
the ideal case of a wire adiabatically connected to the source 
and drain electrodes ($\lambda_{i}$=0). In the presence 
of interaction the tip-source and tip-drain non-linear 
conductances $G_{\ST}$ and $G_{\DT}$ oscillate as a function of 
the tip voltage $u_\text{T}$  already to leading order $\gamma^2$ in the 
tunneling amplitude, independent of contact impurity strengths 
$\lambda_{i}$. This effects holds when the tip acts as 
an electron injector (see Fig.~\ref{Fig-int-G2-1}) as well as when 
it acts as a voltage probe (see Fig.~\ref{fig:seven}), and the 
oscillations vanish for a non-interacting wire [see 
Eqs.~(\ref{IT-nonint-adiab}) and (\ref{IM-nonint-adiab})]. Thus, 
quite differently from a two-terminal set-up, in   three-terminal 
set-ups Andreev-type oscillations become better visible when the 
contact transparency is improved. 

In view of the fact that in realistic experiments the contact 
resistance is always finite, we have quantitatively evaluated the 
influence of the contact resistance on the conductance oscillations [see 
Eq.(\ref{jMT-G2L})] showing that additional Fabry-P\'erot-type 
oscillations superimpose with the Andreev-type ones (see 
Fig.\ref{fig:nine}). We have thus put forward criteria for 
observing the interaction induced Andreev-type oscillations. At 
least two experimental situations are promising: For the 
conventional case of symmetric tunneling form the tip, we have 
determined typical values of the contact resistance below which 
the oscillations in the current-voltage characteristics can 
essentially be attributed to Andreev-type phenomena. The result, 
shown in Table \ref{tab:one}, indicates that the stronger the 
interaction of the wire the larger are the contact resistances 
that are tolerable to still observe Andreev-type oscillations. 
Furthermore, in case that the set-up allows for fully asymmetric 
tunneling,  the leading order correction (\ref{jMT-G2L}) 
competing with the Andreev-type term is  vanishing, {\em even in 
the presence of interaction}. 
\\In summary, in systems like carbon nanotubes where the interaction 
strength is typically strong, $g \simeq 0.2 - 0.3$, while 
electron injection from an STM tip is typically symmetric, 
Andreev-type oscillations may be observed by achieving a high  
quality of the contacts to the leads.  In contrast, in 
semiconductor quantum wires, where the interaction strength is 
usually moderate $g \simeq 0.6-0.7$, asymmetric tunneling induced 
by a magnetic field is more suitable to observe 
Andreev-type oscillations.\\


\noindent {\it The effects of asymmetric tunneling.}
The above-mentioned case of asymmetric tunneling deserves   some further remarks. Recent experiments by Yacoby and co-workers\cite{yacoby,yacoby:2}
have shown that fully asymmetric tunneling  into
semiconductor-based quantum wires can be realized by appropriate
tuning of a magnetic field. Inspired by these
experiments, we have considered the possibility of an
asymmetry in electron tunneling from the tip. Before
discussing our results we would like to point out the relation
between our model and Yacoby's experimental set-up.
While Yacoby \emph{et al.} study  electron tunneling between two parallel wires where momentum conservation is required, our model considers injection from a point-like tip.
Although these two
situations may at first seem incompatible, a regime can be
determined where they are equivalent. In  the experiments of
Refs.~\onlinecite{yacoby,yacoby:2} electrons are
injected from an upper shorter wire with length $L_u$ into a lower
longer wire with length $L_l$. Since the tunneling region
reasonably coincides with the length of the short wire, momentum
conservation only holds up to an uncertainty $\delta k \sim 1/
L_u$. Although this uncertainty is small enough to
select a specific electron momentum state in the upper
wire, $\delta k $ may be much bigger than the mean level spacing
of the lower wire, if the latter is much longer than the former
($L_l \gg L_u$). In this regime, while the electron wave
function behaves like a plane wave for the short wire, for the
long wire it can effectively be considered as a localized wave
packet, and our model applies.

Under these conditions  several interesting effects emerge. In the
first instance, by using the tip as an electron injector, the 
tunneling asymmetry can be exploited to gain the transmission 
coefficient of each contact by measuring the current asymmetry 
(\ref{cur-asym-def}) in the two cases of tunneling purely to the 
right ($\chi=+1$) and to the left ($\chi=-1$), as has been shown 
in Eq.~(\ref{trans-coeff-from-asym}). Secondly, when the tip is 
used in the configuration of a voltage probe, fully asymmetric 
tunnelling allows to eliminate the suppression of the 
source-drain conductance $G_{\SD}$, which occurs for symmetric 
tunneling. Similarly,  $G_{\SD}$ becomes independent of the tip 
position.

When electron-electron interaction is taken into account, the
scenario is even richer. Luttinger liquid theory predicts that 
electron-electron interaction induces a current asymmetry which 
depends on the interaction strength $g$. The appealing question 
arises whether this effect is observable in experiments, where 
currents are measured not directly in the interacting wire but in 
metallic electrodes connected to it. The investigation carried 
out in Ref.~\onlinecite{yacoby-lehur}, based on the assumption 
that the interfaces between the interacting wire and the 
electrodes  can be treated phenomenologically with a  
transmission coefficient \`{a} la Landauer-B\"uttiker, 
 has led these authors to the claim that the interaction 
strength can be observed via the current asymmetry.   Here we 
have scrutinized this prediction by taking the presence of source 
and drain electrodes into account fully consistently within the 
inhomogeneous Luttinger liquid model. Considering as a test bench 
the case of a wire adiabatically contacted to source and drain 
electrodes, we have proven that, although charge 
fractionalization does occur in the bulk of the wire, the sum of 
Andreev-type reflection processes at the contacts leads to a 
current asymmetry $\mathcal{A}$ that is {\em independent} of the 
electron-electron interaction strength, just as it is the case 
with the two terminal conductance $G_\text{2t}$. Thus, already for 
this ideal case, no proof of charge fractionalization can be 
gained from the analysis of $\mathcal{A}$, or from the ration 
$e^2 \mathcal{A}/(h G_\text{2t} )$. We have also shown that, 
nevertheless, interaction effects do appear in the behavior of 
the nonlinear conductance, where interaction induced oscillations 
arise as a function of the tip-source and tip-drain bias. It is 
worth emphasizing that this feature is   due to the 
three-terminal set-up, since   the two-terminal conductance of a 
Luttinger liquid ideally contacted to leads is independent of the 
source-drain bias.

\acknowledgements The authors acknowledge stimulating discussions
with R.~Fazio, T.~Martin, P.~Recher, and B. Trauzettel, and   computational support by D.~Passerone. Funding was provided by the Deutsche Forschungsgemeinschaft (DFG), by the NANOFRIDGE EU Project,  by the ``Rientro dei Cervelli'' MIUR Program, as well as by the Italian-German collaboration program Vigoni.

\appendix
\section{Keldysh formalism and perturbative evaluation of the current}\label{AppA}

\begin{figure}[h]
\centering
\includegraphics[width=7cm,height=1.5cm,clip=]{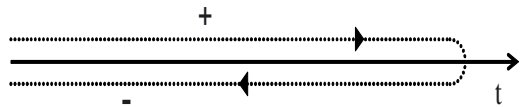}
\caption{\label{Fig-Kel-cont} Keldysh contour along the time axis.}
\end{figure}

In order to compute the current in the three terminal
set-up, we adopt the Keldysh formalism,\cite{KEL} suitable to
account for out-of-equilibrium properties. According to
Eq.~(\ref{I-W}), the current at position $x$ (located in the
source or in drain leads) and time $t$ can be written as
\begin{equation}
I(x,t) = \frac{e v_\W}{2} \sum_{\eta=\pm} \left\langle
j^{(\eta)}(x,t) \right\rangle
\end{equation}
where
\begin{equation}
j^{(\eta)}(x,t) =\sum_{r=\pm} r  :\Psi^{\dagger\, (\eta)}_{r}(x,t)
\Psi^{(\eta)}_{r}(x,t): \,.
\end{equation}
Here $\eta=+$ ($\eta=-$) corresponds to the upper (lower) branch
of the Keldysh contour depicted in Fig.~\ref{Fig-Kel-cont}. The
current $I(x,t)$ and various other quantities introduced below
also depend on the injection point $x_0$ and the impurity
positions $x_1$ and $x_2$. These variables will frequently be
suppressed to simplify notation.

In the Keldysh interaction picture with
\begin{equation}\label{hzero}
  \mathcal{H}_0=\mathcal{H}_\text{kin,W}+\mathcal{H}_{U}+\mathcal{H}_\text{kin,T}
\end{equation}
and
\begin{equation}\label{hint}
  \mathcal{H}_\text{I}=\mathcal{H}_{\lambda}
  +\mathcal{H}_\text{tun}+\mathcal{H}_{\mu_\W}+\mathcal{H}_{\mu_\T}\,
  ,
\end{equation}
one obtains
\begin{eqnarray}
\lefteqn{I(x,t) } & & \label{I-prel} \\
\nonumber\\
&=&\frac{e v_\W}{2} \sum_{\eta=\pm} \left\langle
T_\text{K} \left[ j^{(\eta)}(x,t) \ee^{ -\frac{\ii}{\hbar}
\sum_{\eta^\prime=\pm} \eta^\prime \int_{-\infty}^{\infty}
dt^\prime \mathcal{H}^{(\eta^\prime)}_\text{I}(t^\prime) } \right]
\right\rangle_0 \nonumber
\end{eqnarray}
where $\langle \ldots \rangle_0$ denotes the average with respect
to the equilibrium state determined by the Hamiltonian
$\mathcal{H}_0$, and $T_\text{K}$ is the Keldysh time-ordering
operator. Expanding the exponent in Eq.~(\ref{I-prel})
perturbatively in terms of $\gamma$ and $\lambda$, one obtains the
current to the desired order. Below we sketch the calculation of
$I_{\gamma^2 \lambda}$, \emph{i.e.}, the contribution of order
$\gamma^2 \lambda$ to $I(x,t)$. With the abbreviations
\begin{eqnarray}
\mathcal{U}_{\W} &=& \ee^{ -\frac{\ii}{\hbar}
\sum_{\eta^\prime=\pm} \eta^\prime \int_{-\infty}^{\infty}
dt^\prime \mathcal{H}^{(\eta^\prime)}_{\mu_{\W}}(t^\prime) }
\nonumber \\
\nonumber\\
\mathcal{U}_{\T} &=& \ee^{ -\frac{\ii}{\hbar}
\sum_{\eta^\prime=\pm} \eta^\prime \int_{-\infty}^{\infty}
dt^\prime \mathcal{H}^{(\eta^\prime)}_{\mu_{\T}}(t^\prime) }
\label{O-def}
\end{eqnarray} one obtains\cite{NOTA-G2L}
\begin{widetext}
\begin{eqnarray}
I_{\gamma^2 \lambda}(x,t) &=& \frac{\ii e v_\W^3 v_\T}{4}
\sum_{i=1,2}\, \sum_{r_1,r_2=\pm}\lambda_i \gamma_{r_1} \gamma_{r_2} \iiint
dt_1 dt_2 dt_3 \sum_{\eta, \eta_1, \eta_2, \eta_3=\pm} \eta_1
\eta_2 \eta_3 \Bigg\langle T_\text{K} \Bigg\{ j^{(\eta)}(x,t)
\, \mathcal{U}_\W \, \mathcal{U}_\T \nonumber \\
&\times& \left[ \ee^{-\ii(r_1-r_2) k_\W x_0} \Psi^{\dagger\,
(\eta_1)}_{r_1}(x_0,t_1)\, c^{(\eta_1)}(0,t_1)\,
c^{\dagger\, (\eta_2)}(0,t_2)\, \Psi^{(\eta_2)}_{r_2}(x_0,t_2) \right. \nonumber \\
&& + \left. \ee^{\ii(r_1-r_2) k_\W x_0} c^{\dagger\,
(\eta_1)}(0,t_1)\, \Psi^{(\eta_1)}_{r_1}(x_0,t_1)\,
\Psi^{\dagger\, (\eta_2)}_{r_2}(x_0,t_2)\,
c^{(\eta_2)}(0,t_2) \right] \nonumber \\
& \times & \sum_{r_3=\pm}
\ee^{-2\ii r_3 k_\W x_i} \Psi^{\dagger\, (\eta_3)}_{r_3}(x_i,t_3)\, \Psi^{(\eta_3)}_{-r_3}(x_i,t_3) \Bigg\} \Bigg\rangle_0\nonumber \\
\nonumber\\
&=&-\frac{e v^3_\W v_\T}{2} \sum_{i=1,2}\, \sum_{r_1,r_2=\pm}
\lambda_i \gamma_{r_1} \gamma_{r_2} \iiint dt_1 dt_2 dt_3
\hspace{-.4cm} \sum_{\eta, \eta_1, \eta_2, \eta_3=\pm} \eta_1
\eta_2 \eta_3 \ \Im
\Bigg\langle T_\text{K} \Bigg\{ j^{(\eta)}(x,t)\, \mathcal{U}_\W \, \mathcal{U}_\T \nonumber \\
& \times & \left[ \ee^{-\ii(r_1-r_2) k_\W x_0} \Psi^{\dagger\,
(\eta_1)}_{r_1}(x_0,t_1)\, c^{(\eta_1)}(0,t_1)
\, c^{\dagger\, (\eta_2)}(0,t_2)\, \Psi^{(\eta_2)}_{r_2}(x_0,t_2) \right] \nonumber \\
&\times & \sum_{r_3=\pm} \ee^{-2\ii r_3 k_\W x_i} \Psi^{\dagger\,
(\eta_3)}_{r_3}(x_i,t_3)\, \Psi^{(\eta_3)}_{-r_3}(x_i,t_3) \Bigg\}
\Bigg\rangle_0
\end{eqnarray}
where we have used the properties
\begin{eqnarray}
\left\langle T_\text{K} \left[A^{(\eta_A)}(t_A)
B^{(\eta_B)}(t_B)\ldots Z^{(\eta_Z)}(t_Z) \right] \right\rangle^*
= \left\langle T_\text{K} \left[ Z^{\dagger\, (-\eta_Z)}(t_Z)
\ldots B^{\dagger\, (-\eta_B)}(t_B) A^{\dagger\,
(-\eta_A)}(t_A)\right] \right\rangle \label{prop-KEL}
\end{eqnarray}
and
\begin{equation}
\mathcal{U}_{\W,\T} = \left(\ee^{ -\frac{\ii}{\hbar}
\sum_{\eta^\prime=\pm} \eta^\prime \int_{-\infty}^{\infty}
dt^\prime \mathcal{H}^{(-\eta^\prime)}_{\mu_{\W,\T}}(t^\prime) }
\right)^\dagger \quad . \label{prop-O}
\end{equation}
Since the electron-electron interaction (\ref{HU}) contains only
forward scattering terms, all non-vanishing wire correlation
functions must involve an even number of operators with a given
chirality $r$. This yields $r_2=-r_1=r_3$, so that
\begin{eqnarray}
\lefteqn{I_{\gamma^2 \lambda}(x,t) =-\frac{e v^3_\W v_\T}{2}
\gamma_{+} \gamma_{-} \sum_{i=1,2} \lambda_i \iiint dt_1 dt_2 dt_3
\sum_{\eta, \eta_1, \eta_2, \eta_3=\pm} \eta_1 \eta_2 \eta_3
\sum_{r_3=\pm} \Im \bigg\{ \ee^{2\ii r_3 k_\W (x_0-x_i)} \label{G2L-pre-1} } & & \\
& & \left\langle T_\text{K} \left[ j^{(\eta)}(x,t) \,
\mathcal{U}_\W \, \mathcal{U}_\T \Psi^{\dagger\,
(\eta_1)}_{-r_3}(x_0,t_1)\, c^{(\eta_1)}(0,t_1) \, c^{\dagger\,
(\eta_2)}(0,t_2)\, \Psi^{(\eta_2)}_{r_3}(x_0,t_2) \,
\Psi^{\dagger\,(\eta_3)}_{r_3}(x_i,t_3)
\Psi^{(\eta_3)}_{-r_3}(x_i,t_3) \right] \right\rangle_0 \bigg\}
\nonumber
\end{eqnarray}
The term with $r_3=+$ can be shown to yield the same contribution
as the term with $r_3=-$. To see this explicitly, one makes use of
$\Im(z)=-\Im(z^*)$, exploits Eqs.~(\ref{prop-KEL}) and
(\ref{prop-O}), and renames variables according to $\eta
\rightarrow -\eta$, $\eta_i \rightarrow -\eta_i$ ($i=1,2,3$) and
$t_1 \leftrightarrow t_2$. One can then write
\begin{eqnarray}
\label{G2L-pre-2}
\lefteqn{I_{\gamma^2 \lambda}(x,t)= } & & \\
&=& -e v^3_\W v_\T \, \gamma_{+} \gamma_{-} \sum_{i=1,2} \lambda_i
\iiint dt_1 dt_2 dt_3 \hspace{-.4cm} \sum_{\eta_1, \eta_2,
\eta_3=\pm} \eta_1 \eta_2 \eta_3 \ \Im \left\{ \ee^{-2 \ii k_W
(x_0-x_i)} \, {\rm W}^{\eta_1\eta_2 \eta_3}_{\gamma^2
\lambda,i}(t_1, t_2, t_3) \, {\rm
T}^{\eta_1\eta_2}_{\gamma^2}(t_1, t_2) \right\} \nonumber
\end{eqnarray}
where
\begin{equation}
{\rm W}^{\eta_1\eta_2 \eta_3}_{\gamma^2 \lambda,i}(t_1, t_2,
t_3)=\sum_{\eta=\pm} \left\langle T_\text{K} \left[
j^{(\eta)}(x,t) \, \mathcal{U}_\W \, \Psi^{\dagger\,
(\eta_1)}_{+}(x_0,t_1)\, \Psi^{(\eta_2)}_{-}(x_0,t_2) \,
\Psi^{\dagger\, (\eta_3)}_{-}(x_i,t_3) \,
\Psi^{(\eta_3)}_{+}(x_i,t_3)\right] \right\rangle_0
\label{WG2L-def}
\end{equation}
contains correlation functions of wire operators, while
\begin{equation}
{\rm T}^{\eta_1\eta_2}_{\gamma^2}(t_1, t_2)= \left\langle
T_\text{K} \left[ \mathcal{U}_{\T} \, c^{(\eta_1)}(t_1) \,
c^{\dagger\, (\eta_2)}(t_2) \right] \right\rangle_0
\label{TG2L-def}
\end{equation}
is a correlation function of the tip. These correlation functions
are evaluated in App.~\ref{AppB} starting with
Eq.~(\ref{W-G2L-prel}) and (\ref{T-G2L-prel}), respectively.
Inserting these results, one obtains
\begin{eqnarray}
\lefteqn{I_{\gamma^2 \lambda}(x,t)=2 e v_\W \left(\frac{v_\W}{2
\pi a_\W}\right)^2 \frac{v_\T}{2 \pi a_\T} \, \gamma_{+}
\gamma_{-} \sum_{i=1,2} \lambda_i \iiint
dt_1 dt_2 dt_3 } & & \nonumber \\
& & \Re \bigg\{ \ee^{i e [ ( V_{\s}+V_{\D} -2V_{\rm T}) (t_1-t_2)
+ (V_\s- V_\D) (t_1-t_3+t_2-t_3) ]/2\hbar } \,
\ee^{-2 \ii [k_\W +g^2 e(V_\s+V_\D-2 V_\G)/\hbar v_\W](x_0-x_i)} \, \nonumber \\
&& \times \sum_{\eta_1, \eta_2, \eta_3=\pm} \eta_1 \eta_2 \eta_3
{\rm F}^{\eta_1\eta_2\eta_3}_{\W}(t_1-t_2,
t_2-t_3) \, {\rm F}^{\eta_1\eta_2}_{\T}(t_1-t_2) \, {\rm
b}^{\eta_1\eta_2 \eta_3}_{\W,\gamma^2 \lambda,i}(t_1-t_3, t_2-
t_3) \,
{\rm b}^{\eta_1\eta_2}_{\T,\gamma^2}(t_1-t_2) \, \label{IG2L-3} \\
& & \times \bigg[ \langle \partial_x \Theta(x,t)
\Phi_{+}(x_0,t_1)\rangle^{\rm Kel}_0 + \eta_1 \langle
\partial_x \Theta(x,t) \Phi_{+}(x_0,t_1)\rangle^{\rm ret}_0 +
\langle
\partial_x \Theta(x,t) \Phi_{-}(x_0,t_2)\rangle^{\rm Kel}_0 +
\eta_2 \langle \partial_x \Theta(x,t) \Phi_{-}(x_0,t_2)\rangle^{\rm ret}_0 \nonumber \\
& & \hspace{0.6cm} - \langle \partial_x \Theta(x,t)
\Phi(x_i,t_3)\rangle^{\rm Kel}_0 - \eta_3 \langle \partial_x
\Theta(x,t) \Phi(x_i,t_2)\rangle^{\rm ret}_0 +\frac{1}{ \pi \hbar}
\iint dx^\prime dt^\prime \mu_\W(x^\prime) \langle T_\text{K}
\left[ \partial_x \Theta(x,t) \partial_{x^\prime}
\Phi(x^\prime,t^\prime) \right] \rangle^{\rm ret}_0 \bigg] \bigg\}
\nonumber
\end{eqnarray}
where, for any pair of bosonic operators $A$ and $B$, the
following definitions hold
\begin{align}
\langle A(t_A) B(t_B) \rangle^{\rm Kel} & = \langle \{ A(t_A), B(t_B) \} \rangle \label{kel-def} \\
\langle A(t_A) B(t_B) \rangle^{\rm ret} & = \theta(t_A-t_B) \langle [ A(t_A), B(t_B) ] \rangle \label{ret-def} \\
\langle A(t_A) B(t_B) \rangle^{\rm adv} & = -\theta(t_B-t_A)
\langle [ A(t_A), B(t_B) ] \rangle \quad. \label{adv-def}
\end{align}
We now observe that the last term in Eq.~(\ref{IG2L-3}) can be
dropped. Indeed, since it depends neither on $\eta_i$ nor on $t_i$
($i=1,2,3$), it can be singled out of the sums $\sum_{\eta_i}$ and
integrals $\int dt_i$; the remaining sums and integrations yield a
vanishing result, since the corresponding expression equals the
term of order $\gamma^2 \lambda$ of an expansion of
\begin{equation}
\left\langle T_\text{K} \left[ \ee^{ -\frac{\ii}{\hbar}
\sum_{\eta^\prime=\pm} \eta^\prime \int_{-\infty}^{\infty}
dt^\prime \mathcal{H}^{(\eta^\prime)}_\text{I} (t^\prime) }
\right] \right\rangle_0 \equiv 1\, .
\end{equation}
Simple transformations of the integration variables of
Eq.~(\ref{IG2L-3}), and use of the relations
\begin{eqnarray}
\int_{-\infty}^{\infty} dt \,\langle \partial_x \Theta(x,t) \Phi_{r}(x_0,0) \rangle^{\rm Kel}_0 \, &=& \, 0 \\
\int_{-\infty}^{\infty} dt \, \langle \partial_x \Theta(x,t)
\Phi_{r}(x_0,0) \rangle^{\rm ret}_0 \, &=& \, \frac{\ii}{4 v_\W}
\left[1+r \, \mbox{sgn}(x-x_0)\right]
\end{eqnarray}
obtained from the correlation functions provided in
App.~\ref{AppC}, yield
\begin{eqnarray}
I_{\gamma^2 \lambda}(x,t)
&=& -\frac{e v^3_\W v_\T }{16 \pi^3 a_\T a_\W^2 } \, \gamma_{+} \gamma_{-} \sum_{i=1,2} \lambda_i \iint dt_1 dt_2 \nonumber \\
& & \times \Im \bigg\{ \ee^{\ii e [ ( V_{\s}+V_{\D} -2V_{\rm T})
(t_1-t_2) + (V_\s- V_\D) (t_1-t_3+t_2-t_3) ] /2\hbar} \,
\ee^{-2\ii [k_\W +g^2 e (V_\s+V_\D-2 V_\G)/\hbar v_\W](x_0-x_i)} \, \nonumber \\
&& \times \sum_{\eta_1, \eta_2, \eta_3=\pm}
\eta_1 \eta_2 \eta_3 \, \, {\rm F}^{\eta_1\eta_2 \eta_3}_{\W}(t_1, t_2) \,
{\rm F}^{\eta_1\eta_2}_{\T}(t_1-t_2) \,
{\rm b}^{\eta_1\eta_2 \eta_3}_{\W,\gamma^2 \lambda,i}(t_1, t_2) \,
{\rm b}^{\eta_1\eta_2}_{\T,\gamma^2 \lambda,i}(t_1-t_2) \, \label{IG2L-4} \\
& & \times \left[ \eta_1+\eta_2-2 \eta_3 +\mbox{sgn}(x-x_0)
(\eta_1-\eta_2) \right] \bigg\} \nonumber\, .
\end{eqnarray}
Taking into account Eqs.~(\ref{FW-G2L}), (\ref{bWG2L}),
(\ref{FT-G2}), and (\ref{BT-G2}), we now observe that upon
reversal of Keldysh contour indices $\eta_i \rightarrow -\eta_i$
($i=1,2,3$),
\begin{eqnarray}
{\rm F}^{\eta_1\eta_2 \eta_3}_{\W}(t_1, t_2) &
\rightarrow & {\rm F}^{\eta_1\eta_2 \eta_3}_{\W}(t_1, t_2) \\
{\rm F}^{\eta_1\eta_2}_{\T}(t_1-t_2) &
\rightarrow & - {\rm F}^{\eta_1\eta_2}_{\T}(t_1-t_2) \\
{\rm b}^{\eta_1\eta_2 \eta_3}_{\W,\gamma^2 \lambda,i}(t_1, t_2) &
\rightarrow & \left[ {\rm b}^{\eta_1\eta_2 \eta_3}_{\W,\gamma^2 \lambda,i}(t_1, t_2) \right]^{*} \\
{\rm b}^{\eta_1\eta_2}_{\T,\gamma^2}(t_1-t_2) & \rightarrow &
\left[ {\rm b}^{\eta_1\eta_2}_{\T,\gamma^2 }(t_1-t_2) \right]^{*}
\, ,
\end{eqnarray}
implying that in Eq.~(\ref{IG2L-4}) the contribution for
$\eta_3=-$ is conjugate to the one stemming from $\eta_3=+$. Thus
\begin{eqnarray}
I_{\gamma^2 \lambda}(x,t)
&=& -\frac{e v^3_\W v_\T }{8 \pi^3 a_\T a_\W^2 } \, \gamma_{+} \gamma_{-} \sum_{i=1,2} \lambda_i \iint dt_1 dt_2 \nonumber \\
& & \Im \bigg\{\ee^{\ii e [ ( V_{\s}+V_{\D} -2V_{\rm T}) (t_1-t_2)
+ (V_\s- V_\D) (t_1+t_2) ]/2\hbar }
\, \ee^{-2\ii [k_\W +g^2 e(V_\s+V_\D-2 V_\G)/\hbar v_\W](x_0-x_i)} \, \label{IG2L-5} \\
& & \times \sum_{\eta_1, \eta_2 =\pm} \,{\rm F}^{\eta_1\eta_2
+}_{\W}(t_1, t_2) \, {\rm F}^{\eta_1\eta_2}_{\T}(t_1-t_2) \, {\rm
b}^{\eta_1\eta_2 +}_{\W,\gamma^2 \lambda,i}(t_1, t_2) \, {\rm
b}^{\eta_1\eta_2}_{\T,\gamma^2 }(t_1-t_2) \nonumber\\
&&\times \left[ \eta_2+\eta_1-2 \eta_1 \eta_2 +\mbox{sgn}(x-x_0)
(\eta_2-\eta_1) \right] \nonumber \bigg\}\, .
\end{eqnarray}
The term $\mbox{sgn}(x-x_0)$ appearing in the last line is
positive (negative) for a measurement point $x$ located in the
drain (source) lead. Recalling that the current can be written as
in Eqs.~(\ref{ISDT-1}), (\ref{ISDT-2}), it is easily seen that
those terms that are multiplied by $\mbox{sgn}(x-x_0)$ yield
$I_{\T}/2$, whereas the other ones yield $I_{\M}$. Inserting
Eqs.~(\ref{FW-G2L}), (\ref{bWG2L}), (\ref{FT-G2}), and
(\ref{BT-G2}) into Eq.~(\ref{IG2L-5}), and changing to
dimensionless integration variables $\tau_i= t v_\W/g L$, the
result (\ref{jMT-G2L}) is obtained.\\

Similar procedures can be applied to evaluate the terms of order
$\lambda^2$, $\lambda^3$ and $\gamma^2$. We find
\begin{eqnarray}
I_{\lambda^2}(x,t)= -\frac{e v^3_\W}{2} \sum_{i,j=1,2} \lambda_i
\lambda_j
\iint dt_1 dt_2 \sum_{\eta_1, \eta_2 =\pm}
\eta_1 \eta_2 \ \Re \left\{ \, {\rm W}^{\eta_1\eta_2}_{\lambda^2,ij}(t_1, t_2)
e^{-2i k_\W (x_i-x_j)} \right\} , \label{L2-pre-1}
\end{eqnarray}
\begin{eqnarray}
I_{\lambda^3}(x,t)= -\frac{e v^4_\W}{2} \sum_{i,j,k=1,2} \lambda_i
\lambda_j \lambda_k
\iiint dt_1 dt_2 dt_3 \sum_{\eta_1, \eta_2 =\pm}
\eta_1 \eta_2 \ \Im \left\{ \, {\rm W}^{\eta_1\eta_2}_{\lambda^3,ijk}(t_1, t_2,t_3)
 e^{-2i k_\W (x_i-x_j)} \right\} \label{L3-pre-1}
\end{eqnarray}
and
\begin{eqnarray}
I_{\gamma^2}(x,t)= -\frac{e v^2_\W v_\T}{2} \sum_{r=\pm}
\gamma^2_{r}
\iint \! dt_1 dt_2 \sum_{\eta_1, \eta_2 =\pm}
\eta_1 \eta_2 \
\Re \left\{ \, {\rm W}^{\eta_1\eta_2}_{\gamma^2,r}(t_1, t_2) \,
{\rm T}^{\eta_1\eta_2}_{\gamma^2}(t_1, t_2) \right\} \label{G2-pre-1}
\end{eqnarray}
where
\begin{equation}
{\rm W}^{\eta_1\eta_2}_{\lambda^2,ij}(t_1, t_2)=\sum_{\eta=\pm}
\left\langle T_\text{K} \left[ j^{(\eta)}(x,t)\, \mathcal{U}_\W \,
\Psi^{\dagger\, (\eta_1)}_{+}(x_i,t_1) \, \Psi^{(\eta_1)}_{-}(x_i,t_1) \,
\Psi^{\dagger\, (\eta_2)}_{-}(x_j,t_2) \, \Psi^{(\eta_2)}_{+}(x_j,t_2)\right]
\right\rangle_0 \, , \label{WL2-pre-1}
\end{equation}
\begin{equation}
{\rm W}^{\eta_1\eta_2}_{\lambda^3,ijk}(t_1,
t_2,t_3)=\sum_{\eta,\eta_3,r=\pm} \eta_3 \left\langle T_\text{K}
\left[ j^{(\eta)}(x,t)\, \mathcal{U}_\W \,
\Psi^{\dagger\, (\eta_1)}_{+}(x_i,t_1) \, \Psi^{(\eta_1)}_{-}(x_i,t_1) \,
\Psi^{\dagger\, (\eta_2)}_{-}(x_j,t_2) \, \Psi^{(\eta_2)}_{+}(x_j,t_2) \rho^{(\eta_3)}_{r}(x_k,t_3) \, \right]
\right\rangle_0\, ,\label{WL3-pre-1}
\end{equation}
and
\begin{equation}
{\rm W}^{\eta_1\eta_2}_{\gamma^2,r}(t_1, t_2)=\sum_{\eta=\pm}
\left\langle T_\text{K} \left[ j^{(\eta)}(x,t)\, \mathcal{U}_\W \,
\Psi_r^{\dagger\, (\eta_1)} (x_0,t_1) \Psi_r^{(\eta_2)} (x_0,t_2)
\,\right] \right\rangle_0\, \quad.\label{WG2-pre-1}
\end{equation}
\end{widetext}
\section{Evaluation of ${\rm W}$ and ${\rm T}$-factors by Bosonization}
\label{AppB}

The Hamiltonian (\ref{hzero}) of the interaction picture
decomposes into commuting wire and tip parts, {\em i.e.},
$\mathcal{H}_0=\mathcal{H}_{0,\W} + \mathcal{H}_{0,\T}$. For a
non-interacting wire $\mathcal{H}_{0,\W} =
\mathcal{H}_\text{kin,W}$, and the wire correlation functions
${\rm W}$ introduced in Eqs.~(\ref{WG2L-def}), (\ref{WL2-pre-1}),
(\ref{WL3-pre-1}) and (\ref{WG2-pre-1}) can be factorized into products of
single-particle electron correlators using Wick's theorem. In this
case the W's can be evaluated straightforwardly, and the results
for the contributions (\ref{G2L-pre-2}), (\ref{L2-pre-1}),
(\ref{L3-pre-1}) and (\ref{G2-pre-1})
to the current coincide with the corresponding
terms of an expansion of the current obtained from the scattering
matrix formalism. In the interacting case, however,
$\mathcal{H}_{0,\W} = \mathcal{H}_\text{kin,W}+\mathcal{H}_{U}$,
and Wick's theorem cannot be applied. In this appendix we evaluate
the wire correlators W using the bosonization technique.\cite{BOS}
The wire field operators can be represented as
\begin{equation}
\Psi_r(x)=\frac{\kappa_r}{\sqrt{2 \pi a_\W}} \, \ee^{\ii r \sqrt{4
\pi} \Phi_r(x)} \label{bos-trans}
\end{equation}
where the fields $\Phi_{\pm}$ describe particle-hole excitations,
and $\kappa_r$ are Klein factors represented as Majorana
fermions.\cite{BOS} Finally, $a_\W$ is a cut-off length of order
the lattice spacing.
\\
Introducing Eq.~(\ref{bos-trans}) into Eqs.~(\ref{HW}) and
(\ref{Hgamma}), one obtains
\begin{equation}
\mathcal{H}_{0,\W} = \frac{\hbar v_\W}{2} \int_{-\infty}^{\infty}
\! \!\!\! dx \left\{ \!:\! \Pi^2(x) +\frac{1}{g^2(x)} \left[\partial_x
\Phi(x)\right]^2\!\! :\! \right\}
\end{equation}
where $\Phi=\Phi_+ + \Phi_-$ and $\Pi=-\partial_x(\Phi_+ -\Phi_-)$
are conjugate bosonic fields, \emph{i.e.}, $[ \Phi(x,t),
\Pi(y,t)]= \ii \delta(x-y)$. Finally,
\begin{equation}
g(x)= \left\{
\begin{array}{ccr}
1 & \mbox{for} & |x| > L/2 \\
\left(1 +\frac{U}{\pi \hbar v_\W} \right)^{-1/2} & \mbox{for} & |x| < L/2 \\
\end{array}
\right.
\end{equation}
is the inhomogeneous interaction parameter. Notice that $0 \le g
\le 1$, where $g=1$ describes the non-interacting case present in
the leads. The limit $g \rightarrow 0$ corresponds to strongly
repulsive interaction. The wire current operator Eq.~(\ref{I-W}) is
expressed in terms of the dual field $\Theta=\Phi_{+} - \Phi_{-}$
as
\begin{equation}
I(x,t) = e v_\W \langle \partial_x \Theta(x,t) \rangle \, .
\label{I-W-BOS}
\end{equation}
Further, with the help of the relation
\begin{equation}
\rho_r(x,t) \, =\frac{\partial_x \Phi_r(x,t)}{\sqrt{\pi}}\, , \label{bos-dens}
\end{equation}
the term (\ref{HmuW}) of the Hamiltonian can be written as
\begin{equation}
\mathcal{H}_{\mu_\W}=\frac{1}{\sqrt{\pi}}\int_{-\infty}^{+\infty}
dx \mu_\W(x) \partial_x \Phi(x) \, .
\end{equation}

We start by discussing the derivation of ${\rm W}^{\eta_1 \eta_2
\eta_3}_{\gamma^2 \lambda,i}(t_1, t_2, t_3)$. Inserting
Eqs.~(\ref{bos-trans}), (\ref{I-W-BOS}) and (\ref{bos-dens}) into
Eq.~(\ref{G2L-pre-2}), one obtains
\begin{eqnarray}
\lefteqn{{\rm W}^{\eta_1\eta_2 \eta_3}_{\gamma^2 \lambda,i}(t_1,
t_2, t_3) =}
& & \label{W-G2L-prel} \\
&=& \frac{1}{(2 \pi a_\W)^2} {\rm F}^{\eta_1\eta_2
\eta_3}_{\W}(t_1-t_3, t_2- t_3) \, {\rm B}^{\eta_1\eta_2
\eta_3}_{\W,\gamma^2 \lambda,i}(t_1, t_2, t_3) \nonumber
\end{eqnarray}
Here
\begin{widetext}
\begin{eqnarray}
\lefteqn{{\rm F}^{\eta_1\eta_2 \eta_3}_{\W}(t_1-t_3, t_2- t_3) =
\left\langle T_\text{K} \left[ \kappa^{(\eta_1)}_{+}(t_1)
\kappa^{(\eta_2)}_{-}(t_2)
\kappa^{(\eta_3)}_{-}(t_3) \kappa^{(\eta_3)}_{+}(t_3) \right] \right\rangle_0 } \label{FW-G2L} \\
&=& \theta(t_3-t_1) \theta(t_3-t_2) \eta_1 \eta_2 +
\theta(t_1-t_3) \theta(t_2-t_3) - \theta(t_2-t_3) \theta(t_3-t_1)
\eta_1 \eta_3 -\theta(t_1\!-t_3) \theta(t_3-t_2) \eta_2 \eta_3 \,
\nonumber
\end{eqnarray}
accounts for the correlation function of fermionic Klein factors,
whereas
\begin{eqnarray}
\lefteqn{{\rm B}^{\eta_1\eta_2 \eta_3}_{\W,\gamma^2 \lambda,i}(t_1, t_2, t_3) =} & & \nonumber \\
&=& \sum_{\eta=\pm} \frac{\delta}{\delta J^{(\eta)}_\Theta(x,t)}
\Bigg\langle T_\text{K} \Bigg\{ \exp \Bigg(
-\frac{\ii}{\hbar}\sum_{\eta^\prime=\pm} \!
\eta^\prime \int d\mathbf{x}^\prime \mu_\W(x^\prime) \,
\frac{\partial_{x^\prime} \Phi^{(\eta^\prime)}(\mathbf{x}^\prime)}{\sqrt{\pi}}
 + \sum_{\eta^{''}=\pm} \int d\mathbf{x}^{''}
  J^{(\eta^{\prime \prime})}_\Theta(\mathbf{x}^{\prime \prime})
  \frac{\partial_x \Theta^{(\eta^{\prime \prime})}(\mathbf{x}^{\prime \prime})}{\sqrt{\pi}}
     \nonumber \\
& & \hspace{3.5cm} \left. -i\sqrt{4 \pi} \,
\left[ \Phi^{(\eta_1)}_{+}(x_0,t_1)+\Phi^{(\eta_2)}_{-}(x_0,t_2)-\Phi^{(\eta_3)} (x_i,t_3) \right]
\Bigg) \Bigg\} \Bigg\rangle_0 \right|_{J_{\Theta} \equiv 0} \label{G2L-pre-3}
\end{eqnarray}
correlates bosonic vertex operators. Also, we have introduced the
notation $\mathbf{x}=(x, t)$. The expression (\ref{G2L-pre-3}) can
straightforwardly be evaluated taking into account that for a
functional
\begin{equation}
\zeta[J]= \left\langle T_\text{K} \left\{ \exp
\left(A+\sum_{\eta=\pm} \int d\mathbf{x} J^{(\eta)}(\mathbf{x})
B^{(\eta)}(\mathbf{x}) \right) \right\} \right\rangle_0\, ,
\end{equation}
where $A$ and $B$ are linear combinations of bosonic operators,
one has\cite{BOS}
\begin{equation}
\left. \frac{\delta \zeta[J]}{\delta
J^{(\eta)}(\mathbf{x})}\right|_{J=0} = \left\langle T_\text{K}
\left[ A \, B^{(\eta)} (\mathbf{x}) \right]\right\rangle_0 \, \,
\exp\left\{{\left\langle T_\text{K}
\left(A^2\right)\right\rangle_0}\right\} \, .
\end{equation}
Furthermore, it can be shown that $\mathcal{H}_{\mu_\W}$, \emph{i.e}. the
first term appearing in the exponent of Eq.~(\ref{G2L-pre-3}),
simply yields a shift in the operators $\Phi_{\pm}$ according to
\begin{equation}
\Phi^{(\eta)}_r(x,t) \rightarrow \Phi^{(\eta)}_r(x,t) +
\Phi_{0,r}(x,t) \hspace{1cm} r=\pm \quad,
\end{equation}
where the zero modes 
\begin{eqnarray}
\lefteqn{\Phi_{0,r}(x,t) = -\frac{1}{4\sqrt{\pi}} \frac{e\left[V_\s-V_\D+r (V_\s+V_\D)\right]}{\hbar }\, t } & & \label{zero-mode} \\
\nonumber\\
&+& \frac{e}{4 \sqrt{\pi}\, \hbar v_\W } \left\{
\begin{array}{lcl}
\displaystyle -(V_\s-V_\D) \left[(1-r) x +\frac{L}{2}\right]-g^2 (V_\s+V_\D-2V_\G) \frac{L}{2} & & \mbox{for } x \le -L/2 \\
&&\\
\displaystyle [ g^2 (V_\s+V_\D-2V_\G) +r (V_\s-V_\D) \, ] \, x & & \mbox{for } |x| \le L/2 \\
&&\\
\displaystyle (V_\s-V_\D) \left[(1+r) x -\frac{L}{2}\right]+g^2
(V_\s+V_\D-2V_\G) \frac{L}{2} & & \mbox{for } x \ge L/2 
\end{array} \right. \nonumber
\end{eqnarray}
fulfill  the equation
\begin{equation}
 \Phi_{0,r}(\mathbf{x})-\Phi_{0,r}(\mathbf{y}) = \frac{-i}{\sqrt{\pi} \hbar}\int   d\mathbf{x}^\prime \, \mu_\W(\mathbf{x}^\prime) \, \left[ \langle \Phi_r(\mathbf{x}) \partial_x \Phi(\mathbf{x}^\prime) \rangle^{\rm ret} - \langle \Phi_r(\mathbf{y}) \partial_x \Phi(\mathbf{x}^\prime) \rangle^{\rm ret} \right] \quad.   \label{zero-mode-eq}
\end{equation}
After lengthy but straightforward algebra one obtains
\begin{eqnarray}
\lefteqn{{\rm B}^{\eta_1\eta_2 \eta_3}_{\W,\gamma^2 \lambda,i}(t_1, t_2, t_3) =} & & \nonumber \\
&=& -2 \ii \, e^{(\ii e/\hbar) \left[ V_\s (t_1-t_3) - V_\D
(t_2-t_3) -g^2 (V_\s+V_\D-2 V_\G)(x_0-x_i)/v_\W\right] }
\, b^{\eta_1\eta_2 \eta_3}_{\W,\gamma^2 \lambda,i}(t_1-t_3, t_2- t_3) \nonumber \\
& & \times \bigg\{ \langle \partial_x \Theta(x,t)
\Phi_{+}(x_0,t_1)\rangle^{\rm Kel}_0 + \eta_1 \langle \partial_x
\Theta(x,t) \Phi_{+}(x_0,t_1)\rangle^{\rm ret}_0 + \langle
\partial_x \Theta(x,t) \Phi_{-}(x_0,t_2)\rangle^{\rm Kel}_0
+ \eta_2 \langle \partial_x \Theta(x,t) \Phi_{-}(x_0,t_2)\rangle^{\rm ret}_0 \nonumber \\
& & \hspace{0.6cm} - \langle \partial_x \Theta(x,t)
\Phi(x_i,t_3)\rangle^{\rm Kel}_0 - \eta_3 \langle \partial_x
\Theta(x,t) \Phi(x_i,t_2)\rangle^{\rm ret}_0 +\frac{1}{ \pi \hbar}
\int d{\mathbf{x}^\prime} \mu_\W(x^\prime) \langle T_\text{K}
\left[ \partial_x \Theta(x,t) \partial_{x^\prime}
\Phi(\mathbf{x}^\prime) \right] \rangle^{\rm ret}_0 \bigg\}\, ,
\label{G2L-pre-4}
\end{eqnarray}
where
\begin{eqnarray}
b^{\eta_1\eta_2 \eta_3}_{\W,\gamma^2 \lambda,i}(t_1, t_2) &=&
\exp{\left\{ - 2 \pi \left\langle T_\text{K} \left[ \left(
\Phi^{(\eta_1)}_{+}(x_0,t_1)+\Phi^{(\eta_2)}_{-}(x_0,t_2)
-\Phi^{(\eta_3)}(x_i,0)\right)^2 \right] \right\rangle_0 \right\}} \label{bWG2L} \\
\nonumber\\
&=& \exp \left\{ 4 \pi \left[
\mathcal{R}_{\W}^{}(\xi_0;\xi_i;\tau_1;\tau_2) \, + \, \ii \,
\mathcal{I}^{\eta_1 \eta_2 \eta_3}_{\W}(\xi_0;\xi_i;\tau_1;\tau_2)
\right] \right\}\, . \nonumber
\end{eqnarray}
The correlation functions
$\mathcal{R}_{\W}^{}(\xi_0;\xi_i;\tau_1;\tau_2)$ and
$\mathcal{I}^{\eta_1 \eta_2
\eta_3}_{\W}(\xi_0;\xi_i;\tau_1;\tau_2)$ are defined in
App.~\ref{AppC} [see Eqs.~(\ref{R_W}) - (\ref{c6})] and also given
explicitly there in the zero temperature limit. The arguments
$\tau_i=t_i v_\W/g L=t_i \omega_L^*$ and $\xi_j=x_j/L$ ($j=0,1,2$) are
dimensionless time and space variables. In deriving
Eqs.~(\ref{G2L-pre-4}) and (\ref{bWG2L}) we have used the
equalities
\begin{equation}
\left\langle T_\text{K} \left[ A^{(\eta_A)}(t_A) B^{(\eta_B)}(t_B)
\right]\right\rangle=\frac{1}{2} \left[ \langle A(t_A) B(t_B)
\rangle^{\rm Kel}+\eta_A \langle A(t_A) B(t_B) \rangle^{\rm adv}
+\eta_B \langle A(t_A) B(t_B) \rangle^{\rm ret} \right]
\end{equation}
\\and
\begin{eqnarray}
\langle A(t_A) B(t_B) \rangle^{\rm Kel}
&=& \frac{1}{2} \sum_{\eta_A, \eta_B=\pm} \langle A^{(\eta_A)}(t_A) B^{(\eta_B)}(t_B) \rangle = 2 \Re \langle A(t_A) B(t_B) \rangle \label{kel-prop} \\
\langle A(t_A) B(t_B) \rangle^{\rm ret}
&=& \frac{1}{2} \sum_{\eta_A, \eta_B=\pm} \eta_B \langle A^{(\eta_A)}(t_A) B^{(\eta_B)}(t_B) \rangle = 2 \ii \theta(t_A-t_B) \, \Im \langle A(t_A) B(t_B) \rangle \label{ret-prop} \\
\langle A(t_A) B(t_B) \rangle^{\rm adv} &=& \frac{1}{2}
\sum_{\eta_A, \eta_B=\pm} \eta_A \langle A^{(\eta_A)}(t_A)
B^{(\eta_B)}(t_B) \rangle = -2 \ii \theta(t_B-t_A) \, \Im \langle
A(t_A) B(t_B) \rangle \label{adv-prop}
\end{eqnarray}
valid for any pair $A$ and $B$ of real Bose operators.\\

As far as the tip correlators $\rm T$ are concerned, see
Eqs.~(\ref{TG2L-def}), (\ref{G2-pre-1}), and (\ref{L2-pre-1}),
Wick's theorem might be applied, since the tip is supposed to be
non-interacting, and the use of bosonization is unnecessary.
However, to have a uniform formalism and notation throughout the
paper, we prefer to utilize a bosonized approach for the tip as
well. The tip electron field and density are written as
\begin{equation}
c(y)= \frac{\kappa_\T}{\sqrt{2 \pi a_\T}} \, \ee^{\ii \sqrt{4 \pi}
\varphi(y)} \label{c-op-bos}
\end{equation}
and
\begin{equation}
:c^\dagger(y) c^{}(y): = \frac{\partial_y \varphi(y)}{\sqrt{\pi}}
\, , \label{dens-T-bos}
\end{equation}
where $\varphi(y)$ is a chiral (right-moving) boson field, and
$\kappa_\T$ and $a_\T$ are the Klein factor and cutoff length of
the tip, respectively. By way of example, we evaluate here the
T-factor (\ref{TG2L-def}) appearing in the calculation of
$I_{\gamma^2 \lambda}$. Inserting Eqs.~(\ref{c-op-bos}) and
(\ref{dens-T-bos}) into Eqs.~(\ref{O-def}) and (\ref{TG2L-def}),
one obtains
\begin{eqnarray}
{\rm T}^{\eta_1\eta_2}_{\gamma^2}(t_1-t_2) =
\frac{1}{2 \pi a_\T} {\rm F}^{\eta_1\eta_2}_{\T}(t_1-t_2) \, {\rm B}^{\eta_1\eta_2}_{\T,\gamma^2}(t_1- t_2) \label{T-G2L-prel}
\end{eqnarray}
where, similar to the wire case,
\begin{eqnarray}
{\rm F}^{\eta_1\eta_2}_{\T}(t_1-t_2)= \left\langle T_\text{K}
\left[ \kappa^{(\eta_1)}_\T(t_1) , \kappa^{(\eta_2)}(t_2) \right]
\right\rangle_0 =
-\eta_1 \theta(t_2-t_1) + \eta_2 \theta(t_1-t_2) \label{FT-G2}
\end{eqnarray}
accounts for the correlation function of fermionic Klein factors,
whereas the correlator of bosonic vertex operators reads
\begin{eqnarray}
{\rm B}^{\eta_1\eta_2 }_{\T,\gamma^2}(t_1-t_2) = \left\langle
T_\text{K} \left[ \exp \left( -\frac{\ii}{\hbar}
\sum_{\eta^\prime=\pm}
\! \eta^\prime \int d\mathbf{y}^\prime \mu_{\T}(y^\prime) \,
 \frac{\partial_{y^\prime} \varphi(\mathbf{y}^\prime)}{\sqrt{\pi}}
   +\ii \sqrt{4 \pi} \left[\varphi^{(\eta_1)}(0,t_1)-\varphi^{(\eta_2)}(0,t_2)\right] \right) \right] \right\rangle_0 \, . \label{BT-G2-pre-1}
\end{eqnarray}
It is easily verified that the first term in the exponential
function, which originates from the term (\ref{Hmu-T}) in the
Hamiltonian, merely yields a time-dependent phase factor, so that
\begin{equation}
{\rm B}^{\eta_1\eta_2 }_{\T,\gamma^2}(t_1-t_2)=\ee^{-(\ii e/\hbar)
V_\T (t_1-t_2)} \, b^{\eta_1 \eta_2}_{\T, \gamma^2} (t_1-t_2)
\label{BT-G2}
\end{equation}
where
\begin{eqnarray}
b^{\eta_1 \eta_2}_{\T, \gamma^2} (t_1-t_2)&=&
\exp \left\{-2 \pi \left\langle T_\text{K} \left[ \left(\varphi^{(\eta_1)}(0,t_1) - \varphi^{(\eta_2)}(0,t_2)\right)^2 \right]\right\rangle_0 \right\} \label{bT-G2} \\
&=& \exp \left\{ 4 \pi \left[ \mathcal{R}_{\T}(\tau_1-\tau_2) \, +
\, \ii \, \mathcal{I}^{\eta_1 \eta_2}_{\T}(\tau_1-\tau_2) \right]
\right\} \, . \nonumber
\end{eqnarray}
The functions $\mathcal{R}_{\T}^{}(\tau)$ and
$\mathcal{I}_{\T}^{\eta_1\eta_2}(\tau)$ are given in
App.~\ref{AppC}, see Eqs. (\ref{R_T}) and (\ref{I_T}).
\section{Correlation functions}
\label{AppC} This appendix collects properties of correlation
functions appearing in Eqs.~(\ref{jMT-G2}) and (\ref{jMT-G2L}), as
well as in Eqs.~(\ref{bWG2L}) and (\ref{bT-G2}). The transport
properties of the wire are expressed in terms of the functions
\begin{eqnarray}
\mathcal{R}_{\W}(\xi_0;\xi_i;\tau_1;\tau_2) &=&
\mathcal{R}^{\Phi_{+} \Phi_{+}}_\text{reg}(\xi_0;\xi_i;\tau_1) +
\mathcal{R}^{\Phi_{-} \Phi_{-}}_\text{reg}(\xi_0;\xi_i;\tau_2) +
\mathcal{R}^{\Phi_{+} \Phi_{-}}(\xi_0;\xi_i;\tau_1)
+ \mathcal{R}^{\Phi_{-} \Phi_{+}}(\xi_0;\xi_i;\tau_2) \label{R_W} \\
& & - \mathcal{R}^{\Phi_{+} \Phi_{-}}(\xi_0;\xi_0;\tau_1-\tau_2)
- \mathcal{R}^{\Phi_{-} \Phi_{+}}(\xi_i;\xi_i;0) \nonumber \\
\mathcal{I}^{\eta_1 \eta_2 \eta_3}_{\W}(\xi_0;\xi_i;\tau_1;\tau_2)
&=& \sum_{r=\pm} \left\{ \left[\eta_3 \theta(\tau_1)-\eta_1
\theta(-\tau_1) \right] \,
\mathcal{I}^{\Phi_{+} \Phi_r}(\xi_0;\xi_i;\tau_1) + \left[\eta_3\theta(\tau_2)-\eta_2 \theta(-\tau_2) \right] \,
\mathcal{I}^{\Phi_{-} \Phi_r}(\xi_0;\xi_i;\tau_2) \right\} \nonumber \\
& & \hspace{1cm} -\left[\eta_2 \theta(\tau_1-\tau_2)-\eta_1
\theta(\tau_2-\tau_1) \right] \,
\mathcal{I}^{\Phi_{+} \Phi_{-}}(\xi_0;\xi_0;\tau_1-\tau_2) \, .\label{I_W}
\end{eqnarray}
where the functions $\mathcal{R}_\text{reg}^{\Phi_r
\Phi_{r}}(\xi;\xi^\prime;\tau)$ and $\mathcal{I}^{\Phi_r
\Phi_{r}}(\xi;\xi^\prime;\tau)$ are the real and imaginary parts,
respectively, of the auto-correlation functions of the bosonic
fields $\Phi_r$. Specifically
\begin{eqnarray}
\mathcal{R}^{\Phi_r \Phi_r}_\text{reg}(\xi;\xi^\prime;\tau) &=&\Re
\left\{ \left\langle \Phi_r(x,t) \Phi_r(x^\prime,0)-\frac{1}{2}
\left[\Phi^2_r(x,t)+\Phi^2_r(x^\prime,0)\right] \right\rangle_0 \right\} \label{phireg} \\
\mathcal{I}^{\Phi_r \Phi_r}(\xi;\xi^\prime;\tau)&=&\Im \left\{
\langle \Phi_r(x,t) \Phi_r(x^\prime,0) \rangle_0 \right\}
\label{phiim}
\end{eqnarray}
Likewise, the real and imaginary parts of the cross-correlation
functions of fields with different chirality $r$ read
\begin{eqnarray}
\mathcal{R}^{\Phi_r \Phi_{-r}}(\xi;\xi^\prime;\tau)=\Re \left\{ \langle \Phi_r(x,t) \Phi_{-r}(y,0) \rangle_0 \right\} \\
\mathcal{I}^{\Phi_r \Phi_{-r}}(\xi;\xi^\prime;\tau)=\Im \left\{
\langle \Phi_r(x,t) \Phi_{-r}(y,0) \rangle_0 \right\}\label{c6}
\end{eqnarray}
Notice that the real part of the correlation functions of fields
with the same chirality needs to be defined with an infrared
regularization as in Eq.~(\ref{phireg}). The above equations are
given in terms of the dimensionless time and space variables
$\tau=t v_\W/g L$ and $\xi=x/L$ introduced previously. From the
inhomogeneous Luttinger liquid model one obtains at zero
temperature \small{
\begin{equation}
\begin{array}{lcl}
\mathcal{R}^{\Phi_r \Phi_r}_\text{reg}(\xi,\xi^\prime,\tau)
&=&\displaystyle -\frac{1}{32 \pi} \left\{ \left(g+g^{-1}-2
r\right)
\sum_{m \in Z_{\rm even}} \rho^{|m|} \ln{ \frac{\alpha_{\W}^2+(\tau+\xi_r + m)^2}{\alpha_{\W}^2+m^2} }  \right. \\
& & \displaystyle \hspace{1cm} +\left(g+g^{-1}+2 r\right)
\sum_{m \in Z_{\rm even}} \rho^{|m|}\ln{ \frac{\alpha_{\W}^2+(\tau-\xi_r - m)^2}{\alpha_{\W}^2+m^2} }  \label{rr-ww-GS-Re} \\
& & \displaystyle \hspace{1cm}+ \left(g-g^{-1}\right) \sum_{m \in
Z_{\rm odd}} \rho^{|m|} \left( \ln{
\frac{\alpha_{\W}^2+(\tau+\xi_R + m)^2}{\alpha_{\W}^2+(\xi_R +
m)^2} } + \ln{ \frac{\alpha_{\W}^2+(\tau-\xi_R -
m)^2}{\alpha_{\W}^2+(\xi_R + m)^2} }
\right.  \\
& & \left. \hspace{5cm} \displaystyle \left. +\ln{
\frac{\left[\alpha_{\W}^2+(\xi_R+m)^2\right]^2}{[\alpha_{\W}^2+(2\xi+m)^2]
[\alpha_{\W}^2+(2\xi^\prime+m)^2]} } \right) \right\}\\
\end{array}
\end{equation}
\begin{equation}
\begin{array}{lcl}
\mathcal{I}^{\Phi_r \Phi_r}(\xi,\xi^\prime,\tau)&=&\displaystyle
-\frac{1}{16 \pi} \left\{ \left(g+g^{-1}-2 r\right)
\sum_{m \in Z_{\rm even}} \rho^{|m|} \arctan{ \left( \frac{\tau+\xi_r + m}{\alpha_{\W}} \right) }  \right. \\
& & \displaystyle \hspace{1cm}+ \left(g+g^{-1}+2 r\right)
\sum_{m \in Z_{\rm even}} \rho^{|m|} \arctan{ \left( \frac{\tau-\xi_r - m}{\alpha_{\W}} \right) }  \label{rr-ww-GS-Im} \\
& & \displaystyle \hspace{1cm} \left. +\left(g-g^{-1}\right)
\sum_{m \in Z_{\rm odd}} \rho^{|m|} \left[ \arctan{ \left(
\frac{\tau+\xi_R + m}{\alpha_{\W}}\right) }  + \arctan{
\left(\frac{\tau-\xi_R - m}{\alpha_{\W}}\right)} \right] \right\}
\end{array}
\end{equation}
and
\begin{eqnarray}
\mathcal{R}^{\Phi_r \Phi_{-r}}(\xi;\xi^\prime;\tau) &=& -
\frac{1}{32 \pi} \left\{  \left(g-g^{-1}\right) \sum_{m \in Z_{\rm
even}} \!\!\! \rho^{|m|} \left[ \ln{
\left(\frac{\alpha_{\W}^2+(\tau+\xi_r + m)^2}{\alpha_{\W}^2+m^2}
\right) } + \ln{ \left(\frac{\alpha_{\W}^2+(\tau-\xi_r -
m)^2}{\alpha_{\W}^2+m^2}
\right) } \right] \right.\nonumber \\
& & \hspace{1cm}+ \left(g+g^{-1}-2r\right) \sum_{m \in Z_{\rm
odd}} \!\!\! \rho^{|m|} \ln{\left( \frac{\alpha_{\W}^2+(\tau+\xi_R
+ m)^2}{\alpha_{\W}^2+(\xi_R + m)^2}
\right)} \nonumber  \\
& & \hspace{1cm}+ \left(g+g^{-1}+2r\right) \sum_{m \in Z_{\rm
odd}} \!\!\! \rho^{|m|} \ln{\left( \frac{\alpha_{\W}^2+(\tau-\xi_R
- m)^2}{\alpha_{\W}^2+(\xi_R + m)^2}
\right)} \label{rmr-ww-GS-Re}  \\
& & \hspace{1cm}+ \left(g+g^{-1}\right) \sum_{m \in Z_{\rm odd}}
\rho^{|m|} \ln{ \frac{\left[\alpha_{\W}^2+(\xi_R+m)^2
\right]^2}{[\alpha_{\W}^2+(2\xi+m)^2]
[\alpha_{\W}^2+(2\xi^\prime+m)^2]} }
\nonumber \\
& & \hspace{1cm} - \frac{1}{2}\left(g-g^{-1}\right) \sum_{m \in
Z_{\rm even} } \rho^{|m|} \left[ \ln{ \left(
\frac{[\alpha_{\W}^2+( 2\xi+ 1 + m)^2] [\alpha_{\W}^2+(
2\xi^\prime+ 1 + m)^2]}{(\alpha_{\W}^2+ m^2)^2}
\right) }  \right. \nonumber \\
& & \hspace{4.6cm} + \left. \left. \ln{ \left(
\frac{[\alpha_{\W}^2+( 2\xi- 1 + m)^2] [\alpha_{\W}^2+(
2\xi^\prime- 1 + m)^2]}{(\alpha_{\W}^2+ m^2)^2} \right) } \right]
\right\} \nonumber
\end{eqnarray}
and
\begin{eqnarray}
\mathcal{I}^{\Phi_r \Phi_{-r}}(\xi;\xi^\prime;\tau) &=& -
\frac{1}{16 \pi} \left\{  \left(g-g^{-1}\right) \sum_{m \in Z_{\rm
even}} \!\!\! \rho^{|m|} \left[ \arctan \left( \frac{\tau +\xi_r +
m}{\alpha_{\W}} \right)  +  \arctan \left(\frac{\tau -\xi_r -
m}{\alpha_{\W}} \right) \right]
\right.\nonumber \\
& & \hspace{1cm}+ \left(g+g^{-1}-2r\right) \sum_{m \in Z_{\rm
odd}} \!\!\! \rho^{|m|} \arctan \left( \frac{\tau+\xi_R +
m}{\alpha_{\W}}
\right)  \label{rmr-ww-GS-Im} \\
& & \hspace{1cm}\left. + \left(g+g^{-1}+2r\right) \sum_{m \in
Z_{\rm odd}} \!\!\! \rho^{|m|} \arctan \left( \frac{\tau-\xi_R -
m}{\alpha_{\W}} \right) \right\} \, .\nonumber
\end{eqnarray}
Here we have introduced $\xi_r=\xi-\xi^\prime$,
$\xi_R=\xi+\xi^\prime$, and the dimensionless cutoff length
$\alpha_\W=a_\W/gL$, as well as the Andreev-type reflection
coefficient $\rho=(1-g)/(1+g)$.

The correlation functions for the non-interacting tip can directly
be obtained from the above results. The tip is described by a
single chiral mode, and we need the correlation function only for
coordinates at the injection point $y=0$. From
Eqs.~(\ref{rr-ww-GS-Re}) and (\ref{rr-ww-GS-Im}) we find for
$\xi=\xi^\prime=0$ by taking the limit $g\to 1$ and replacing
$\alpha_\W$ by $\alpha_\T$
\begin{eqnarray}
\mathcal{R}_{\T}(\tau) & =& \mathcal{R}^{\varphi
\varphi}_\text{reg}(0;0;\tau) = \left. \mathcal{R}^{\Phi_+
\Phi_+}_\text{reg}(0;0;\tau)\right|_{\substack{g \rightarrow 1
\\ \alpha_\W \rightarrow \alpha_\T}} = -\frac{1}{8 \pi}
\ln{ \frac{\alpha_{\T}^2+\tau^2}{\alpha_{\T}^2} }\label{R_T} \\
&& \nonumber\\
\mathcal{I}^{\eta_1 \eta_2}_{\T}(\tau) & =& {\rm
F}_{\T}^{\eta_1\eta_2}(\tau) \, \mathcal{I}^{\varphi
\varphi}(0;0;\tau)
= \left. {\rm F}_{\T}^{\eta_1\eta_2}(\tau)\, \mathcal{I}^{\Phi_+ \Phi_+}(0;0;\tau)\right|_{\substack{g \rightarrow 1\\
\alpha_\W \rightarrow \alpha_\T}} = -\frac{\eta_2 \theta(\tau)-\eta_1 \theta(-\tau)}{4\pi} \arctan\left( \frac{\tau}{\alpha_\text{T}} \right) \, . \label{I_T}
\end{eqnarray}}

\end{widetext}


\begin{thebibliography}{99}
\bibitem{nota-short-range} Due to the effectively short-ranged interaction induced by the gate, a Wigner crystal regime does not arise in the situation discussed here.
\bibitem{kane:1992} Note that the renormalization exhibits the same scaling exponent as in the homogeneous Luttinger liquid model (see \emph{e.g.} C. L. Kane, and M. P. A. Fisher, Phys. Rev. B \textbf{46}, 15233
(1992)). In the inhomogeneous Luttinger liquid model used here, however, the renormalization flow driven by the relevant impurity backscattering operator is cut not only by a finite temperature $T$ or bias $V$, but also by the finite length of the wire $L$. For this reason our perturbative treatment is safe also at $V=0$ and $T=0$.
\bibitem{NOTA-G2L} One can prove that only impurity backward scattering contributes to order $\gamma^2 \lambda$.
\end{thebibliography}
\end{document}